\documentclass{emulateapj}

\usepackage{graphicx}
\usepackage{txfonts}
\usepackage{natbib}
\usepackage[dvipdfm,
pdfstartview=FitH,
CJKbookmarks=true,
bookmarksnumbered=true,
bookmarksopen=true,
colorlinks,
pdfborder=001,
linkcolor=blue,
anchorcolor=blue,
citecolor=blue
]{hyperref}

\begin{document}


\title{Searching for bulk motions in the intracluster medium of massive, merging clusters with {\sl Chandra} CCD data}


\author{Ang Liu\altaffilmark{1,2}, Heng Yu\altaffilmark{1,3}, Paolo Tozzi\altaffilmark{4,1} and Zong-Hong
Zhu\altaffilmark{1}}


\altaffiltext{1}{Department of Astronomy, Beijing Normal University,
    Beijing 100875, China; yuheng@bnu.edu.cn, zhuzh@bnu.edu.cn}
\altaffiltext{2}{Homer L. Dodge Department of Physics and Astronomy, University of Oklahoma, Norman, OK 73019, USA}
\altaffiltext{3}{Dipartimento di Fisica, Universit\'a di Torino, Via P. Giuria 1, I-10125 Torino, Italy}
\altaffiltext{4}{INAF -- Osservatorio Astrofisico di Arcetri, Largo E. Fermi, I-50122 Firenze, Italy}


\begin{abstract}

We search for bulk motions in the intracluster medium (ICM) of massive clusters showing evidence of
an ongoing or recent major merger with spatially resolved spectroscopy in {\sl Chandra} CCD
data. We identify a sample of 6 merging clusters with $>$150 ks {\sl Chandra} exposure in the
redshift range $0.1 < z < 0.3$. By performing X-ray spectral analysis of
projected ICM regions selected according to their surface brightness,
we obtain the projected redshift maps for all of these clusters. After performing a robust analysis
of the statistical and systematic uncertainties in the measured X-ray redshift $z_{\rm X}$,
we check whether or not the global $z_{\rm X}$ distribution differs from that expected when the ICM is at rest.
We find evidence of significant bulk motions at more than 3$\sigma$
in A2142 and A115, and less than 2$\sigma$ in A2034 and A520.
Focusing on single regions, we identify significant localized velocity differences in all of the merging clusters.
We also perform the same analysis on two relaxed clusters
with no signatures of recent mergers, finding no signs of bulk motions, as expected.
Our results indicate that deep {\sl Chandra} CCD data enable us to identify the presence of bulk
motions at the level of $v_{\rm BM} >$ 1000\ ${\rm km\ s^{-1}}$ in the ICM of massive merging clusters
at $0.1<z<0.3$. Although the CCD spectral resolution is not sufficient for
a detailed analysis of the ICM dynamics, {\sl Chandra} CCD data
constitute a key diagnostic tool complementing X-ray
bolometers on board future X-ray missions.

\end{abstract}



\section{Introduction}

Clusters of galaxies are the largest virialized systems in the universe, and the strong interplay between
their baryonic and dark components is such that they are at the crossroads between
astrophysics and cosmology.  It is widely accepted that clusters form in a
hierarchical manner from a primordial density perturbation field, and evolve through the accretion and
merging of other virialized halos. In particular, merging between halos frequently happens at
early epochs and still occurs today.  Such merging processes strongly affect the dynamic structure
of clusters, often producing bulk motions in the hot intracluster medium (hereafter ICM).
On one hand, ICM bulk motions could potentially provide an important diagnostic of the formation and
evolution of clusters. On the other hand, the ICM velocity field is extremely difficult to measure with our
present facilities, and, more importantly, since measurements of the cluster properties rely on the
assumption of hydrostatic equilibrium, unnoticed bulk motions may significantly affect observables
such as total mass proxies and baryon mass fractions \citep{nagai2013,Nelson2014}.
Consequently, this will also affects studies of the formation of large-scale structure and constraints
on cosmological parameters \citep[see][for a review]{allen2011}.

Bremsstrahlung emission from the optically thin, hot ICM is observed in the X-ray band,
and it constitutes
a powerful diagnostics of the thermodynamics of the ICM itself, allowing one to measure the temperature
and density of the emitting gas. In addition, thanks to the almost perfect collisional equilibrium, the
measurement of the line emission from highly ionized heavy elements constitutes a direct probe of
the metal content of the ICM.  Merging processes can introduce several clear signatures in the ICM,
such as the irregular distribution of the X-ray surface brightness, the inhomogeneous distribution
of temperatures, and fluctuations
in the ICM velocity field, as already mentioned. These signatures last for one or more
dynamical times until the relaxation processes bring the ICM back to hydrostatic equilibrium.
The surface brightness distribution and temperature map of the ICM of massive clusters is routinely
obtained from X-ray imaging and spatially resolved spectroscopy.  The direct observation
of bulk motion instead presents several difficulties. In some rare cases, the bulk motion velocities
can be estimated indirectly via the properties of the bow shock resulting from a tangential
merging process, as in the well-known cases of the Bullet cluster \citep{markevitch2002},
A2146 \citep{russell2011}, A520 \citep{markevitch2005}, and a few others. The turbulence expected
as a result of the stirring of the ICM during the merger process can be investigated by
Doppler line broadening in the high-resolution spectra obtained with the Reflection Grating Spectrometer on board
{\sl XMM-Newton}.  However, only upper limits in the range 200--600\ ${\rm km\ s^{-1}}$ have been obtained,
and only for the central regions \citep{2010Sanders,2015Pinto}.  Alternatively, the power spectrum of the velocities can be inferred
from the power spectrum of the gas density fluctuations \citep{2014Zhuravleva}.

A straightforward approach is to directly measure the ICM velocity field along the line of sight
through the Doppler shift of its emission lines.  The most prominent emission line of the ICM X-ray
spectra is the ubiquitous K-shell line complex of H-like and He-like iron
in the 6.7--6.9 keV rest frame, originally detected by \citet{mitchell1976}. Many other emission lines,
including the L-shell
from iron and $\alpha$-element transitions, may also be found, particularly in the soft energy range
(0.5--2.0 keV).  However, the ${\rm K}\alpha$ iron complex is the only one which is detectable in any
cluster observed
with a total number of net counts as low as $\sim 1000$ in the entire X-ray band \citep[see][]{yu2011}.
In fact, iron line emission has been identified in X-ray clusters as distant as $z\sim1.5$
\citep{rosati2004,rosati2009,stanford2005,tozzi2013},
although the detection may be challenging at $z>1.5$ \citep[see, e.g.,][]{tozzi2015}.

Despite this, applying this method to the measurement of ICM bulk motions is extremely
challenging.  The typical gas velocity of 1000\ ${\rm km\ s^{-1}}$ \citep{nagai2003} expected in major
mergers corresponds only to a $\sim$20 eV shift of the ${\rm K}\alpha$ iron line complex in the 6.7--6.9 keV rest
frame.  Such a shift is below the CCD energy resolution of {\sl XMM-Newton}/EPIC, {\sl Chandra}/ACIS,
and {\sl Suzaku}/XIS
($\sim$100 eV). This constitutes a major obstacle to the direct measurement of the line-of-sight
ICM velocity field.  Despite these difficulties, there are a few works that have successfully achieved detections
of bulk motions in massive, nearby clusters using spatially resolved spectral analysis of CCD data.
The first successful detection was made in the Centaurus cluster with {\sl ASCA} \citep{dupke2001a}.
The typical bulk motion velocity was constrained to $(2.4 \pm 0.1)\times 10^3\ {\rm km\ s^{-1}}$, and was
eventually confirmed by {\sl Chandra} data \citep{dupke2006}. Another case is the Perseus cluster, which
was also found to present ICM bulk motions by {\ ASCA} \citep{dupke2001b}.
The first systematic search for ICM bulk motions was performed by \citet{dupke2005}
with {\sl ASCA}, but only  2 out of 12 low-redshift ($z<0.13$) clusters showed reliable signatures of bulk motions.
In addition, a significant velocity difference of $(5.9 \pm 1.6) \times 10^3\ {\rm km\ s^{-1}}$
between two regions of Abell 576 was found by combining {\sl Chandra} and {\sl XMM-Newton}
data \citep{dupke2007}.

Thanks to its lower background at high energies, {\sl Suzaku}/XIS has been used to search for radial bulk
motions in the ICM through X-ray redshift measurements. However, only upper limits of the order of
$2000\ {\rm km\ s^{-1}}$ are derived for AWM7, Abell 2319, and Coma \citep[][respectively]{sato2008,sato2011,sugawara2009}, and
$\sim$300\ ${\rm km\ s^{-1}}$ for Centaurus \citep{tamura2014}.
Only in one case \citep[Abell 2256;][]{tamura2011} has
a robust detection of  a bulk motion of $1500 \pm 300\ {\rm km\ s^{-1}}$ has been found.
Recently, \citet{2015Ota} searched for gas bulk motion in eight nearby clusters with {\sl Suzaku},
finding signs of large bulk velocity in excess of their uncertainties in only two of them
(Abell 2029 and Abell 2255).  Despite the low background of {\sl Suzaku}/XIS,
its low angular resolution is a major limiting factor for this investigation. Surprisingly, {\sl Chandra ACIS}
has not been used often for these studies, despite its high angular resolution which allows one to
focus on smaller regions and, in principle, to detect more efficiently the effects of bulk motions.

Recently, we proposed a strategy to search for bulk motion in the ICM along the line of sight by
performing spatially resolved spectral analysis
in {\sl Chandra} CCD data of the so-called Bullet cluster 1E0657-56 \citep{aliu2015}.
This case turned out to be particularly challenging due to the extremely high
temperatures, which make the iron emission line complex less prominent, and the fact that the merger
is occurring in the plane of the sky.  Nevertheless, we were able to find tantalizing $\sim 2 \sigma$
evidence for a bulk motion of the significant mass of ICM in two regions near the center.
This is consistent with the picture where the ICM is pushed perpendicularly to the bullet trail due to
the extremely large pressure reached when crossing the main cluster center.
Future missions carrying X-ray bolometers \citep[e.g., Astro-H;][]{2014Takahashi} may be able to
confirm or rule out this result \citep[see the recent simulations by][]{2013Biffi}.

In this work, we intend to apply our technique to a sample of massive, merging clusters
imaged by {\sl Chandra}. Our goal is to investigate whether ICM bulk motions in medium-redshift
(z $\sim$ 0.1--0.3) clusters classified as major mergers can be detected in {\sl Chandra}
ACIS data.  We also try to provide robust estimators of the ICM bulk motions whenever
possible, such as the global rms velocity difference across the cluster $v_{\rm BM}$ and the
maximum velocity difference measured across the cluster $\Delta v_{\rm max}$. We also aim to
provide a qualitative description of the dynamical status of the ICM in single clusters through visual
inspection of the redshift map, in order to exploit the angular resolution of {\sl Chandra}.

The paper is organized as follows.   In Section \ref{sec2}, we describe the sample selection.  In Section
\ref{sec3}, we describe the X-ray data reduction and analysis, including a brief description of the strategy already
presented in \citet{aliu2015}.  In Section \ref{sec4}, we present results for the single clusters, while in
Section \ref{sec5} we synthesize the results for the entire sample.  In Section \ref{sec6}, we discuss
future prospects for improvements in this field.  Finally, in Section \ref{sec7}, we summarize our conclusions.
Throughout this paper, we adopt the 7 years {\sl WMAP} cosmology, with $\rm \Omega_{m}$= 0.272,
$\rm \Omega_{\Lambda}$ = 0.728, and $H_{0}$ = 70.4 km\ $\rm s^{-1}$\ $\rm Mpc^{-1}$
\citep{komatsu2011}.  Quoted error bars always correspond to a 1$\sigma$ confidence level.

\section{Sample selection}
\label{sec2}

\begin{deluxetable*}{cccc}
\tablewidth{\textwidth}
\tablecaption{Sample of merging clusters and corresponding {\sl Chandra} observations used in this
work.   }
\tablehead{
\colhead{Cluster Name}   & \colhead{Optical Redshift}    & \colhead{ObsID} & \colhead{Total Exposure (ks)}
}
\startdata
A2142 & 0.090(1)       &  ACIS-S: 15186-16564-16565 & 155.1 \\
\hline
A2034 & 0.113(2) & ACIS-I:  7695-2204-12885           & 254.5 \\
    &           &      13192-12886-13193      &        \\
\hline
A115 & 0.197(3) & ACIS-I: 13458-13459-15578-15581 & 310.6 \\
\hline
A520 & 0.203(4) & ACIS-I: 4215-9424-9425-9426-9430 & 516.4 \\
\hline
1RXS\ J0603.3+4214 & 0.225(5) & ACIS-I:15171-15172-15323 & 235.9 \\
\hline
A2146 & 0.234(6)  & ACIS-I: 12245-12246-12247-13020  & 375.3 \quad  \\
          &              & 13021-13023-13120-13138               & \quad  \\
\hline
\enddata
\tablenote{The observing mode is VFAINT for all the ObsID. The exposure times in the fourth column correspond to the effective values after data reduction. Optical redshift references: (1)\citet{Parekh2015},
(2)\citet{muriel2014}, (3)\citet{hiroi2013}, (4)\citet{cassano2013}, (5)\citet{weeren2012},
(6)\citet{piffaretti2011}.}
\label{tabsample}
\end{deluxetable*}

\begin{deluxetable*}{cccc}
\tablewidth{\textwidth}
\tablecaption{Relaxed clusters and corresponding {\sl Chandra} observations used in this
work.  }
\tablehead{
\colhead{Cluster Name}   & \colhead{Optical Redshift}    & \colhead{ObsID} & \colhead{Total Exposure (ks)}
}
\startdata
A1689 & 0.183(7) & ACIS-I: 5004-6930-7289-7701 & 151.3 \\
\hline
A1835 & 0.252(8) & ACIS-I: 6880-6881-7370 & 193.7
\enddata
\tablenote{The observing mode is VFAINT for all the ObsID. The exposure times in the fourth column correspond
to the effective values after data reduction.  Optical redshift reference: (7)\citet{hiroi2013},
(8)\citet{girardi2014}.}
\label{tabsample2}
\end{deluxetable*}

We search the {\sl Chandra} data archive for clusters classified as mergers based on their X-ray
morphology, radio diffuse emission, and/or galaxy dynamics, in the redshift range $0.1<z<0.3$.
Since our goal is to perform spatially resolved spectral analyses with high signal-to-noise ratios (S/Ns),
we require long exposure times.  We fix the threshold for the minimum total exposure to $150\ {\rm ks}$ on ACIS-I
or ACIS-S.  The signatures of an occurring, or impending, major merger,
are a clear bimodal or multiple stucture in the redshift distribution of the member galaxies,
a strongly disturbed appearance in the X-ray surface brightness,
or the presence of a radio halo or radio relic.  We found six clusters showing at least one of these
prominent properties.  The clusters are listed in Table \ref{tabsample} in order of increasing
optical redshift, together with {\sl Chandra} Obsid and total exposure time (after data reduction).

A2142 is classified as a merger on the basis of its X-ray morphology \citep{Parekh2015} and the
presence of a radio halo \citep{Cuciti2015}.  A2034 is classified as a merger on the basis of a shock
visible in the X-ray surface brightness \citep{owers2014}.
Abell 115 has a clear bimodal appearance in the X-ray and has a radio relic \citep{Cuciti2015}.  A520
shows a clear radio halo \citep{Cuciti2015} and a clearly disturbed appearance in the X-ray band.
1RXS\ J0603.3+4214 has a clear bimodal structure and a bright, peculiar radio relic,
known as the ``Toothbrush" \citep{weeren2012}.
A2146 shows a prominent bow shock in the {\sl Chandra} X-ray image, indicating a violent merger analogous to
the more famous Bullet cluster 1E0657-56, which was the subject of our first investigation
of ICM bulk motion \citep{aliu2015}.  The properties of each cluster will be discussed in more details
in Section 4.

As an immediate check of the robustness of our analysis, we also select two
relaxed clusters in the same redshift range and with comparable exposure depth.
The two clusters in this control sample are Abell 1689 and Abell 1835, and are listed in Table \ref{tabsample2}.
Clearly, we do not expect to detect any
departure from statistical noise in the $z_{\rm X}$ distribution of these relaxed clusters.

\section{Data reduction and analysis}
\label{sec3}

\subsection{Data reduction}

The {\sl Chandra} observations used in this work are listed in Table \ref{tabsample} and \ref{tabsample2}.
All of the observations were taken in VFAINT mode with ACIS-I, except in the case of A2142 which
was observed with ACIS-S.
The data are reduced using the {\sl ciao} software (version 4.7) with CALDB 4.6.8.
The detailed data reduction procedure has been described in \citet{aliu2015}.
The area of the clusters used in this analysis is defined as the circular region which provides the
maximum S/N in the 0.5--10 keV band once the unresolved sources are removed.
Point sources are detected by running the {\sl wavdetect} task and are then visually inspected,
in particular, to remove those deeply embedded in the strong ICM emission.
The background files are extracted from a series of circular regions as far as
possible from the cluster area, but still on the solid angle defined by the overlap of all of the Obsids.
To define the cluster subregions, we apply the contour binning technique \citep{sanders2006} within
the maximum S/N circle. Regions are identified simply on the basis of the surface brightness
contours in the 0.5--10 keV band image. To make sure that each region can be used to constrain the
X-ray redshift with comparable accuracy, we set a criterion in terms of S/N in
the 0.5--10 keV band in each region. Clearly, this criterion does not take into account the expected
gradient in the iron abundance, which significantly affects the visibility of the iron emission line complex.
We set this threshold to S/N$>140$, except in the case of Abell 2146 where we choose a weaker
constraint of S/N$>100$ due to the relatively lower number of counts. The S/N threshold and the
number of regions selected in each cluster are listed in Table \ref{regions}. For each region, the spectrum is extracted from the merged
event file after the removal of unresolved sources, while the calibration files
(response matrix files, RMF, and ancillary response files, ARF) are generated independently
for each ObsID and then combined by weighting them by the corresponding exposure times.
In this way, we keep track of all the differences in the ACIS effective area among the different regions in
the detector and among the ObsIDs taken at different epochs.

\begin{deluxetable}{ccc}
\tablewidth{0.45\textwidth}
\tablecaption{The minimum S/N required in each region and the number of regions
selected for spectral analysis in each cluster.}
\tablehead{
\colhead{Cluster Name}   & \colhead{Minimum S/N}   & \colhead{Number of Regions}\\
\colhead{}   & \colhead{(0.5--10 keV)}   &
}
\startdata
A2142 & 140 & 52  \\
A2034 & 140 & 14 \\
A115 & 140 & 18 \\
A520 & 140 & 20 \\
1RXS\ J0603.3+4214 & 140 & 9 \\
A2146 & 100 & 19 \\
A1689 & 140 & 10 \\
A1835 & 140 & 11

\enddata
\label{regions}
\end{deluxetable}

\subsection{Spectral analysis}

In this work, we use the same analysis strategy presented in \citet{aliu2015} to measure the projected
X-ray redshift $z_{\rm X}$ in each region and estimate the total (statistical and systematic)
uncertainty.  Our analysis is optimized on the basis of the spectral simulations shown in \citet{aliu2015}. Here, we
simply recall the relevant aspects of the adopted spectral analysis.

The spectra are analyzed using {\sl Xspec} v12.8.2 \citep{1996Arnaud}.  To successfully model the X-ray emission, we use two {\tt mekal} plasma emission models
\citep{1985Mewe, 1986Mewe,1995Liedahl} that include thermal bremsstrahlung and line emission, with abundances measured relative to the solar values of
\citet{asplund2005} in which Fe/H $ = 3.6 \times 10^{-5}$.  During the fit, the redshift parameters of the two thermal components are always linked together, while the two temperatures and abundances
are left as free parameters.

The main advantage in using a double-temperature thermal spectrum is reducing the possible bias in the measurement of the iron line centroid due to the presence of unnoticed thermal structure
along the line of sight.  As a matter of fact, different temperature components above 3 keV are not detectable, and a single-temperature thermal component would return an excellent fit even in the presence
of widely different temperatures \citep[see][]{2004Mazzotta}.  However, in \citet{aliu2015}, using spectral simulations we showed that the use of two components provides better results for the measurement of the actual
redshift of the ICM, while the temperature structure, which is not our main interest here, is essentially left unconstrained.  Clearly, this also introduces a larger systematic uncertainty associated with the
temperature structure, which is carefully evaluated during our analysis, as explained below.

Galactic absorption is described by the model {\tt tbabs} \citep{2000Wilms}.
The central values of the Galactic HI column density ${\rm NH_{Gal}}$ are based on \citet{2005LAB}.  We conservatively
allow ${\rm NH_{Gal}}$ to vary by an interval of $\sim 10$\% around the central value.
This has a small effects since our best-fit $z_{\rm X}$ are obtained from spectral fits in the
hard (2--10 keV) band, which is only marginally affected by the Galactic absorption.
Cash statistics \citep{1979Cash} are applied to the unbinned source and background
spectra in order to exploit the full spectral resolution of the ACIS-I and ACIS-S CCD.
Cash statistics are preferred for faint spectra with respect to the canonical $\chi^{2}$ analysis of binned
data \citep{1989Nousek}.  To avoid local minima, we repeat the fit several times before and after
running the {\tt steppar} command on
all of the free parameters, particularly on the redshift, with a step of $\delta z = 10^{-4}$.
Finally, the plots of $\Delta C_{\rm stat}$ versus redshift are visually inspected to investigate
whether or not there are other possible minima around the best-fit values \citep[see also][]{yu2011},
which could indicate a noisy spectrum and therefore an unreliable best-fit value for $z_{\rm X}$.  We
decide to conservatively exclude all of the results with a secondary minimum closer than
$\Delta C = 6.6$, which corresponds to a formal 99\% confidence level for one free parameter.

Our reference spectral analysis consists of two steps. As the first step, we fit the spectra in the
2--10 keV energy range to obtain the best-fit redshift. We choose to use only the 2--10 keV energy range to minimize the effects of the presence of low-temperature components on the centroid of the iron line
emission complex. This conservative choice is also supported by the spectral simulations we presented in \citet{aliu2015}. For each region, we produce a plot showing the
$\Delta C_{\rm stat}$ value versus the redshift, obtained by varying the redshift parameter
and marginalizing the fit with respect to the other parameters.  The minimum is generally well defined
and roughly symmetric, and therefore provides a robust estimate of the statistical uncertainty.

As a second step, we carefully take into account the systematic uncertainties on redshift due to the
unknown thermal structure of the ICM.  We conservatively consider all of the possible
temperature distributions of the ICM by setting a four-dimensional grid of spectral parameters,
namely, temperature and abundance of the two thermal components $T_{1}$, $T_{2}$, $Z_{1}$, and $Z_{2}$.
The steps of the grid are typically 0.5 keV for the temperature and 0.05 for the metal abundance.
Clearly, only a subgrid of temperature and abundance values should be taken into account, since some combination of them provides an unacceptable fit.
To efficiently select the subgrid, we make use of the full 0.5--10 keV energy range in order to exploit the complete information on the thermal structure and the chemical composition.
Therefore, for each region, we measure the best-fit redshift in the full 0.5--10 keV band leaving all of the parameters free, and
collect the absolute minimum $C_{\rm min}$. Then, we select all of the sets of values on the grid
that provide a best fit close enough to the absolute best fit.
The criterion we adopt is given by $\Delta C_{\rm stat} \equiv C_{\rm stat}-C_{\rm min} <  4.72$, which
corresponds to a $1 \sigma$ confidence level for four free parameters.
By applying the criterion $\Delta C_{\rm stat} < 4.72$ for the fits performed in the full
0.5-10 keV band, we select a subgrid of  $T_{1}$, $T_{2}$, $Z_{1}$
and $Z_{2}$ values statistically compatible within $1\, \sigma$ with the spectrum observed in each region.
Then, we run the spectral fit on the subgrid, but focusing again in the 2.0--10 keV band only, and measure
a set of best-fit $z_{\rm X}$ values corresponding to the subgrid parameters.
The distribution of $z_{\rm X}$ obtained in this manner is used to derive a redshift range defined
by the upper and lower 90\% percentiles of the $z_{\rm X}$ distribution on the grid, and thus the
systematic uncertainties on redshift.  Since the systematic uncertainty $\sigma_{\rm syst}$ is due to the
degeneracy of the temperature values, we assume that it is uncorrelated to the
statistical error $\sigma_{\rm stat}$.  Therefore, we compute the
total $1 \sigma$ uncertainty as $\sigma_{\rm tot} = \sqrt{\sigma_{\rm stat}^2+\sigma_{\rm syst}^2}$.
We do not consider other relevant sources of systematic errors.  Systematics related to calibration
issues or the time dependence of the gain are kept under control with a direct check on the fluorescent
lines of Au ${\rm L}\alpha$ and Ni ${\rm K}\alpha$, which are prominent in the ACIS-I background spectrum
at 7.5 and 9.7 keV, respectively.   Typically, uncertainties introduced by calibration variation
(as a function of position on the CCD or as a function of the observation period) on the merged
spectrum of each region correspond to errors $\sim 5$ times smaller than the typical total uncertainty
on $z_{\rm X}$ \citep[see][]{aliu2015}.  Having assessed that  gain-calibration does not significantly affect
our conclusions, we do not perform a full treatment of the gain-calibration uncertainties in ACIS-I
data.

Results are presented in the form of redshift maps and significance maps.
In particular, in the significance maps, we combine the information included in the best-fit redshift and
the corresponding total error by showing  $(z_{\rm X} - \langle z_{\rm X}\rangle)/\sigma_{\rm tot}$ ,
where $\langle z_{\rm X}\rangle$  is the average $z_{\rm X}$ found across the analyzed regions for which we
have a reliable spectral fit.  Significant (larger than 2$\sigma$) deviations in the redshift map are
thus visible as bright blue and bright red regions.
The complete spectral results for all of the clusters are also presented in the appendix, where the best-fit redshifts
and the corresponding lower and upper statistical and systematic error bars are listed for each region of
each cluster (values are rounded to the fourth decimal digit).

\subsection{ICM dynamical analysis}

In this section, we describe how we use the results of the spatially resolved spectral analysis to investigate and possibly characterize
the presence of bulk motions in the ICM of each cluster. The first check, analogous to what we did in
\citet{aliu2015}, is to perform a simple $\chi^2$ test on the distribution of $z_{\rm X}$ under the assumption of a unique value for the ICM redshift.
Therefore, we compute $\chi^2 \equiv  \Sigma (z_{\rm X}-\langle z_{\rm X}\rangle )^2/\sigma_{\rm tot}^2$
and, assuming $N_{\rm reg}-1$ degrees of freedom (dof), where $N_{\rm reg}$ is the number of selected regions,
we evaluate the probability of being inconsistent with the constant redshift hypothesis.  To account
for the small differences in the lower and upper error bars, we use the corresponding total error for
$z_{\rm X}$ values lower or higher than $\langle z_{\rm X}\rangle$.

We also compute the typical bulk motion velocity as the excess of the redshift fluctuations
with respect to the statistical and systematic noise.
To do that, we simply compute the total root mean square deviation of $z_{\rm X}$ with respect to the mean,
defined as
$\sigma^2_{\rm rms}\equiv  \Sigma (z_{\rm X}-\langle z_{\rm X}\rangle )^2$, and compare it to the average uncertainty
$\langle \sigma_{\rm tot}\rangle = \Sigma \sigma_{\rm tot} / N_{\rm reg}$ (where we use the average value of
the lower and upper error bars when they differ).  The average excess, which can be ascribed to
bulk motions, is
computed as $\sigma_{\rm BM}^2 = \sigma_{\rm rms}^2 -\langle \sigma_{\rm tot}^2\rangle$.
Therefore, the  bulk motion velocity averaged over the region included in the maximum S/N radius is
finally estimated as $v_{\rm BM} = c \times \sigma_{\rm BM}/(1+\langle z\rangle)$.
The error on $v_{\rm BM}$ is estimated as $\sqrt{2}$ times the total error on the mean.
Moreover, in those cases where there is some evidence of bulk motions in the ICM, we also provide
the value of the maximum velocity difference $\Delta v_{\rm max}$ found across all of the regions with a reliable
spectral fit.

We visually inspect the histogram distribution of $z_{\rm X}$ and the
redshift and significance maps to infer information on the dynamical status of the
ICM.  This was done in \citet{aliu2015} for the Bullet cluster, where
the two regions with a maximally different redshift are found to be aligned along the bullet trail.
We tentatively interpreted them as two regions pushed  in opposite directions and perpendicularly to the
direction of the bullet at velocities $\sim$5--6 $ \times 10^3\ {\rm km\ s^{-1}}$ with respect to the average
cluster redshift.  Clearly, it is not always possible to draw simple pictures like this one.  However,
we try to achieve a simple qualitative classification as pre-merger, ongoing, or post-merger
according to whether we can clearly identify
two well-defined halos with distinct redshift or a few scattered regions with different redshift.
Along this same line, clusters with disturbed surface brightness but no signs of bulk motion may be
described as post-merger, a phase in which the bulk motions created by the merger already
evolved toward a turbulent velocity field with velocity values below the sensitivity achievable in our data.
As previously discussed, the regime of evolved merger, when most of the initial
kinetic energy due to the merger evolved into a turbulent velocity field of a few $\times100\ {\rm km\ s^{-1}}$,
can be probed only with the much higher spectral resolution of X-ray bolometers.

\section{Results on individual clusters}
\label{sec4}

In this section, we show the results of our analysis. For each of the clusters, we measure the best-fit redshift
in all of the selected regions, and hence obtain the redshift map and the significance map as defined in
Section 3.2. We also measure the global redshift $z_{\rm Xtot}$, which is defined as the best-fit redshift of the total emission of the cluster within the maximum S/N circle.
This value corresponds to the emission-weighted average redshift of the ICM, which in principle may differ significantly from the simple average redshift among the regions $\langle z\rangle$.
The statistical significance of the presence of bulk motions is evaluated in each case.  Whenever possible, we estimate the typical velocity of the bulk motion
$v_{\rm BM}$ and the maximum velocity difference $\Delta v_{\rm max}$.
Moreover, for each cluster, we provide a short summary of the available literature concerning their dynamical status.

\subsection{Abell 2142}

Abell 2142 has already been identified  as a merger cluster already on the basis of {\sl ROSAT} observations
\citep{henry1996}, and more recently has been classified as ``non-relaxed"  on the basis
of the X-ray morphology obtained with {\sl Chandra} data by \citet{Parekh2015}.
The two cold fronts, identified in the northwest and southeast regions, clearly separate cooler gas ($ kT \sim $ 7--9 keV)
from hotter \citep[$kT > 12$ keV;][]{markevitch2000}, making it a promising
target to investigate the onset of Kelvin-Helmholtz instabilities at the cold fronts.   Evidence of dynamical
substructure was also found with optical spectroscopy of the  member galaxies \citep{oegerle1995}.
Interestingly, the
presence of a narrow-angle tailed radio galaxy may be explained by significant bulk motions in the ICM,
as suggested by the statistical analysis of \citet{bilton1998}. Finally, \citet{Cuciti2015} reported A2142 to have a radio halo.

We use the three most recent observation with {\sl Chandra} ACIS-S for a total of
$155.1\ {\rm ks}$, as listed in Table \ref{tabsample}, while we discard
two older observations with ACIS-I (ObsID 5005 and 7692).  This choice will minimize the difference in
calibration between ObsID.  The total number of net counts in the 0.5--10 keV band
within a circle of 3 arcmin in the merged image is about $10^6$.
This makes A2142 an ideal target for our analysis, with 52 potentially useful regions for
spatially resolved spectral analysis.  The best-fit redshift obtained by fitting the total X-ray emission within
$3$ arcmin is $z_{\rm Xtot} = 0.0852_{-0.0009}^{+0.0012}$.  To evaluate the systematic uncertainty on $z_{\rm X}$,
we consider a temperature range from 3 to 16 keV to explore the temperature structure of the
ICM, despite the fact that the projected temperature
shows only a moderate range from 8 to 12 keV. Metal abundances are mildly scattered around
$Z = 0.45 Z_\odot$ in the units provided by \citet{asplund2005}, nevertheless we consider a wide range
$0.15 < Z/Z_\odot <1.5$ in order to account for possible large fluctuations.
In 5 regions, we find that the X-ray spectral analysis does not provide a reliable value of $z_{\rm X}$,
and therefore we use only 47 regions.  The best-fit $z_{\rm X}$ of the regions used in our analysis
with the corresponding statistical and systematic errors are listed in Table
\ref{table:a2142}.  The average redshift is $\langle z_{\rm X} \rangle = 0.0855$, which is
almost coincident with the global, emission-weighted redshift $z_{\rm Xtot}$.

\begin{figure*}
\centering
\includegraphics[width=7.8cm]{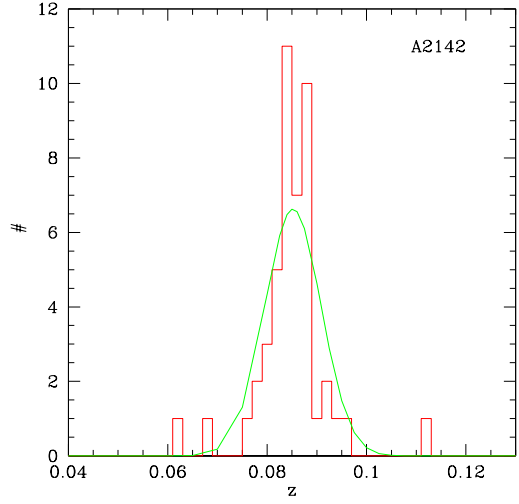}
\caption{Histogram distribution (solid red line) of the best-fit $z_{\rm X}$ for the 47 regions of Abell 2142
with reliable spectral fit.
The  green line is the Gaussian with  $\sigma_G = \langle \sigma_{\rm tot}\rangle$ centered on
$\langle z_{\rm X}\rangle$.}
\label{histo_A2142}
\vfill
\end{figure*}

\begin{figure*}
\centering
\includegraphics[width=0.4\textwidth]{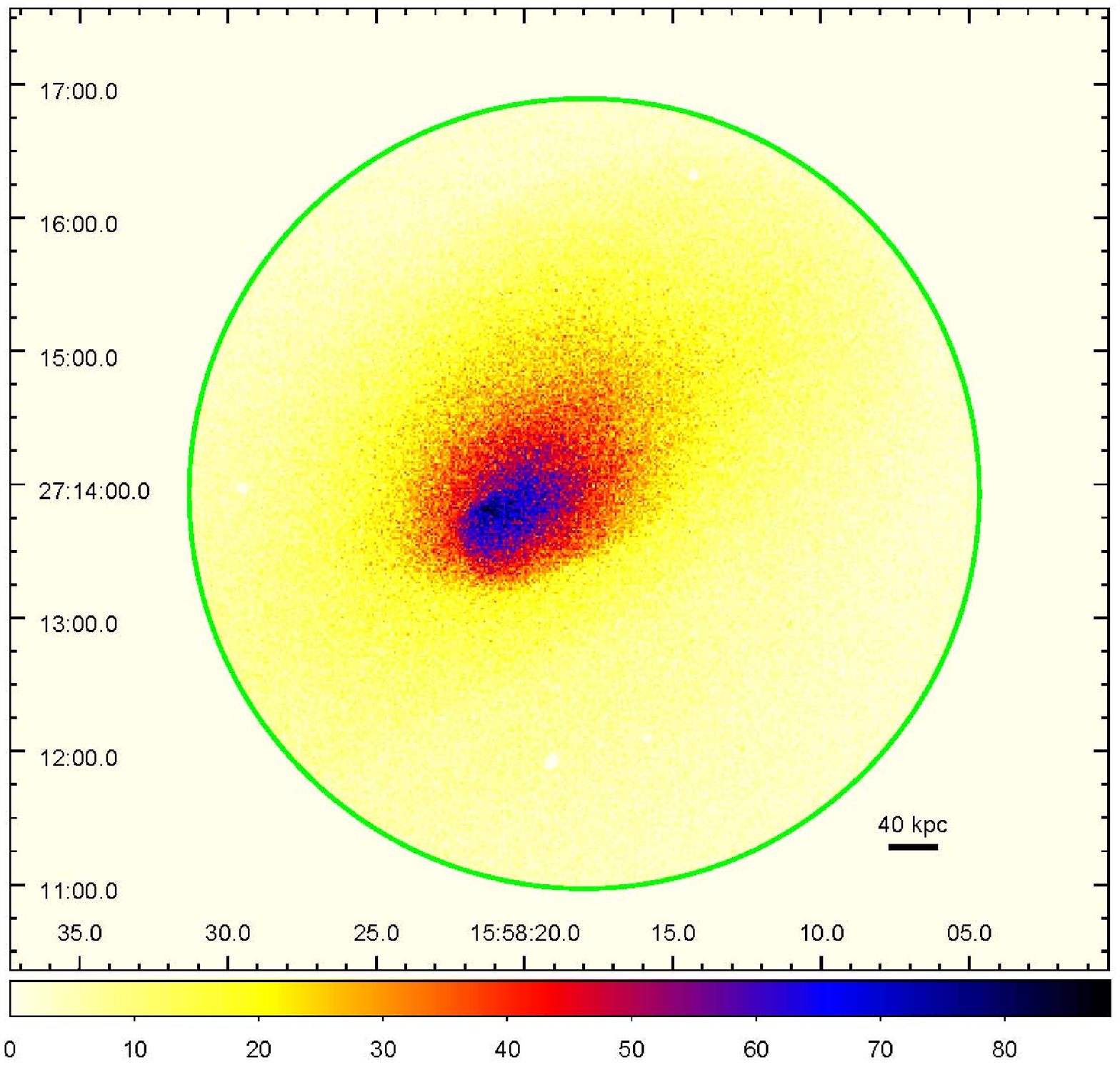}
\includegraphics[width=0.4\textwidth]{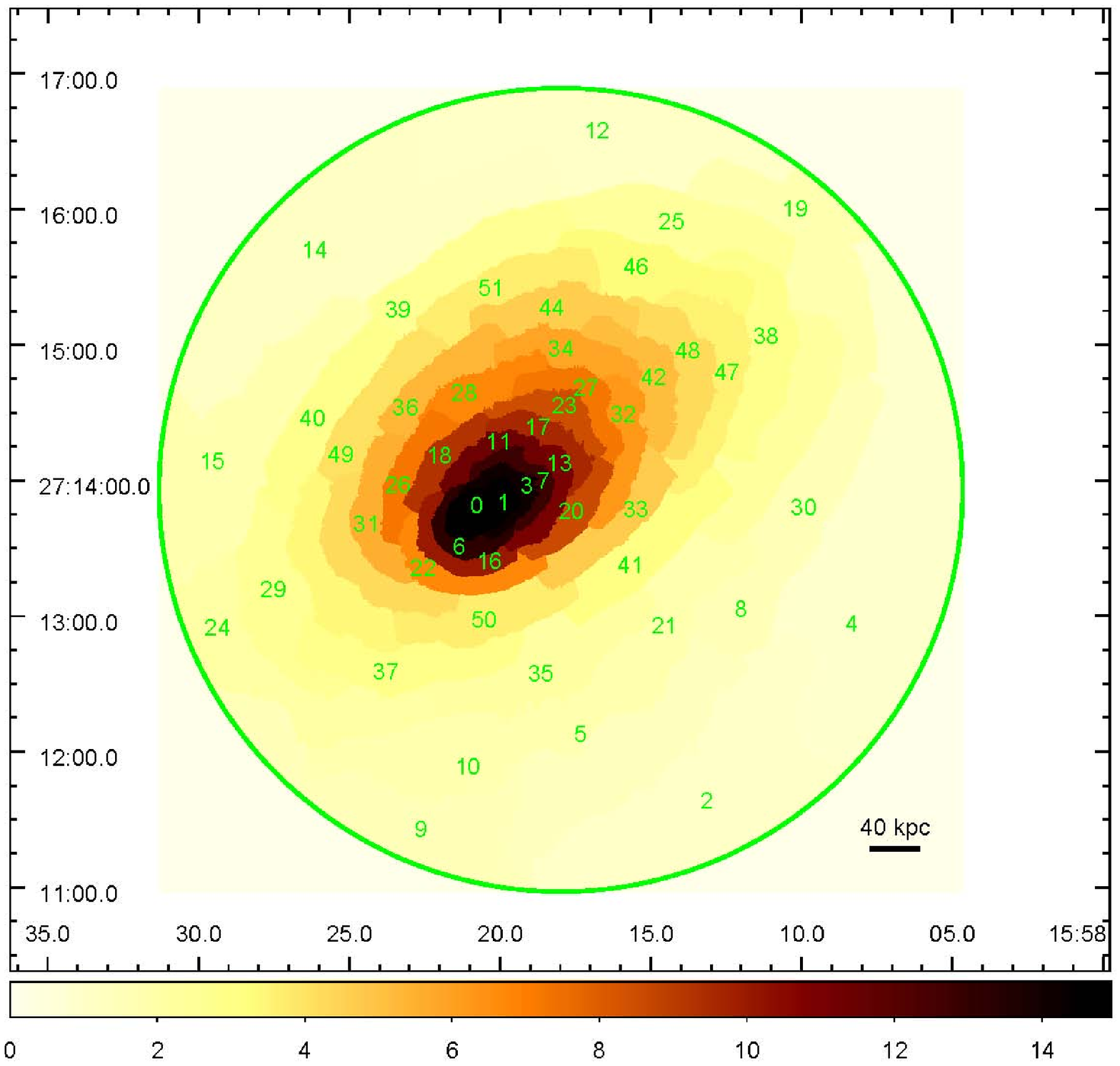}
\includegraphics[width=0.4\textwidth]{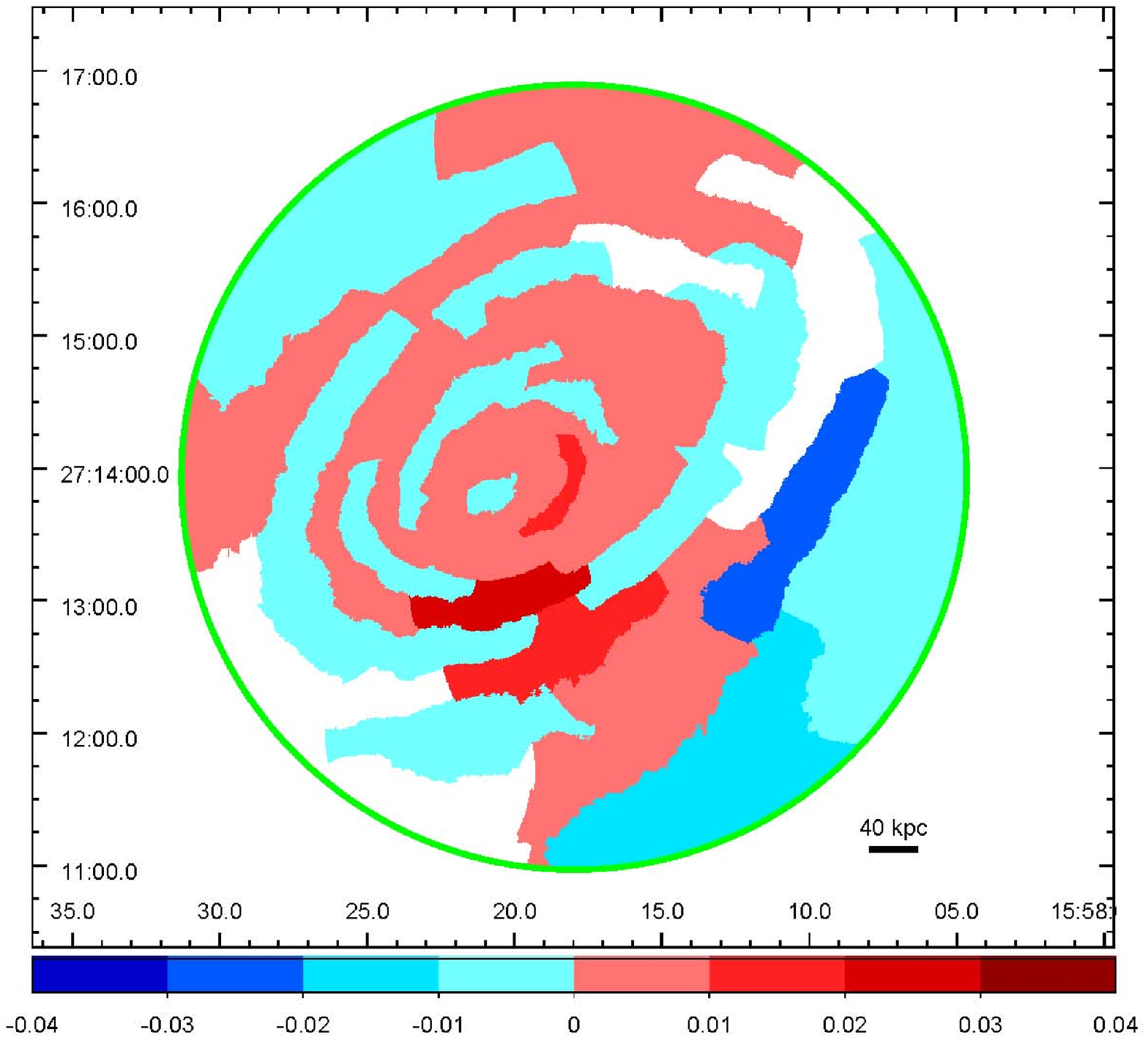}
\includegraphics[width=0.4\textwidth]{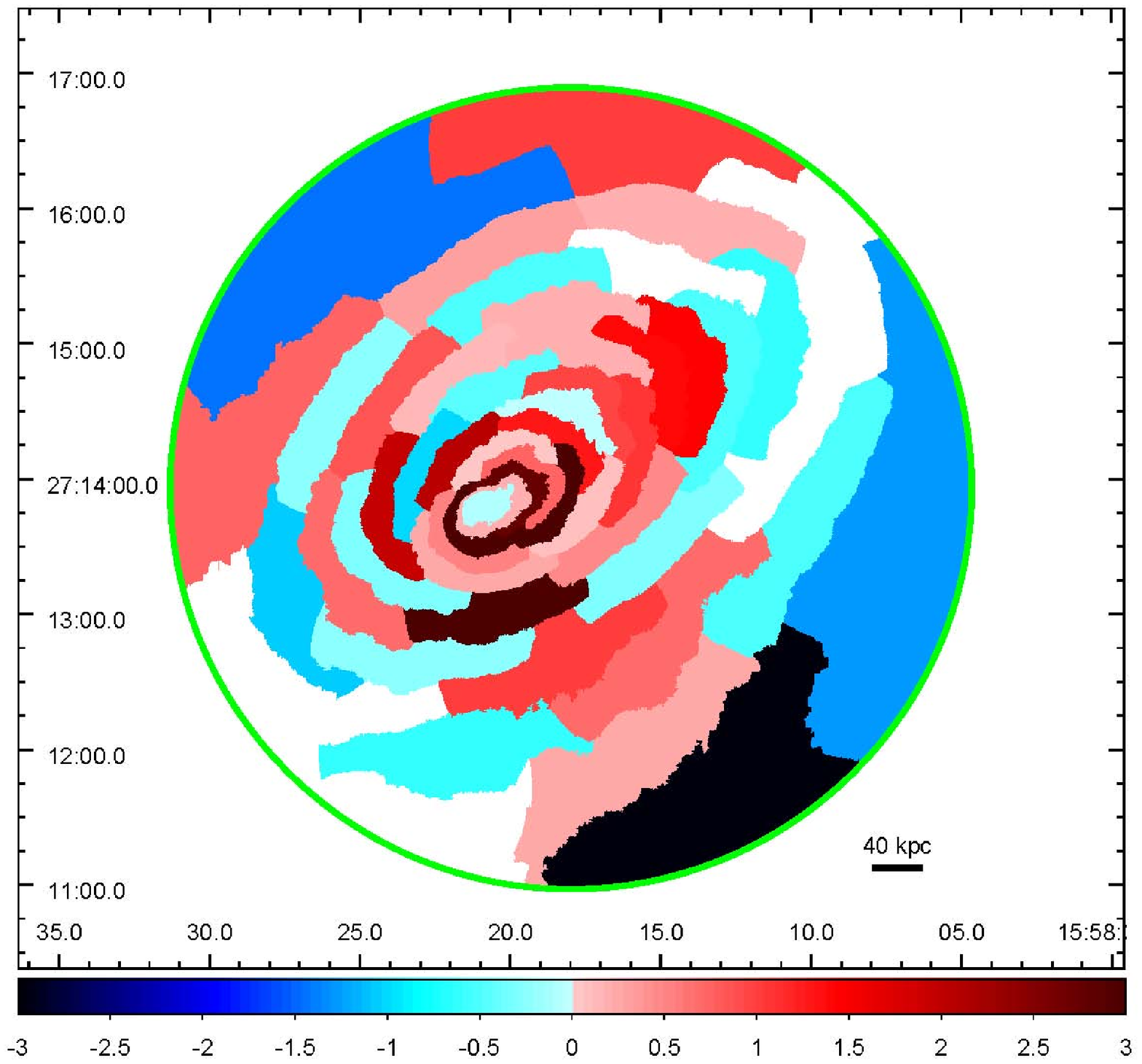}
\caption{Top left panel: surface brightness distribution of Abell 2142 in the 0.5--10 keV band; top right panel: region
map; bottom left panel: redshift map; bottom right panel: significance map. White regions do not have a reliable redshift measurement.}
\label{a2142}
\vfill
\end{figure*}

The hypothesis of a constant redshift is rejected at more than  3$\sigma$, with a
$\chi^2 = 88.02$ for 46 dof, as listed in Table \ref{chi2}.  However, the histogram
distribution of $z_{\rm X}$ (see Figure \ref{histo_A2142}) clearly shows that the evidence of bulk motions comes
from three regions (region 2, 8, and 50) maximally distant from the average redshift.
The rest of the $z_{\rm X}$ values
are distributed according to a Gaussian slightly narrower than the Gaussian whose variance is the average
$\sigma_{\rm tot}^2$.  This is expected since we assumed a very conservative estimate for $\sigma_{\rm tot}$.
The projected X-ray redshift map and significance map are shown in Figure \ref{a2142}.
We measure $v_{\rm BM} = 1400\pm 300\ {\rm km\ s^{-1}}$ for the average bulk motion velocity in the ICM of A2142
within a radius of 3 arcmin.  Excluding the three extreme $z_{\rm X}$ regions, we are able to establish a
hard upper limit of $v_{\rm BM}< 1450 \ {\rm km\ s^{-1}}$.  This result is consistent with the {\sl Suzaku} X-ray
spectral analysis by \citet{2015Ota}, which provided $\Delta v < 4200 \ {\rm km\ s^{-1}}$.  However, thanks to the
{\sl Chandra} angular resolution, we find that region 50 is redshifted with respect to the average
velocity by  $5000 \pm 1500\ {\rm km\ s^{-1}}$, while the contiguous regions 2 and 8 are blueshifted by
$7400 \pm 2200\ {\rm km\ s^{-1}}$.  In particular, we can interpret region 50 as being dominated by a
significant amount of ICM mass pushed at high velocity as a result of the ongoing merger.
The two blueshifted regions can be tentatively associated with some rotation of the ICM.
We note that, given the small field of view of the ACIS-S detector, we repeated the spectral analysis
with a different choice of background, finding the same statistical results.  Therefore, our
conclusions are not affected by uncertainties
in the background, despite the fact that the cluster emission covers almost the entire ACIS-S field of view.

To summarize, as in the case of the Bullet cluster \citep{aliu2015}, the evidence of bulk motions is
coming from a few regions with an extreme redshift difference with respect to the mean after a very
conservative selection of reliable spectra and an accurate estimate of the total uncertainty on $z_{\rm X}$.
This may suggest that the cluster is a post-merger or caught in a stage where the shocks have
developed since $\sim 1$ dynamical time or more and the velocity field of the ICM still shows
a few regions with large velocities with respect to the velocity of the center of mass.
The interpretation in terms of ICM dynamics is nevertheless still very hard, and a significantly higher
spectral resolution with the same angular resolution is needed before the actual velocity
field of the ICM can be accurately measured.

\subsection{Abell 2034}

Abell 2034, at redshift $z_{\rm opt} = 0.113$ \citep{muriel2014}, shows many signatures typically
associated with merging clusters, from a radio relic near the position of a discontinuity in the X-ray
surface brightness, to significant heating of the ICM above its equilibrium temperature, and a
significant offset of the cD galaxy from
the centroid of the X-ray emission \citep{kempner2001,kempner2003,2009Giovannini,2011VanWeeren}.
In a recent detailed study by \citet{owers2014} using a deep {\sl Chandra} observation of $250\ {\rm ks}$, a shock
with a Mach number of $1.59 \pm 0.06$, corresponding to a shock velocity of $\sim$2000\ ${\rm km\ s^{-1}}$, was
clearly observed. Combining {\sl Chandra} data with spectroscopic observations for 328 spectroscopically
confirmed  member galaxies, they show the presence of a substructure located at the front edge of the
shock and that the merger is proceeding along a north--south direction with an inclination angle of
$\sim 23^{\circ}$ with respect to the plane of the sky.  The cluster is included in the
sample of the Merging Cluster
Collaboration.\footnote{http://www.mergingclustercollaboration.org/}

\begin{figure*}
\centering
\epsscale{1}
\includegraphics[width=7.8cm]{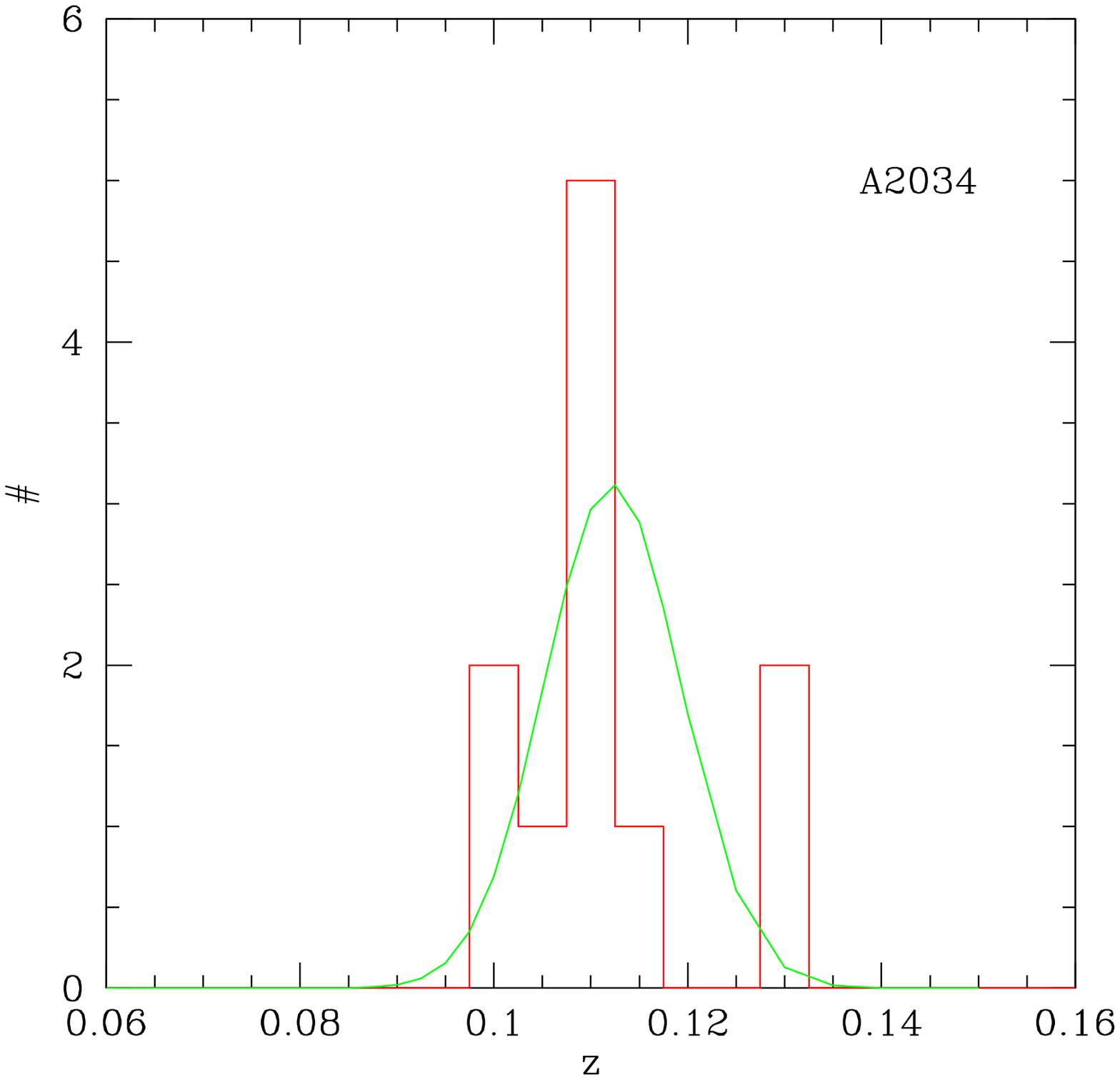}
\caption{Histogram distribution of the best-fit $z_{\rm X}$ for the 11 regions of Abell 2034
with reliable spectral fit.  Lines are as in Figure \ref{histo_A2142}. }
\label{histo_A2034}
\vfill
\end{figure*}

\begin{figure*}
\centering
\includegraphics[width=0.4\textwidth]{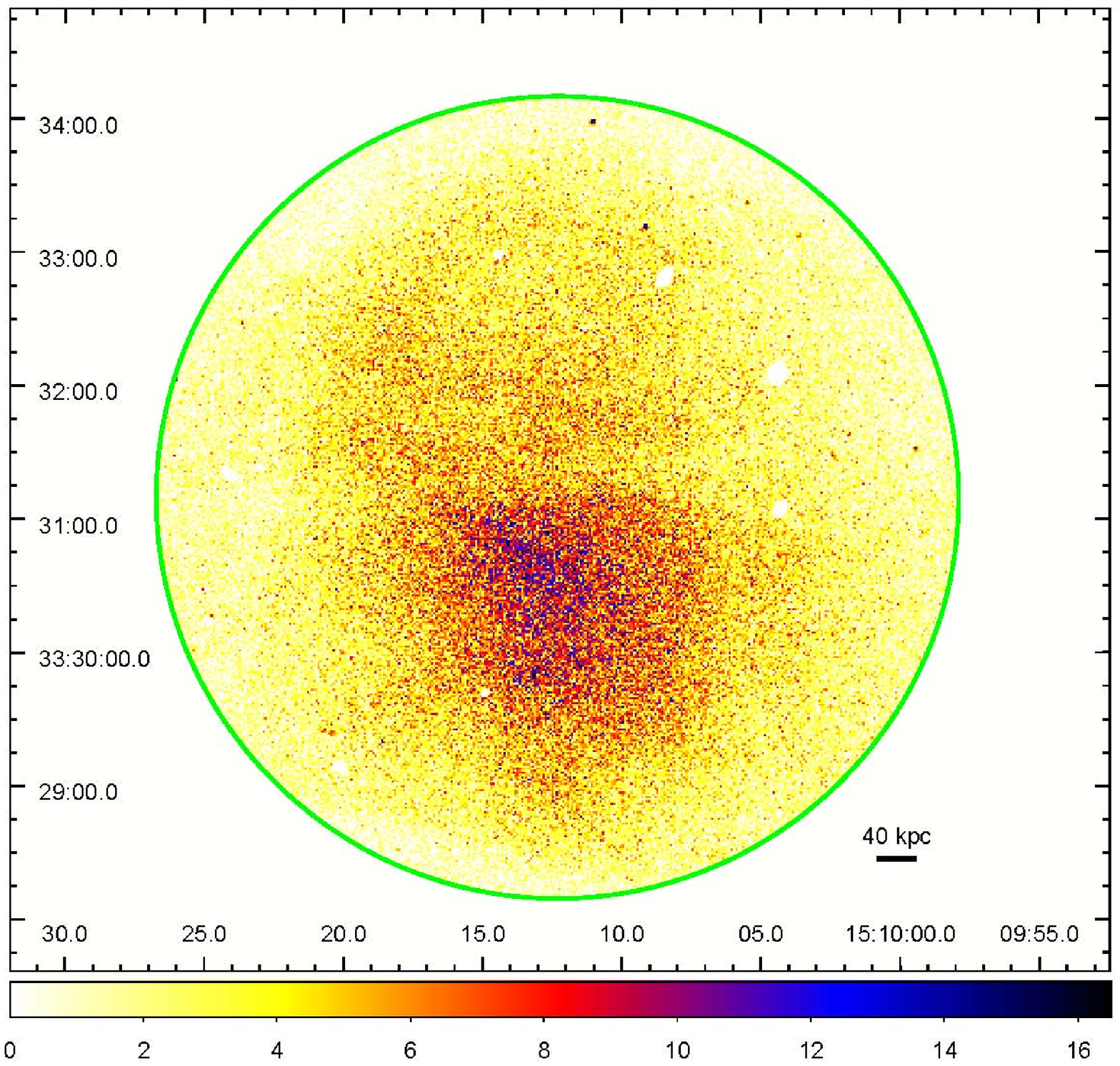}
\includegraphics[width=0.4\textwidth]{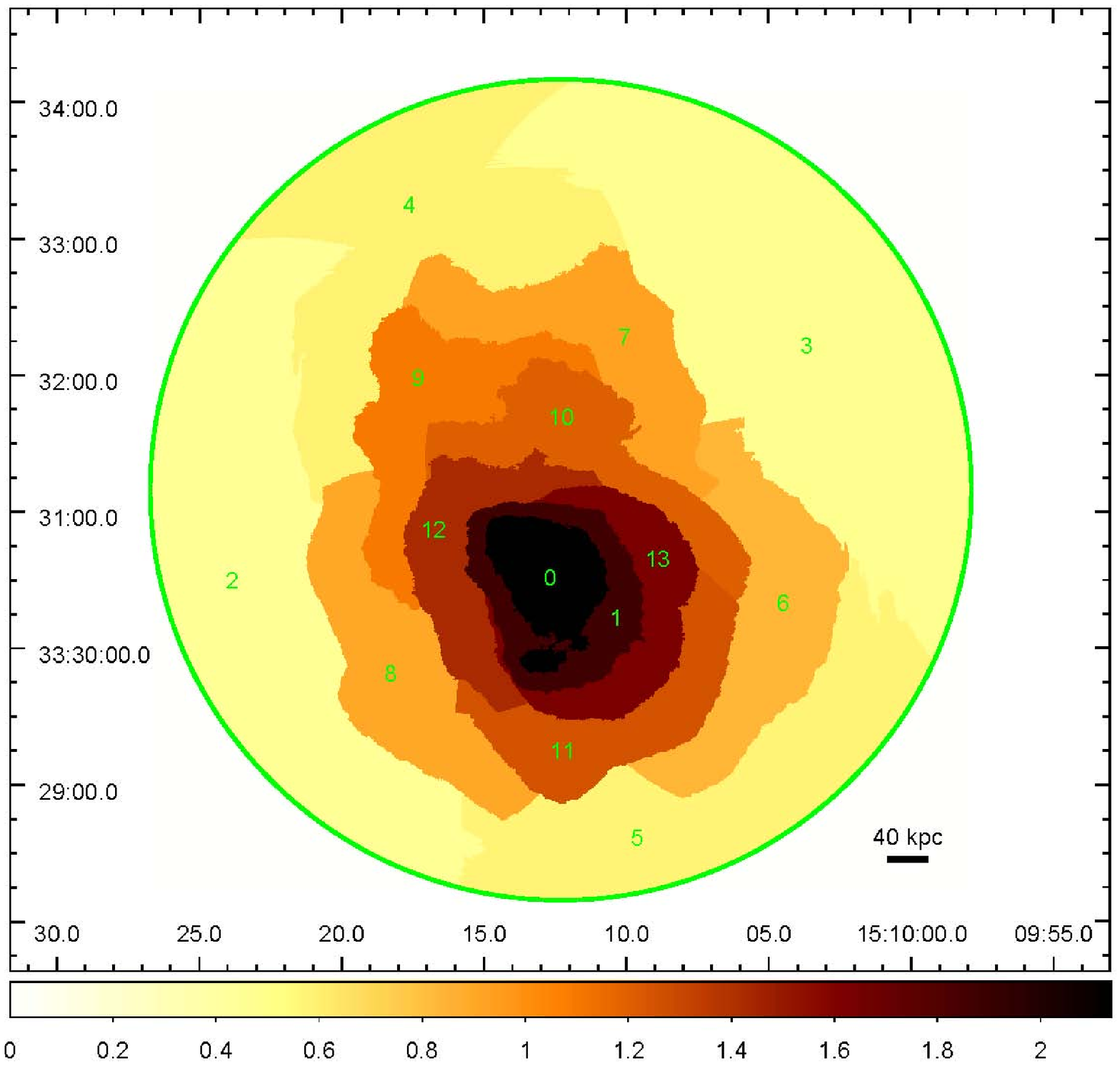}
\includegraphics[width=0.4\textwidth]{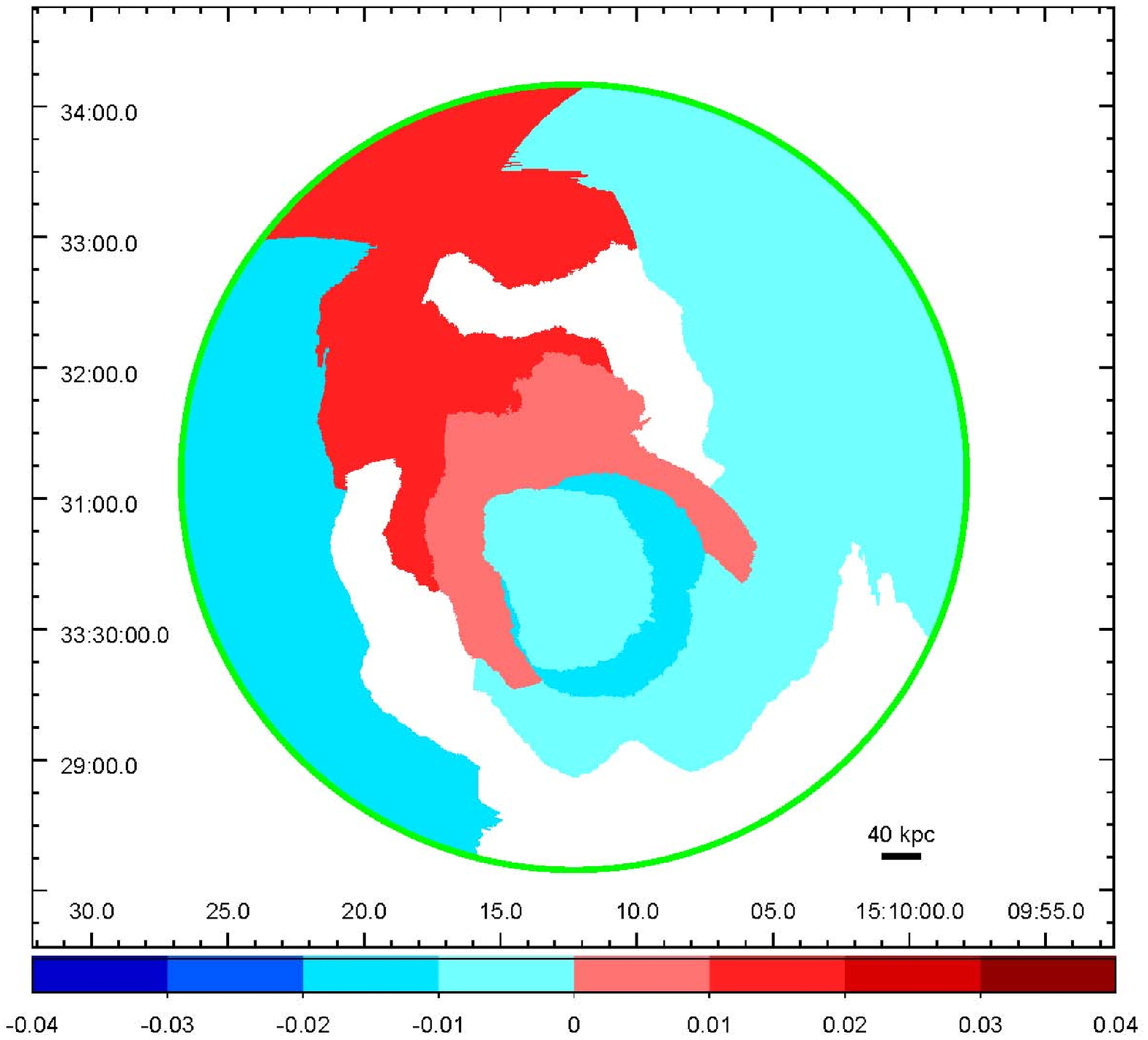}
\includegraphics[width=0.4\textwidth]{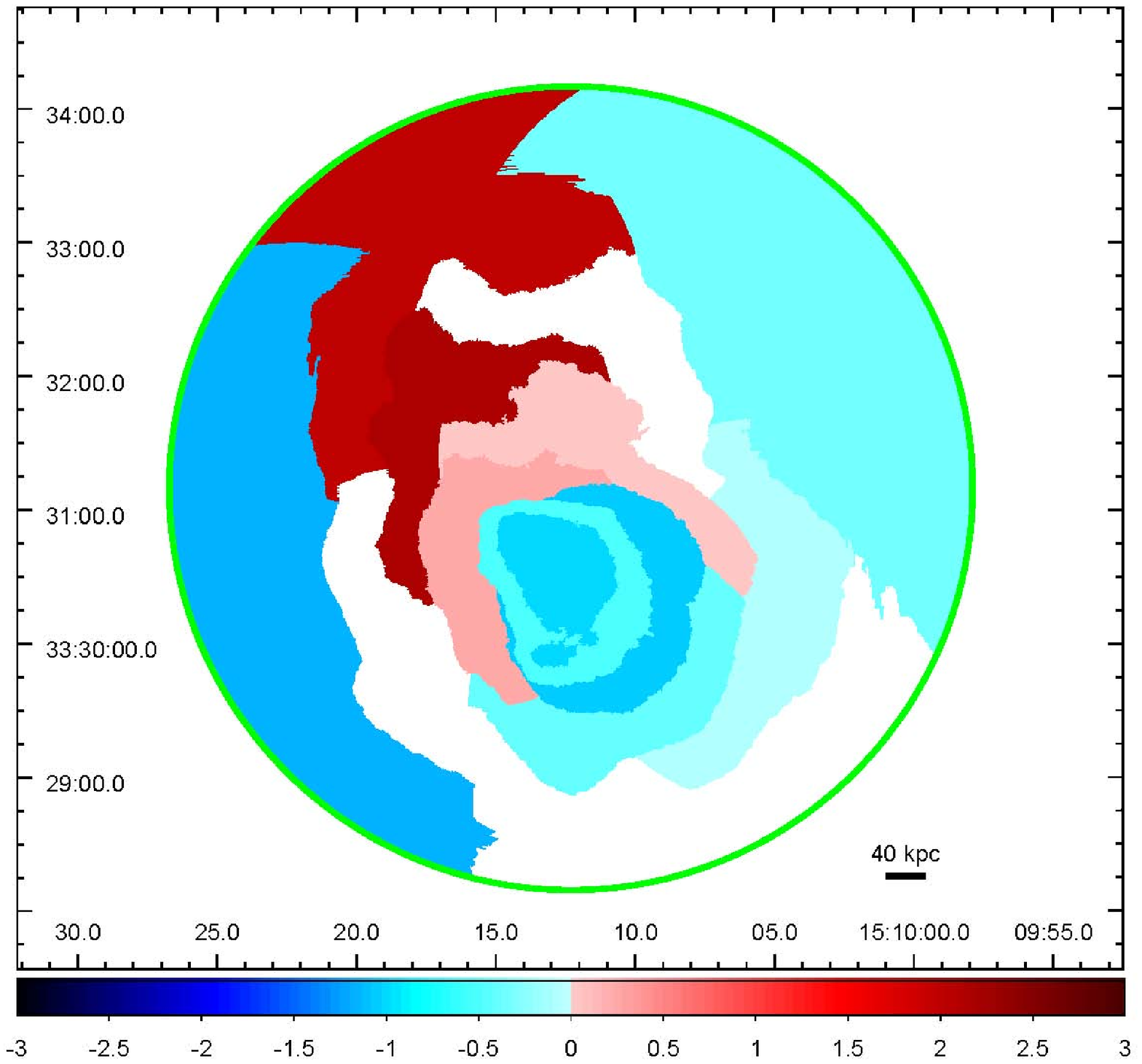}
\caption{Same as Figure \ref{a2142}, but for Abell 2034. }
\label{a2034}
\vfill
\end{figure*}

We use {\sl Chandra} ACIS-I observations for a total of 254.5 ks.  The total number of net counts
in the 0.5--10 keV band is $2.66\times 10^5$  within a circle of 3 arcmin in the merged image.
We select 14 ICM regions for redshift analysis, but we obtained reliable spectral fits only in 11 out of
14.  We find $z_{\rm Xtot} = 0.1124_{-0.0044}^{+0.0039}$, which is in very good agreement with
$z_{\rm opt}$ and $\langle z_{\rm X}\rangle = 0.1122$.  To evaluate the systematic uncertainties on $z_{\rm X}$, we
consider a temperature and
abundance range of 6--12 keV and 0.2--0.8 $Z_\odot$, respectively.  The best-fit $z_{\rm X}$ of the 11 regions
used in our analysis with corresponding errors are listed in Table \ref{table:a2034}.
The histogram distribution of $z_{\rm X}$ shows that most of the best-fit $z_{\rm X}$ values are approximately
distributed as a Gaussian, with the exception of two high-redshift values (see Figure \ref{histo_A2034}).
However, due to the estimated uncertainties, the probability of a uniform redshift distribution across the
ICM is still high at the 24\% level ($\chi^2 =12.70$ for 10 dof), which implies no significant detection
(less than 2$\sigma$) of global bulk motions.
Formally, we measure an average bulk motion of $v_{\rm BM} = 1700\pm 840\ {\rm km\ s^{-1}}$.
In addition, the  projected redshift map shows that the two high-redshift regions 4 and 9
are contiguous (see Figure \ref{a2034}, lower panels), which can be interpreted as an enhanced likelihood of a
redshift difference between the diffuse emission in the north and the main halo.
The average redshift of regions 4 and 9 is 0.1310, while that of the other regions is 0.1081.  This
implies a velocity difference of $\Delta v_{\rm max} = 6200 \pm 3300\ {\rm km\ s^{-1}}$. The presence of two
distinct halos with different redshift may suggest a classification as a pre-merger.
Due to the presence of a well-developed shock, this cluster may be caught in the early stages
of the merger  process, in agreement with the conclusion of \citet{owers2014}, who estimate that the
merger is observed only $\sim$0.3 Gyr after core-passage.

\subsection{Abell 115}

Abell 115 has been classified as a merger on the basis of its double X-ray morphology \citep{forman1981}.
The ICM thermal structure is best described by two halos that host a cool core with significantly
hotter gas between and around them, suggesting that the system is in an advanced stage of merging,
despite the fact that no shock front has been found in this region \citep{shibata1999,guti2005}.
A dynamical analysis based on 88 cluster member spectra \citep{barrena2007} shows a line-of-sight
velocity difference of 1646 ${\rm km\ s^{-1}}$ between the northern and southern subclusters
at a projected separation of $0.89\ {\rm Mpc}$. In addition, \citet{barrena2007}  interpret the system as a
pre-merger based on the location of the BCGs being consistent with the peaks in the X-ray surface brightness
distribution.  Diffuse radio emission north of the cluster has been found by \citet{Govoni2001}
with a 1.4 GHz VLA observation and interpreted as a radio relic that may be associated with the ongoing merger.
Abell 115 is also included in the Merging Cluster Collaboration sample.

We used 4 ACIS-I contiguous-in-time observations for a total of $310.6\ {\rm ks}$. The total number of counts
in the 0.5--10 keV band within a circle of 5.2 arcmin in the merged image is $2.88 \times 10^{5}$.
This allows us to obtain 18 useful regions for spectral analysis. Our spatially resolved spectroscopic analysis provides us with reliable $z_{\rm X}$ only in
12 regions out of 18.  The average redshift $\langle z_{\rm X} \rangle = 0.1979$ is in good agreement
with $z_{\rm opt} = 0.197$ \citep{hiroi2013} while the global redshift
$z_{\rm Xtot} = 0.1892_{-0.004}^{+0.003}$ is significantly lower.
To evaluate systematic uncertainties on $z_{\rm X}$, we consider temperature and
abundance ranges of 2--12 keV and 0.2--0.9 $Z_\odot$, respectively.  The best-fit $z_{\rm X}$ of the 12 regions
used in our analysis with corresponding errors are listed in Table \ref{table:a115}.

\begin{figure*}
\centering
\includegraphics[width=7.8cm]{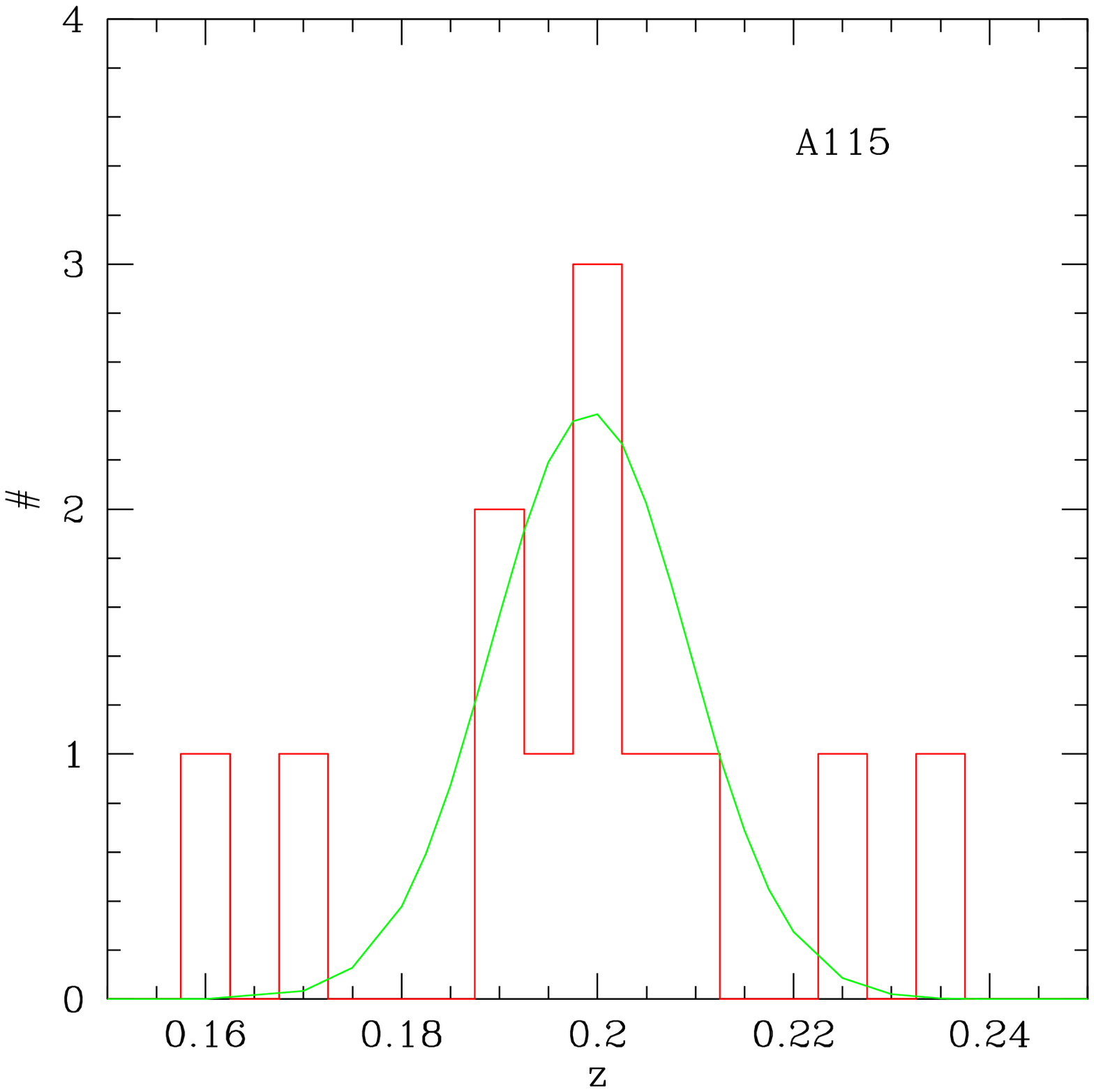}
\caption{Histogram distribution of the best-fit $z_{\rm X}$ for the 12 regions of Abell 115 with reliable
spectral fit.   Lines are as in Figure \ref{histo_A2142}. }
\label{histo_A115}
\vfill
\end{figure*}

The hypothesis of a constant redshift across the 12 regions is rejected at more than 3$\sigma$
($\chi^2 = 30.4$ for 11 dof).  The histogram distribution of $z_{\rm X}$ (see Figure \ref{histo_A115}) shows
the presence of two regions significantly above and below the central value.
We measure an average bulk motion velocity of $v_{\rm BM}  = 4600 \pm 1100\ {\rm km\ s^{-1}}$.

\begin{figure*}
\centering
\includegraphics[width=0.4\textwidth]{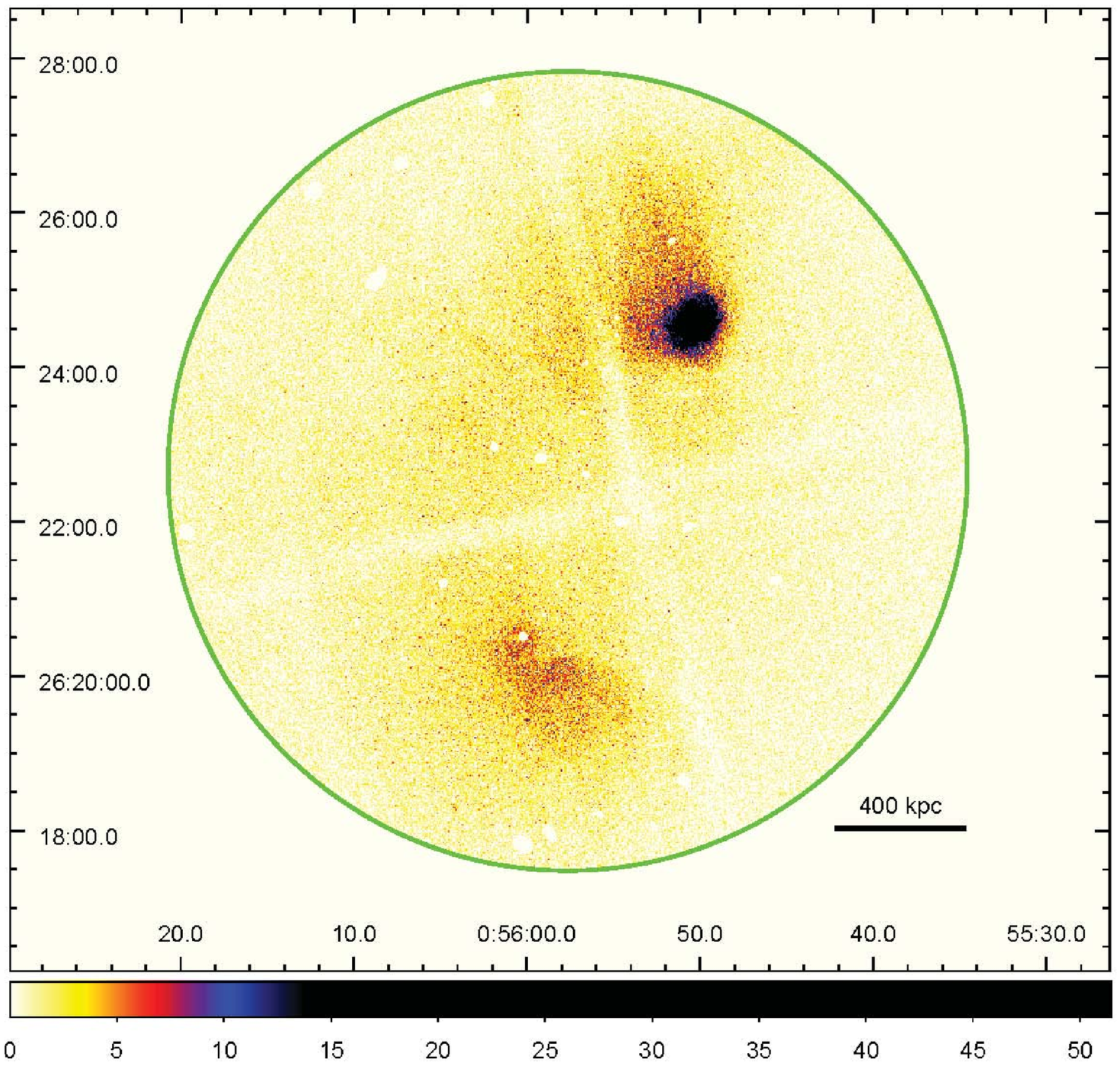}
\includegraphics[width=0.4\textwidth]{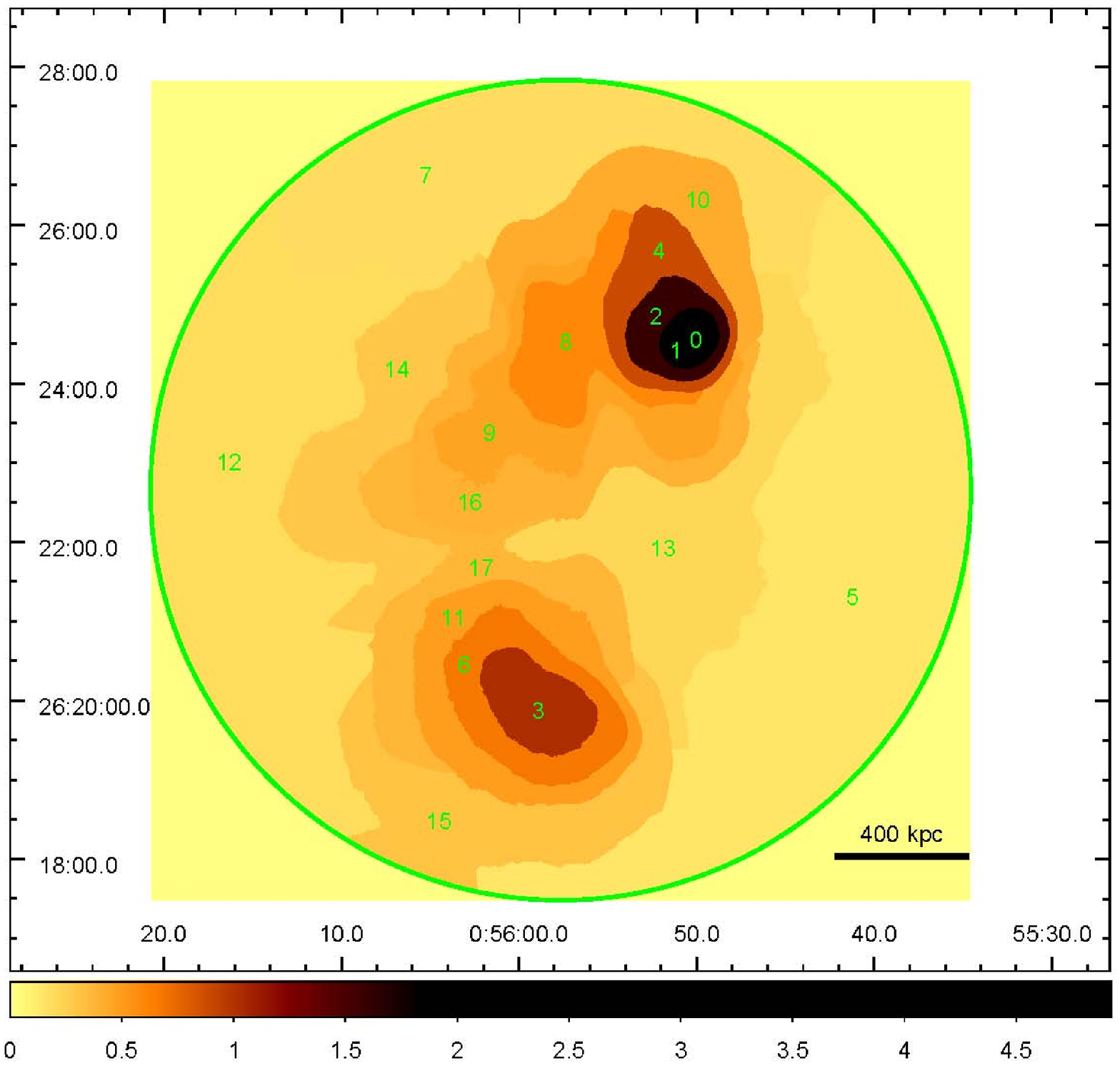}
\includegraphics[width=0.4\textwidth]{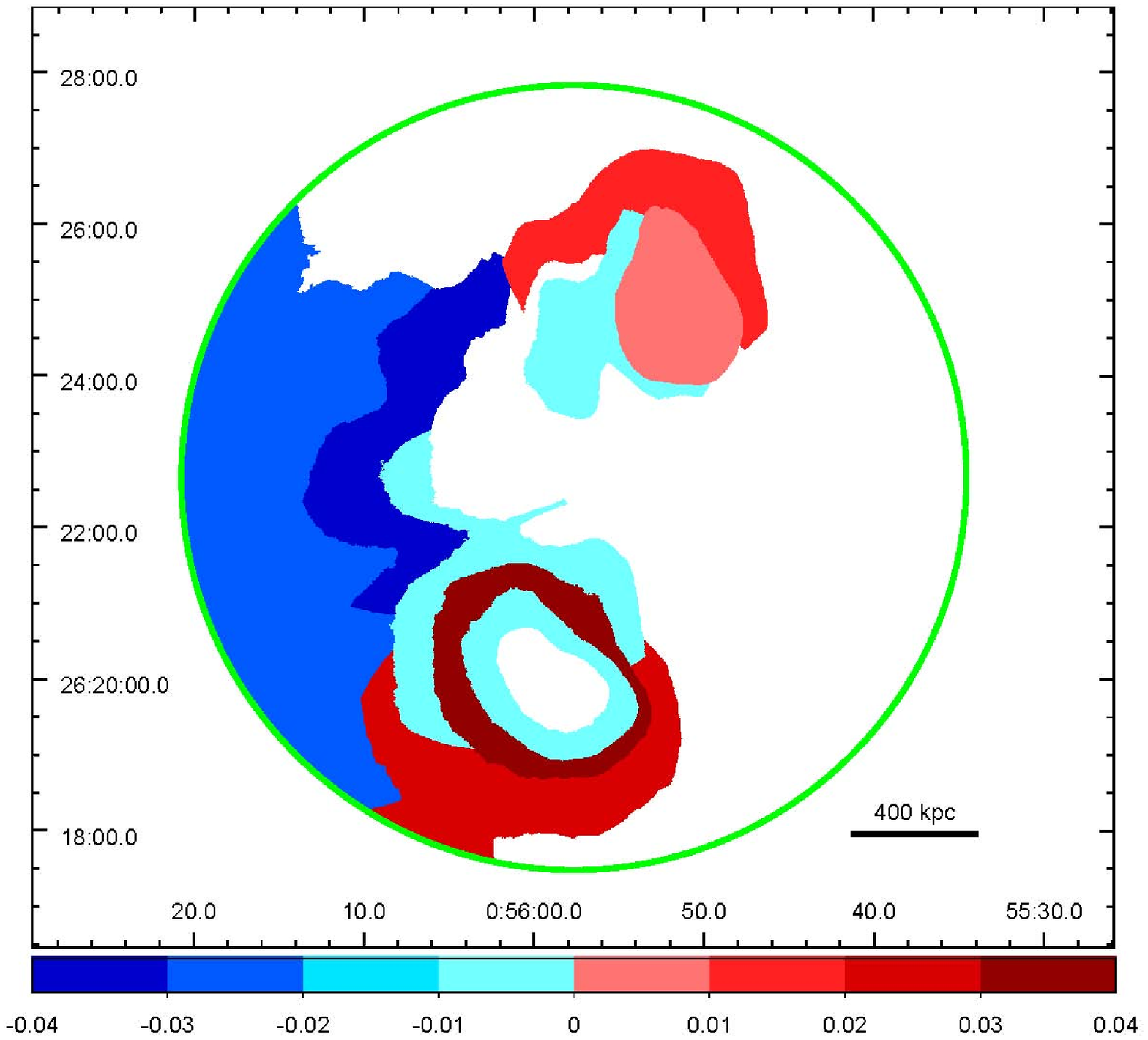}
\includegraphics[width=0.4\textwidth]{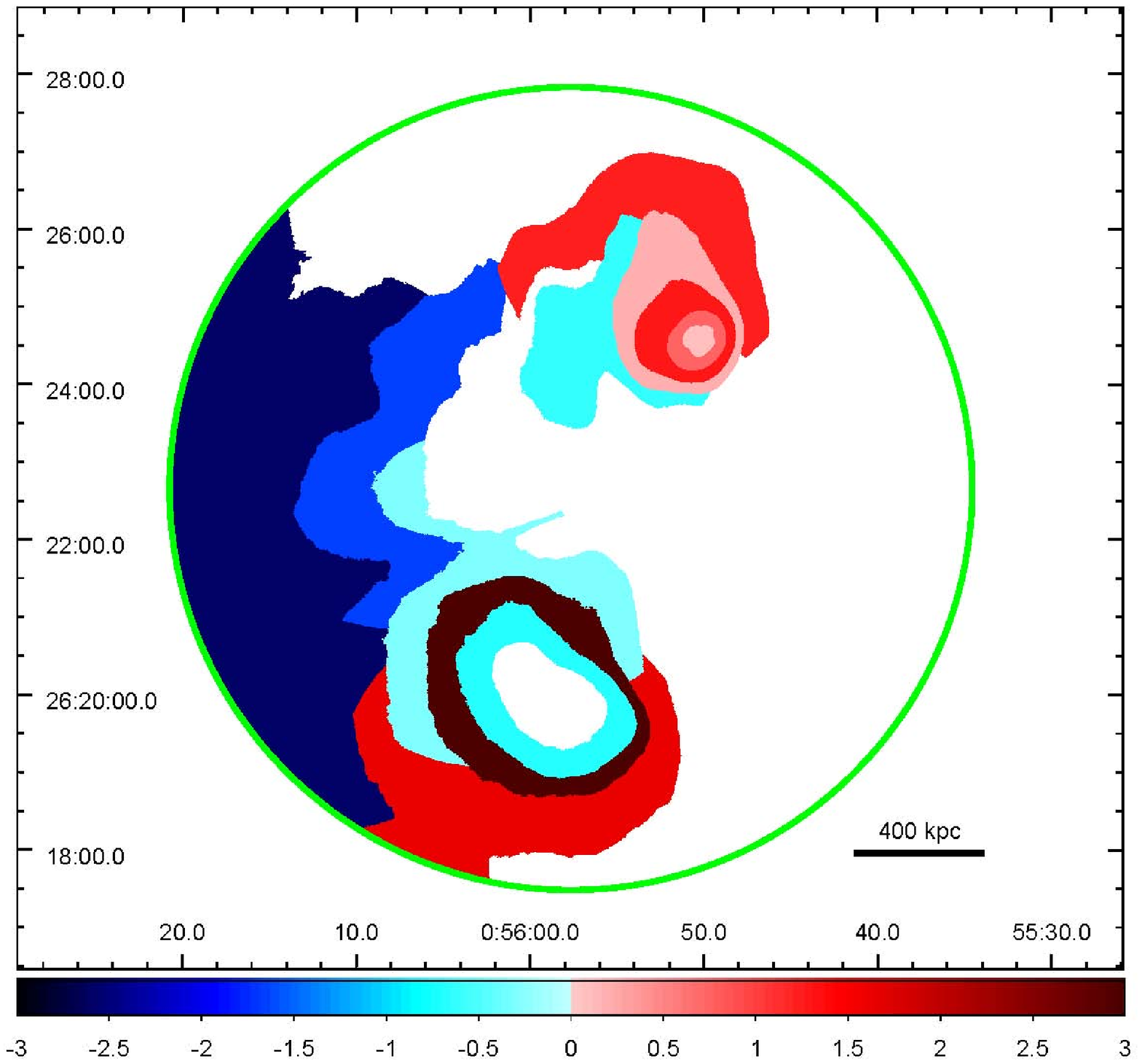}
\caption{Same as Figure \ref{a2142}, but for  Abell 115. }
\label{a115}
\vfill
\end{figure*}

Visual inspection of the redshift and significance maps (see Figure \ref{a115}, lower panels) allows
us to identify three main regions with compatible redshift.  We first isolate regions belonging
to the northern (regions 0, 1, 2, 4, and 10) and southern (3, 6, 11, 17, and 15) clumps.
We find a redshift difference that corresponds to about 1$\sigma$, and is therefore consistent with having the
same redshift.  Formally, translated into a velocity difference, this would imply a relative velocity of
$3300 \pm 3200\ {\rm km\ s^{-1}}$.   If we consider instead regions 12 and 14, then we find an average redshift of
0.1634, corresponding to a velocity difference of $8600 \pm 3000\ {\rm km\ s^{-1}}$ with respect to the
bulk of the ICM.
We note that this is also the gas that is found to be significantly hotter (at least by a factor
of two, as verified in the projected temperature map not discussed here), and therefore it is likely to be
heated and violently displaced by the ongoing merger.
This is consistent with a substantial amount of the shocked gas being compressed between
the two halos and pushed toward the observer, with the two
halos having a minimum in the velocity difference.  Therefore, we conclude that A115 is caught in the
early phase of the merging process, with the two ICM halos still well separated.

\subsection{Abell 520}

Abell 520, at $z_{\rm opt} = 0.203$ \citep{cassano2013}, was found to host a radio halo elongated in the
NE-SW direction, corresponding to the apparent merger axis \citep{1999Giovannini,Govoni2001}.
Its X-ray morphology and the optical spectroscopy suggest a disturbed dynamical state
\citep[][respectively]{Govoni2001,proust2000}.  Despite the recent in-depth studies
\citep{mahdavi2007,girardi2008}, the detailed structure of the merger of Abell 520 is still unclear.
However, a prominent bow shock with $M=2.1_{-0.3}^{+0.4}$ has been identified thanks to
 {\sl Chandra} data analysis \citep{markevitch2005}.
Abell 115 is also included in the Merging Cluster Collaboration sample.

A520 has the longest {\sl Chandra} ACIS-I exposure for a total of 516.4 ks.  The number of net counts
in the 0.5--10 keV band in the merged image is $3.36 \times 10^5$ in a circle of 3.2 arcmin radius.
The best-fit redshift from the global emission is $z_{\rm Xtot} = 0.2082_{-0.0049}^{+0.0046}$.
To estimate the systematic uncertainty on $z_{\rm X}$, we consider ranges of 3--15 keV and 0.2--1
$Z_\odot$ for temperature and abundance, respectively.

\begin{figure*}
\centering
\includegraphics[width=7.8cm]{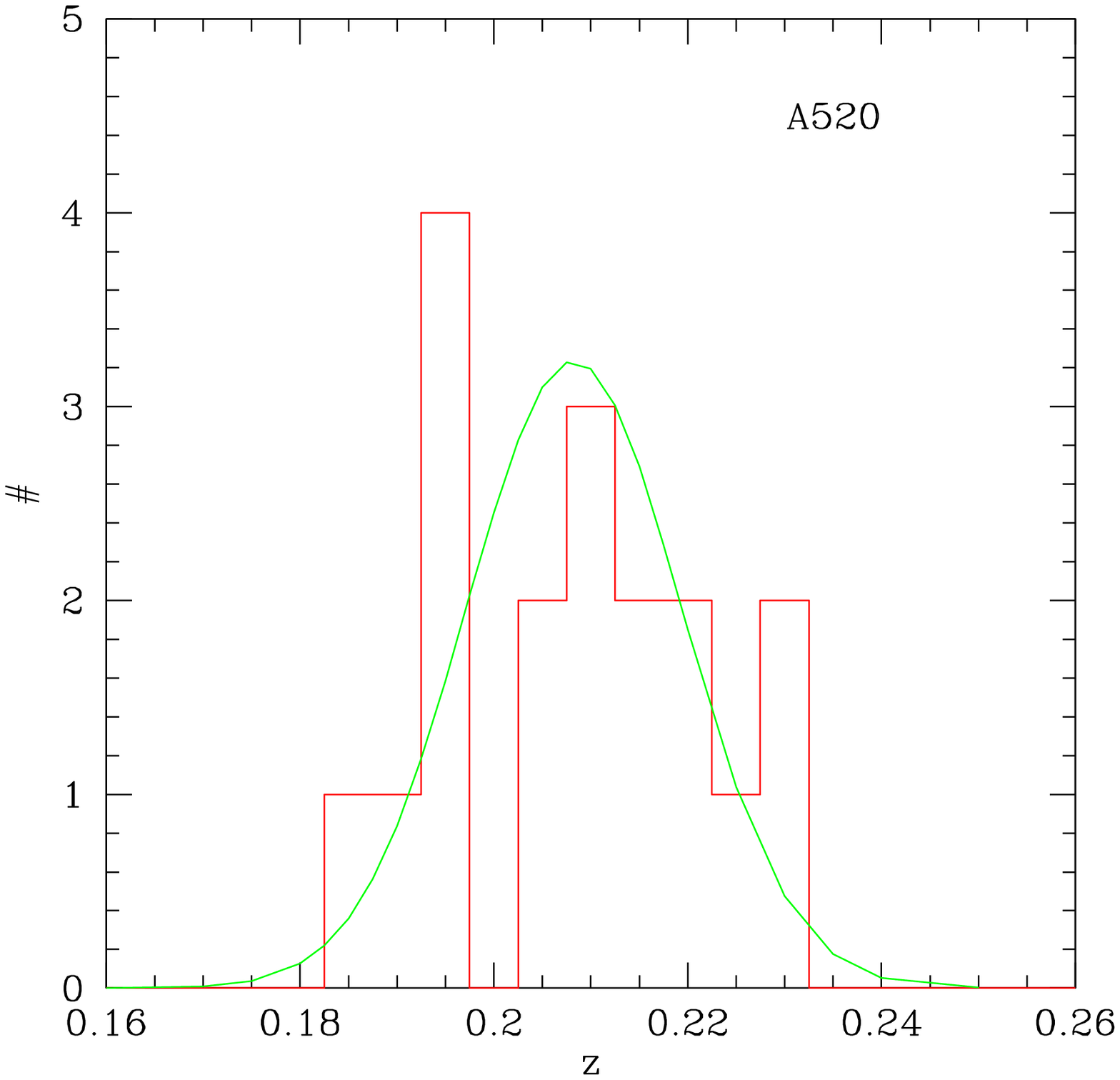}
\caption{Histogram distribution of the best-fit $z_{\rm X}$ for the 18 regions of Abell 520 with reliable
spectral fit.   Lines are as in Figure \ref{histo_A2142}. }
\label{histo_A520}
\vfill
\end{figure*}

\begin{figure*}
\centering
\includegraphics[width=0.4\textwidth]{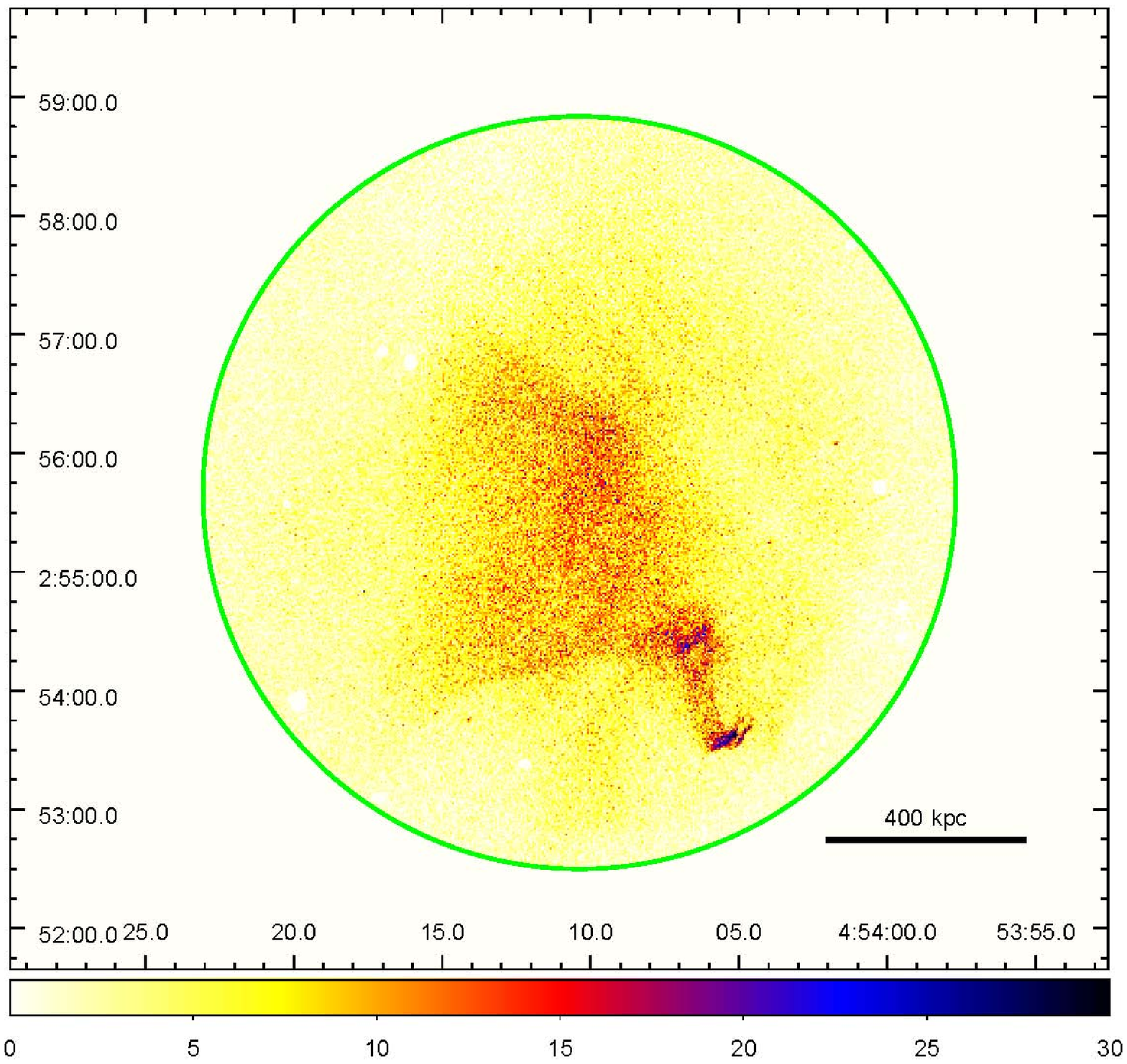}
\includegraphics[width=0.4\textwidth]{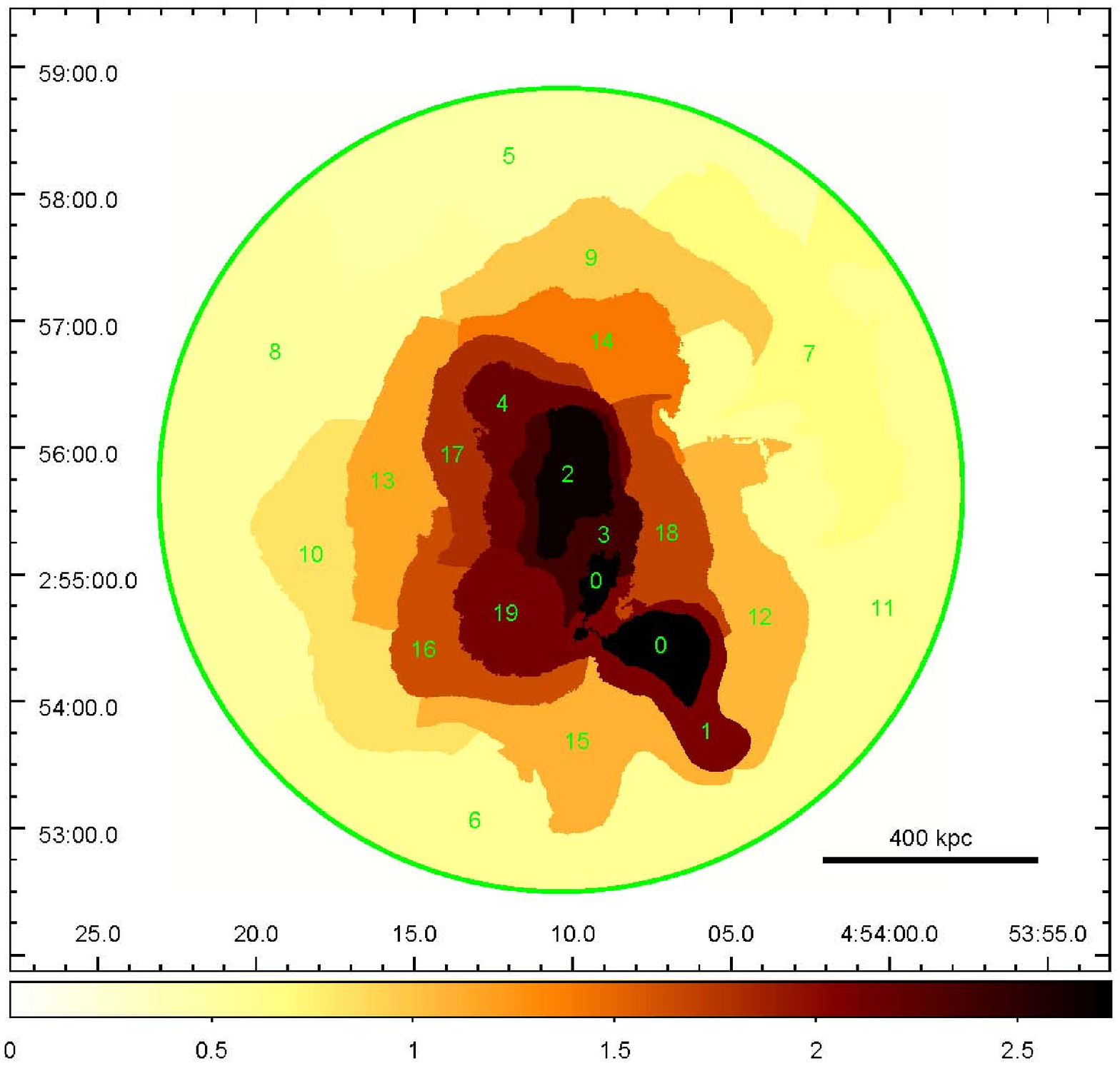}
\includegraphics[width=0.4\textwidth]{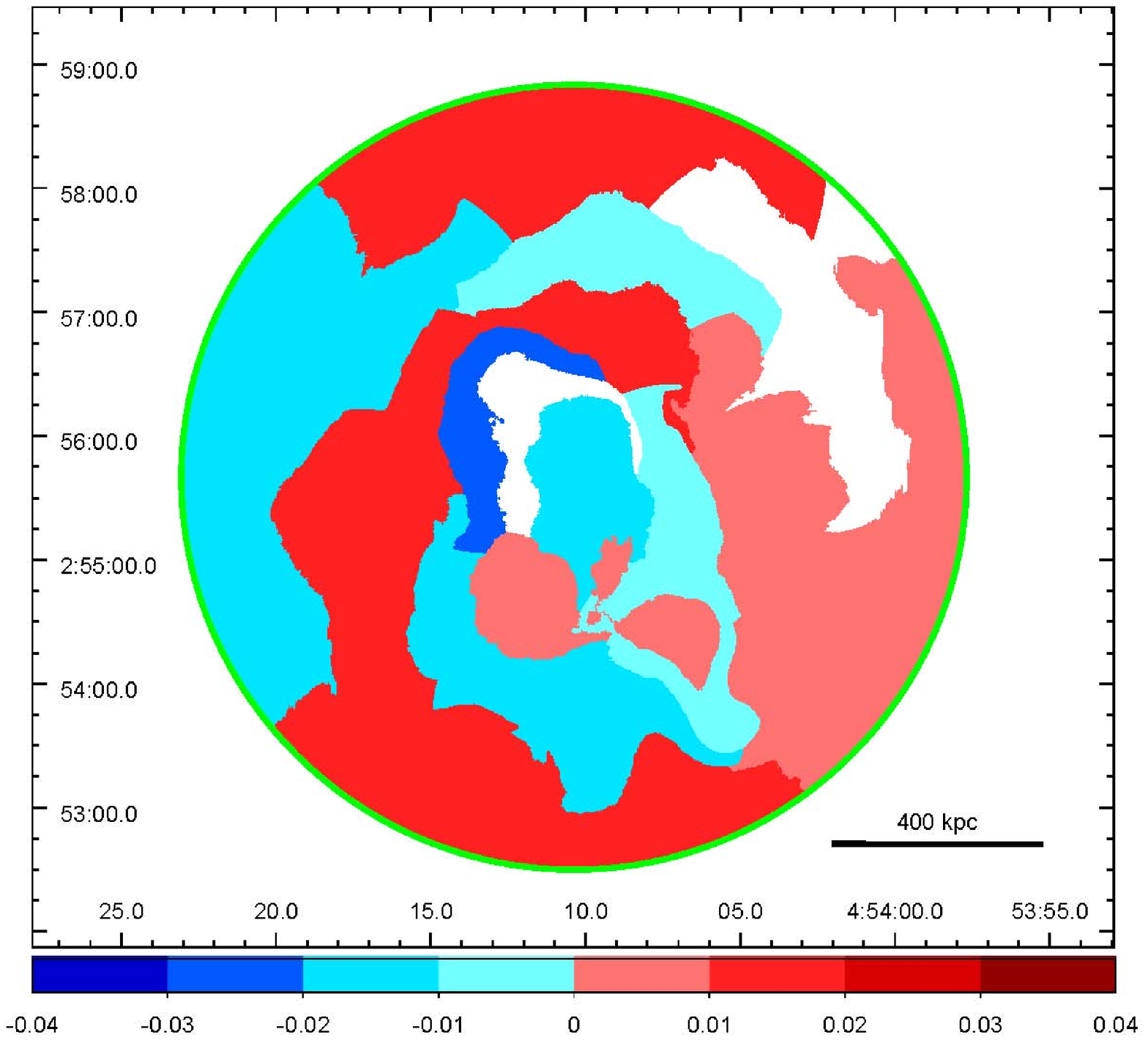}
\includegraphics[width=0.4\textwidth]{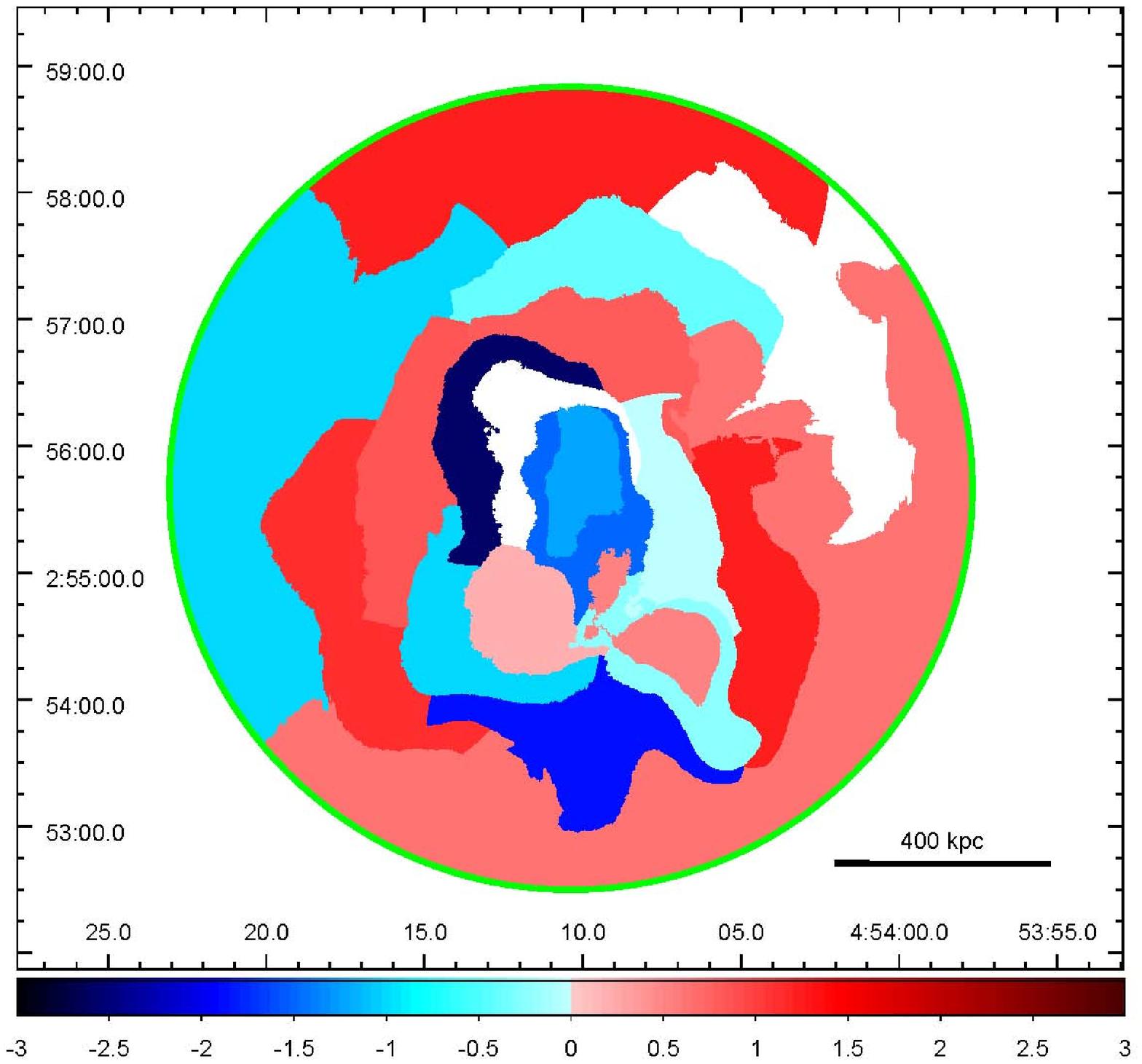}
\caption{Same as Figure \ref{a2142}, but for Abell 520. }
\label{a520}
\vfill
\end{figure*}

We discard only two regions that do not provide reliable redshifts.  Therefore, we base our analysis on
18 regions, whose best-fit $z_{\rm X}$ and associated errors are listed in Table \ref{table:a520}.  The average
redshift among those regions is $\langle z_{\rm X}\rangle = 0.2083$, in good agreement with $z_{\rm Xtot} $
and $z_{\rm opt}$.   The hypothesis of a constant redshift is rejected at an $\sim$87\% confidence level
($\chi^2 = 23.4$ for 17 dof), which is therefore less than 2$\sigma$.
We find $v_{\rm BM}$$\sim$$1800 \pm 900\ {\rm km\ s^{-1}}$.  This is mostly contributed by the presence
of blueshifted regions, mostly in the central 1 arcmin, and the surrounding higher-redshift regions,
as can be seen by comparing the histogram distribution (see Figure \ref{histo_A520}) with the
redshift and significance maps (see Figure \ref{a520}).  If we compute the difference between the
two sets of regions with comparable redshifts, then we find $\Delta v_{\rm max} = 5900 \pm 4000\ {\rm km\ s^{-1}}$.
The overall redshift map suggests a merger with a halo moving toward the observer that was
caught shortly after the first pericentric passage.

\subsection{1RXS\ J0603.3+4214}

At redshift $z_{\rm opt} =0.225$ \citep{weeren2012}, 1RXS\ J0603.3+4214 (the ``Toothbrush" cluster)
was discovered in the radio band thanks to its prominent linear radio relic and radio halo plus several
additional features which suggest that this is a complex merger.   Simulations also show that the merger
might be a triple process: in addition to northern and southern subclusters with similar mass,
there might be another smaller subcluster infalling in the south
\citep{bruggen2012}.  Measurement of the radio spectral
index allowed \citet{weeren2012} to constrain the merger Mach number to be in the range 3.3--4.6. However,
X-ray observation by {\sl XMM-Newton} returned  a Mach number of $<$2 \citep{ogrean2013}, which is inconsistent with
the radio constraints.  1RXS J0603.3+4214 is also included in the Merging Cluster Collaboration sample

\begin{figure*}
\centering
\includegraphics[width=7.8cm]{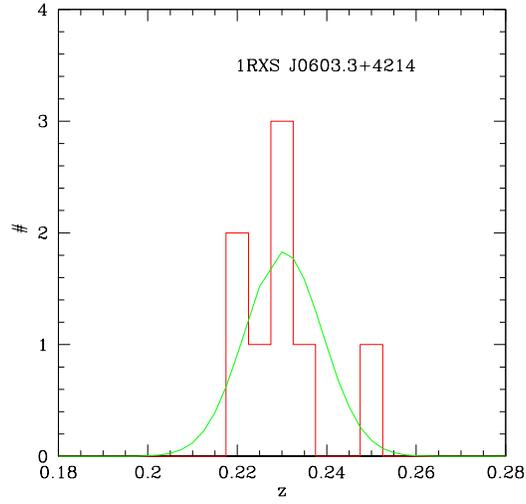}
\caption{Histogram distribution of the best-fit $z_{\rm X}$ for the 8 regions of 1RXS\ J0603.3+4214
with reliable spectral fit.  Lines are as in Figure \ref{histo_A2142}. }
\label{histo_RXJS0603}
\vfill
\end{figure*}

\begin{figure*}
\centering
\includegraphics[width=0.4\textwidth]{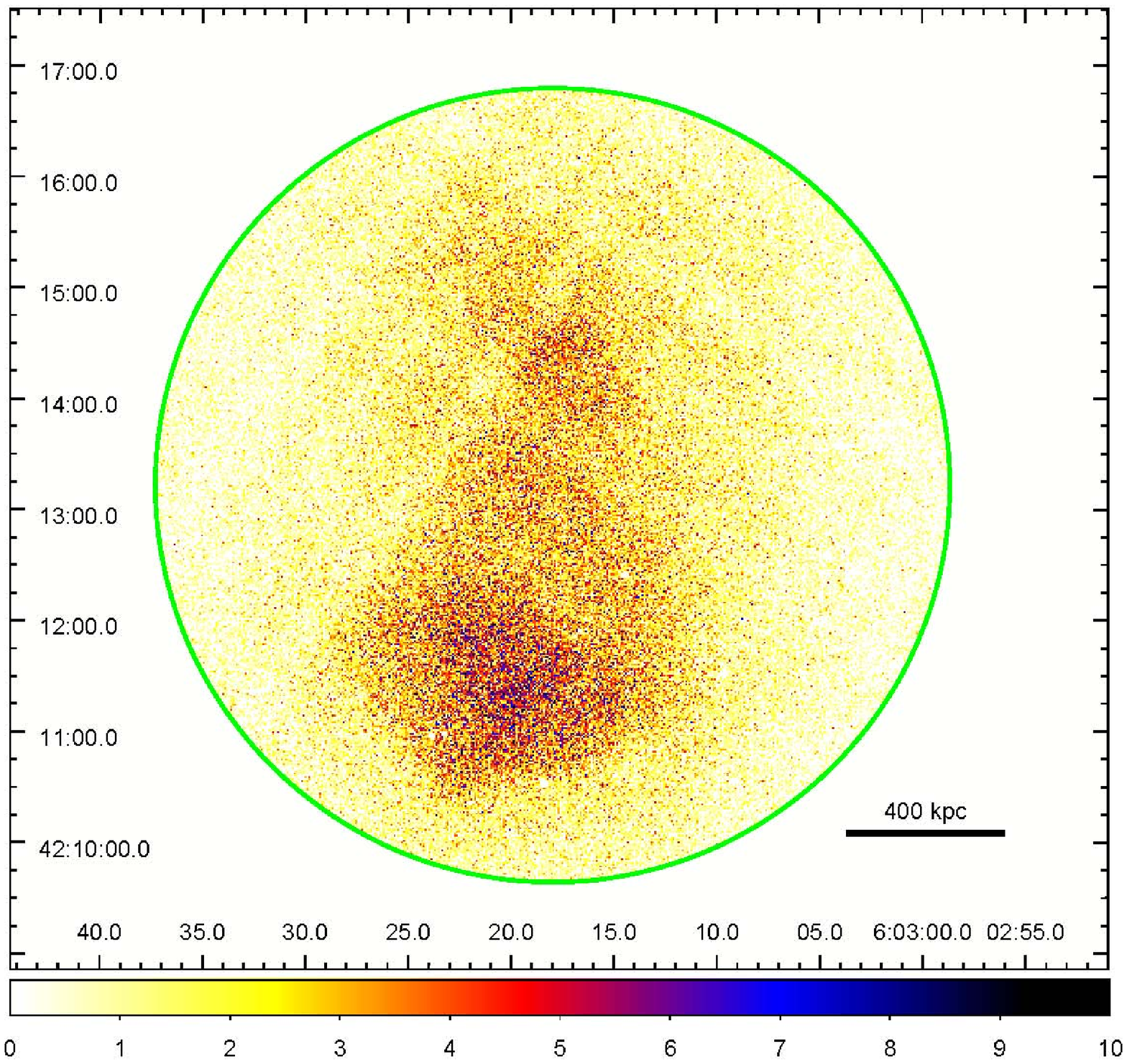}
\includegraphics[width=0.4\textwidth]{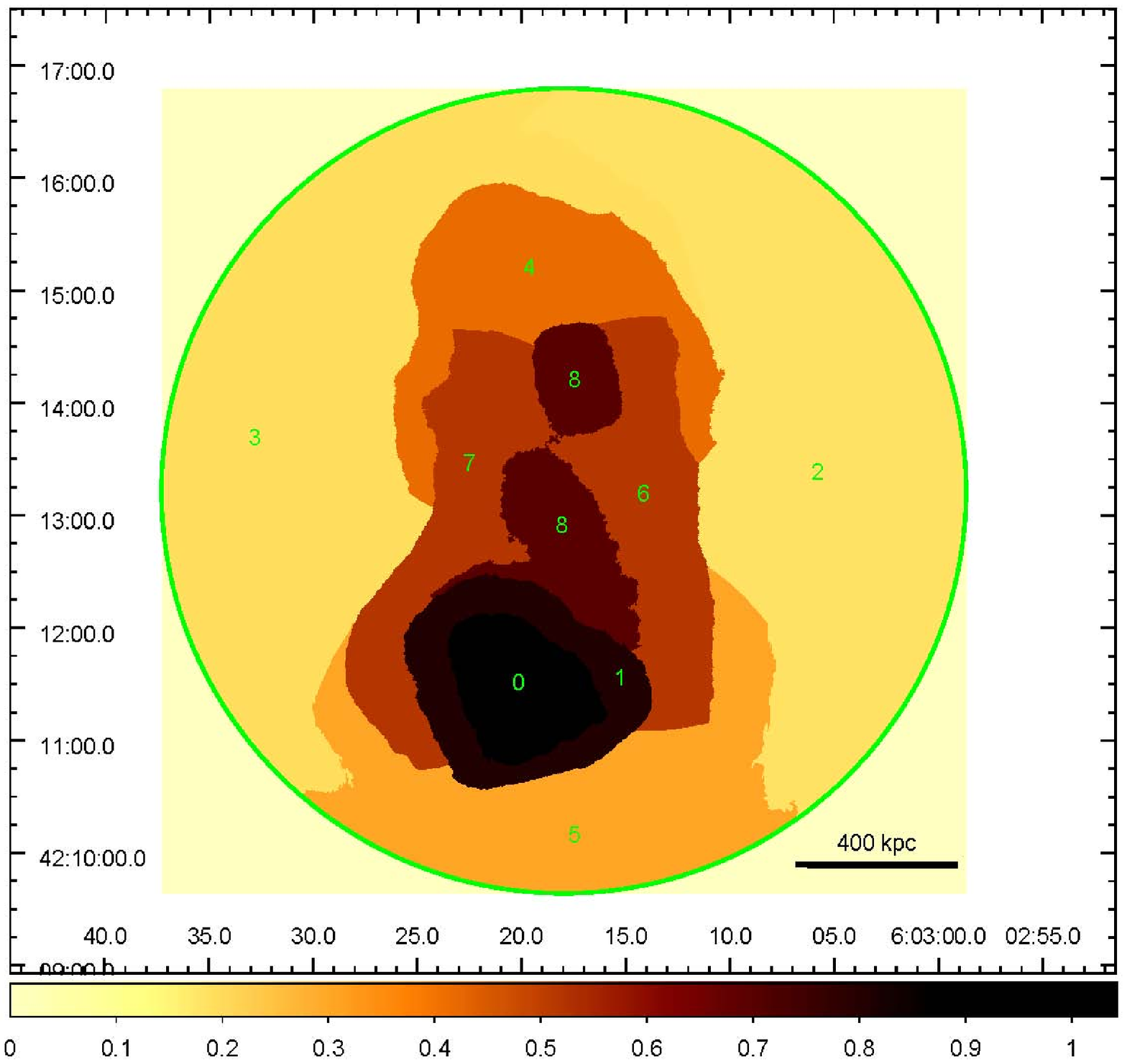}
\includegraphics[width=0.4\textwidth]{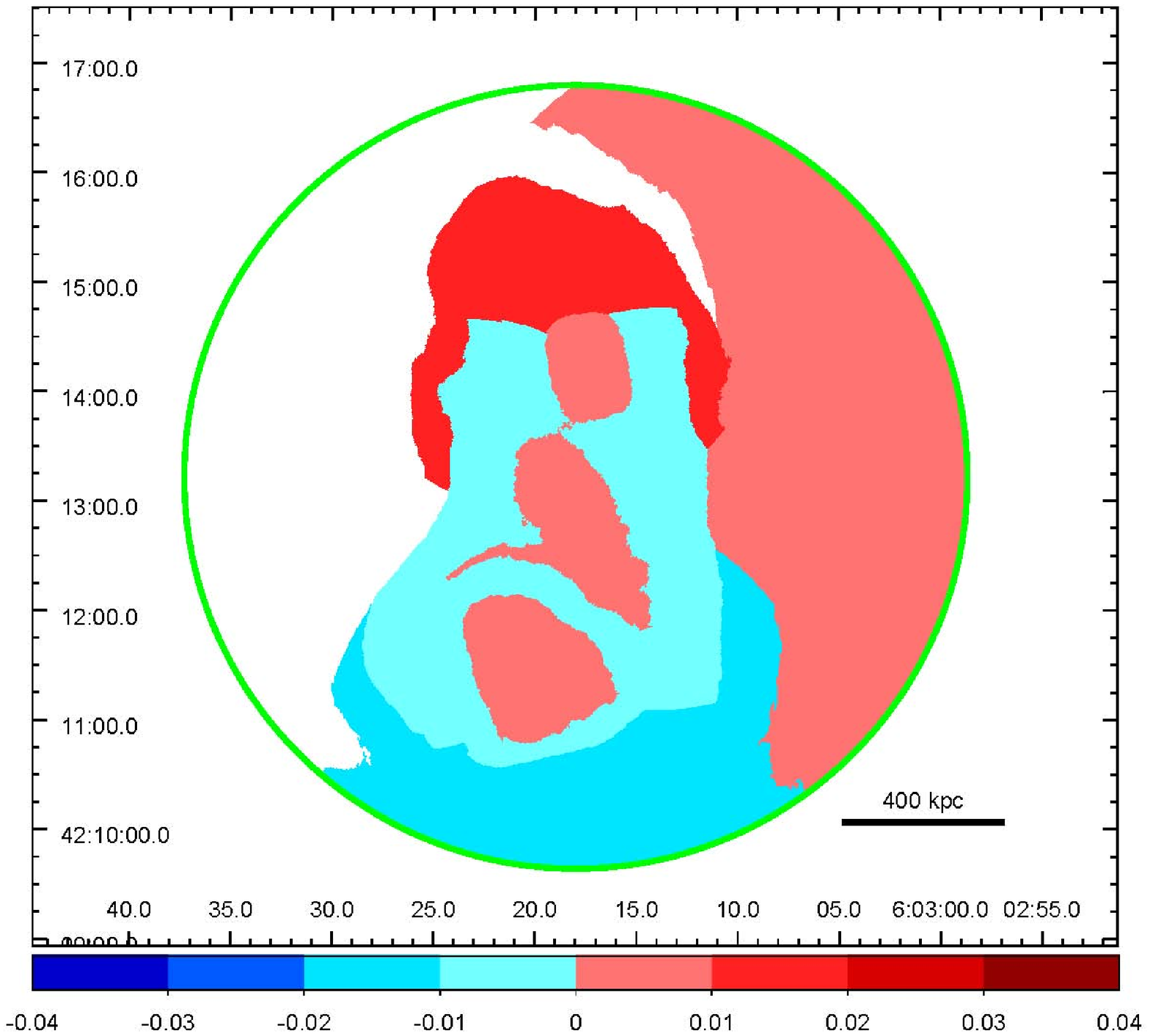}
\includegraphics[width=0.4\textwidth]{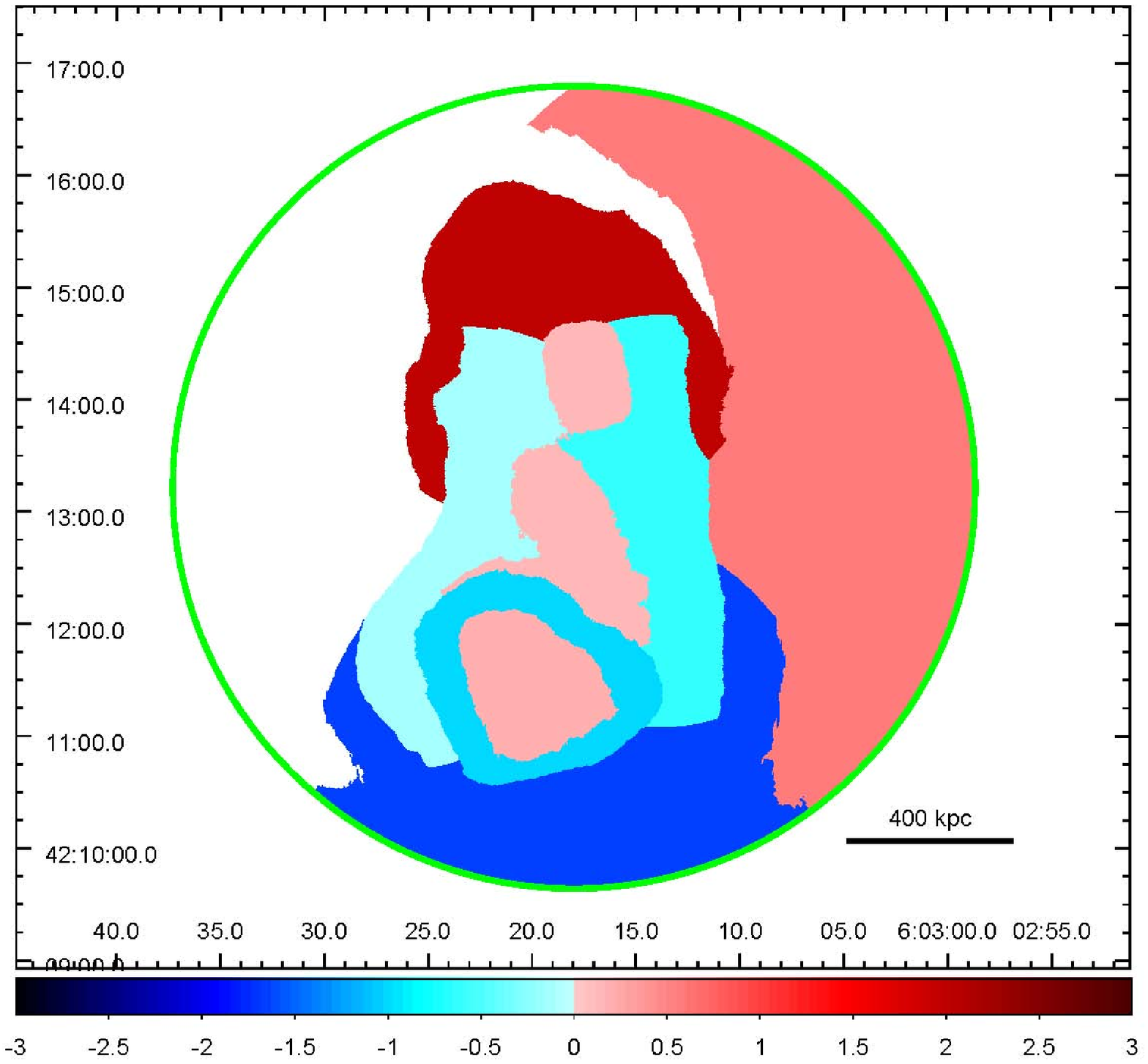}
\caption{Same as Figure \ref{a2142}, but for 1RXS\ J0603.3+4214. }
\label{rxsj0603}
\vfill
\end{figure*}

We use three {\sl Chandra} ACIS-I observations for a total of 235.9 ks.  The total number of counts
in the 0.5--10 keV band within a circle of $3.5 {\rm arcmin}$ is $1.74 \times 10^{5}$ in the merged image.
To evaluate the
systematic uncertainty on $z_{\rm X}$, we consider ranges of 8--16 keV and 0.3--0.8 $Z_\odot$ for the temperature
and abundance, respectively.  We are able to obtain reliable spectral fits in 8 out of the 9 regions
originally considered for analysis.  The best-fit $z_{\rm X}$ and corresponding errors are listed in
Table \ref{table:rxsj}.  The average redshift across the regions
is $\langle z\rangle = 0.2303$, which is in good agreement with $z_{\rm Xtot} = 0.2316_{-0.0033}^{+0.0029}$
and in reasonable agreement with $z_{\rm opt}$.
The hypothesis of a constant redshift is acceptable ($\chi^2 =  8.7$ for 7 dof), as also shown in Figure \ref{histo_RXJS0603} by the histogram distribution of the best-fit
$z_{\rm X}$ values. The upper limit to the
average bulk motion velocity is $v_{\rm BM} < 2100\ {\rm km\ s^{-1}}$.  However, there is a significant
uniform gradient in the best-fit redshift from the northern clump (high-z) to the southern clump (low-z; see Figure\ref{rxsj0603}), which corresponds to a maximum difference of
$\sim$$0.0295 \pm 0.0115$, or a maximum velocity difference of  $\Delta v_{\rm max} \sim 7200\ \pm\ 2800 \ {\rm km\ s^{-1}}$.  On the basis of these results, this cluster can be classified as a pre-merger.

\subsection{Abell 2146}

Despite the lack of  diffuse radio emission, A2146 is classified as an extreme binary, head-on cluster
merger in \citet{mann2012}.  A deep {\sl Chandra} study reveled the presence of two shock fronts
with Mach numbers of $M = 2.2 \pm 0.8$  and $M \sim 1.7 \pm 0.3$ \citep{russell2010}.
Using the radial velocity of BCGs in the subcluster and the main cluster, \citet{canning2012} find
that the merger axis has an angle of 17$^\circ$ with respect to the plane of the sky.
Based on these properties and its appearance in the X-ray, Abell 2146 is often compared to the Bullet cluster.

We use observations with {\sl Chandra}  ACIS-I for a total of 375.3 ks,
as listed in Table \ref{tabsample}.  For simplicity, we do not consider observations with ACIS-S for a total of
45 ks in order to retain uniform calibration throughout our analysis, given the smaller amount of
observing time contributed by the ACIS-S exposures.  The total number of net counts in the 0.5-10 keV
band within a circle of 3.6 arcmin is about $1.99 \times 10^5$.
The best-fit redshift obtained by fitting the total X-ray emission is $z_{\rm Xtot} = 0.2310_{-0.0010}^{+0.0022}$.
Measured temperatures range from the inner $\sim$3 keV to the outer $\sim$8 keV in our analysis
of the temperature structure (not reported in this work).
To evaluate the systematic uncertainty on $z_{\rm X}$, we consider
ranges from 1 to 12 keV and 0.15 to 1.5 $Z/Z_\odot$ to explore the temperature and metal abundance
throughout the ICM, respectively.

\begin{figure*}
\centering
\includegraphics[width=7.8cm]{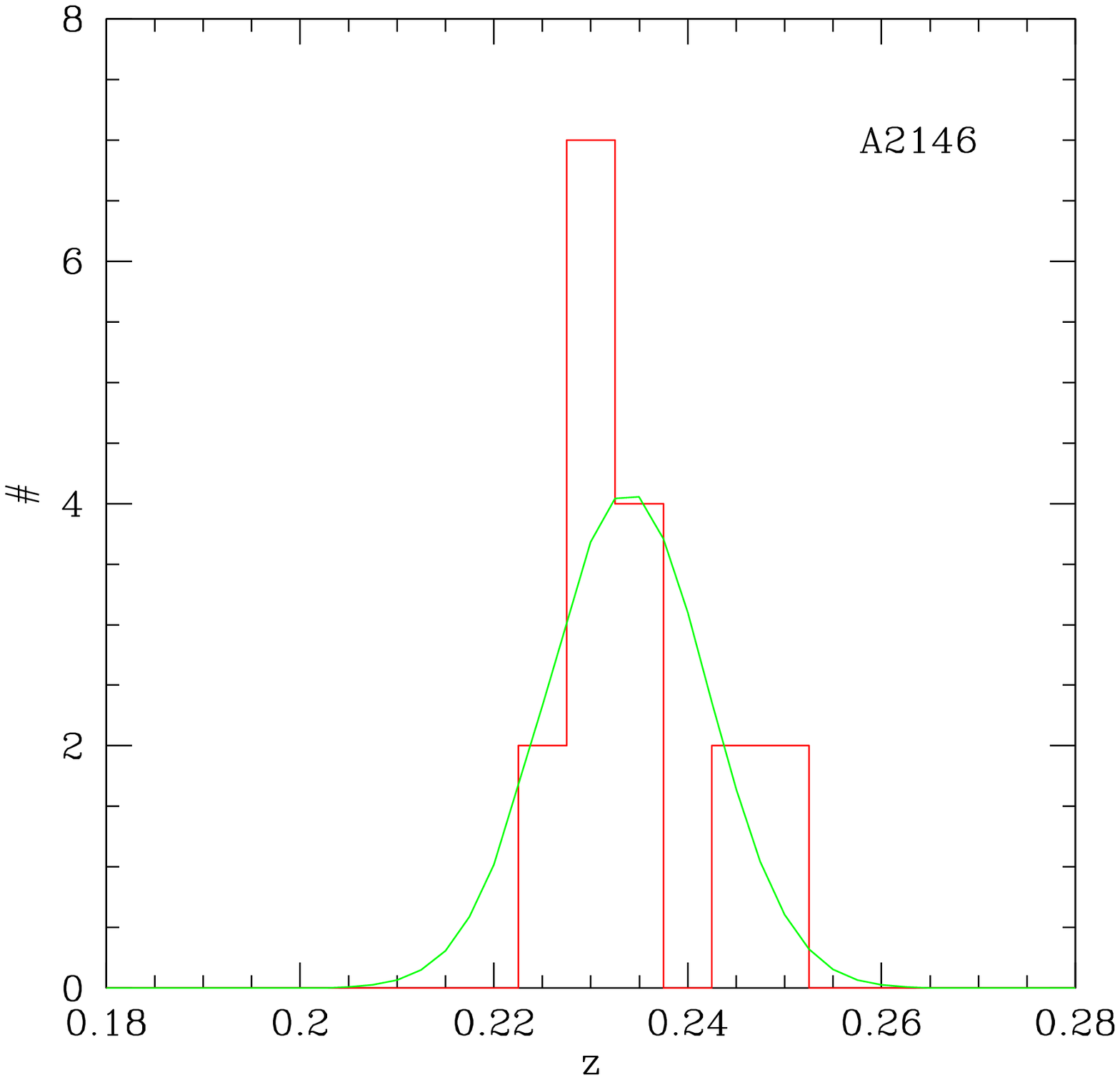}
\caption{Histogram distribution of the best-fit $z_{\rm X}$ for the 17 regions of A2146 with reliable spectral fit.
Lines are as in Figure \ref{histo_A2142}. }
\label{histo_A2146}
\vfill
\end{figure*}

\begin{figure*}
\centering
\includegraphics[width=0.4\textwidth]{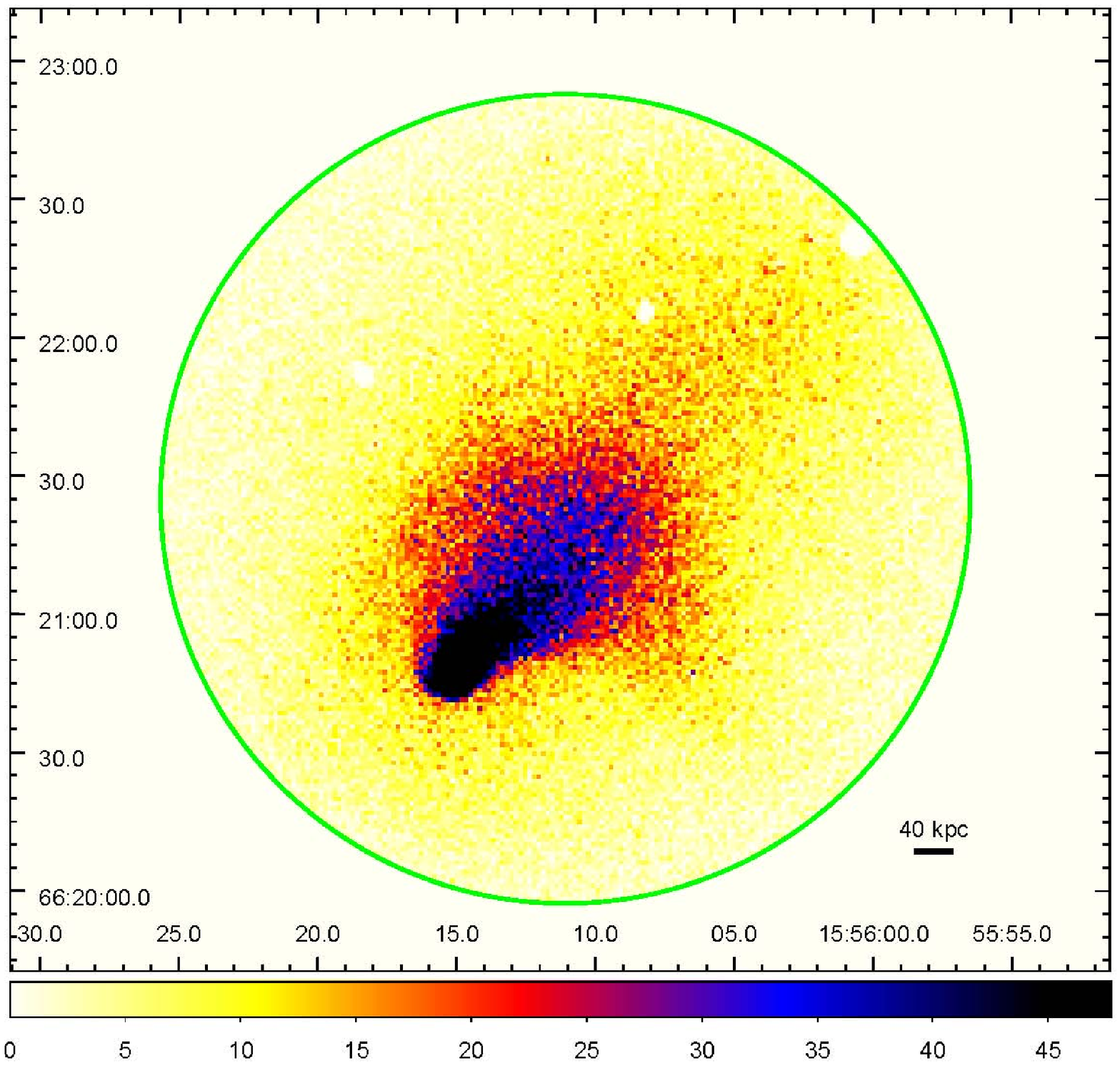}
\includegraphics[width=0.4\textwidth]{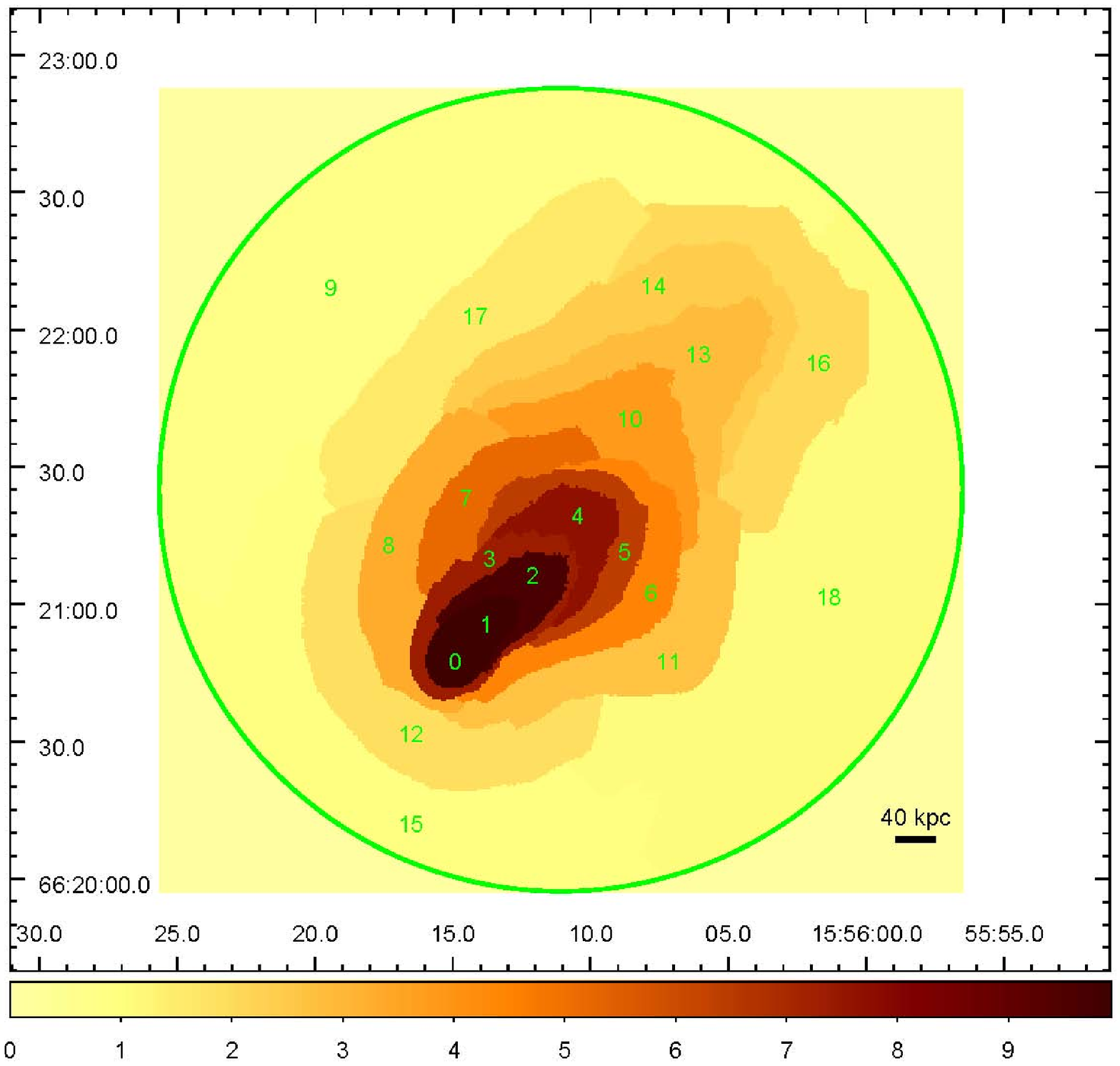}
\includegraphics[width=0.4\textwidth]{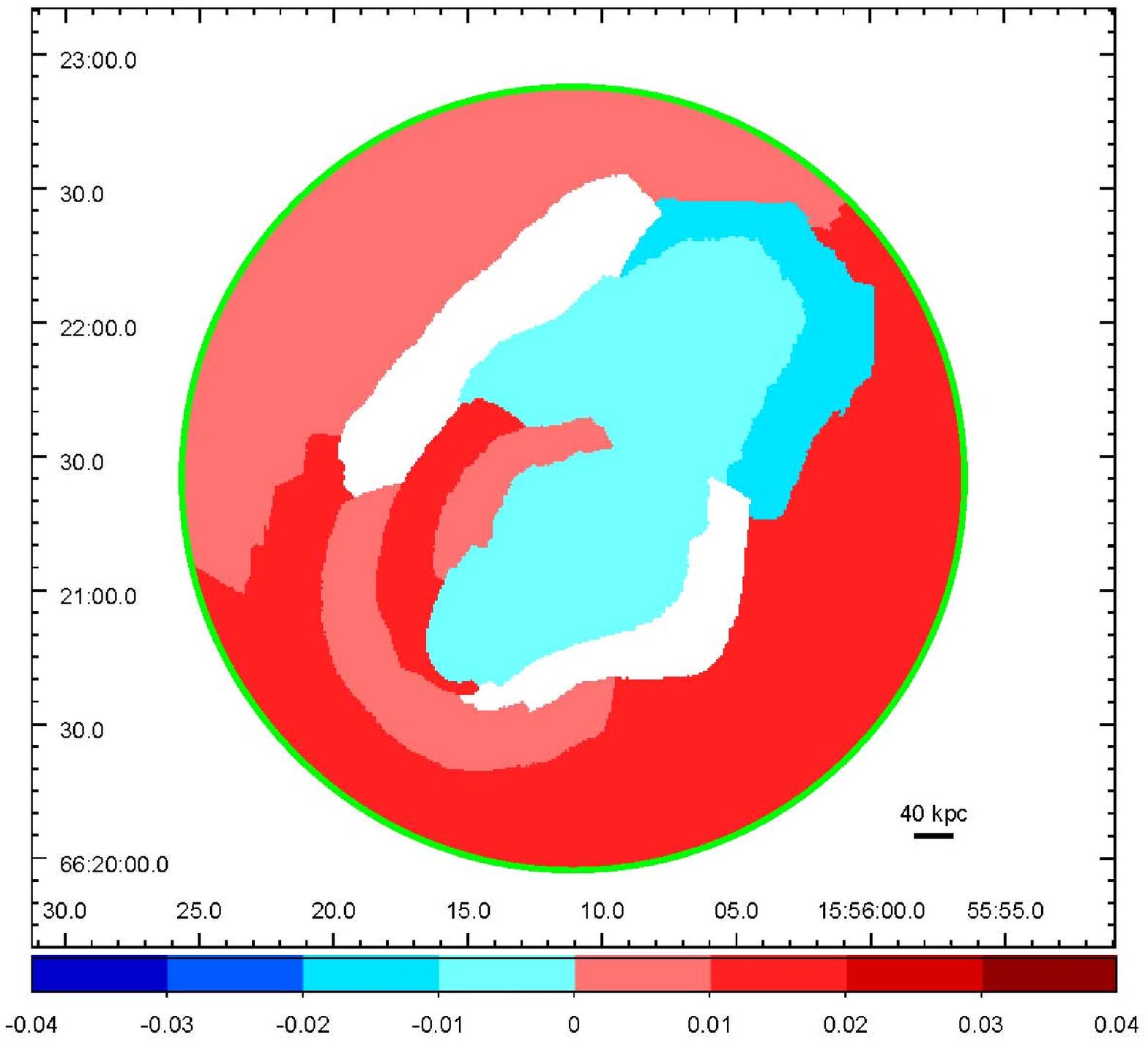}
\includegraphics[width=0.4\textwidth]{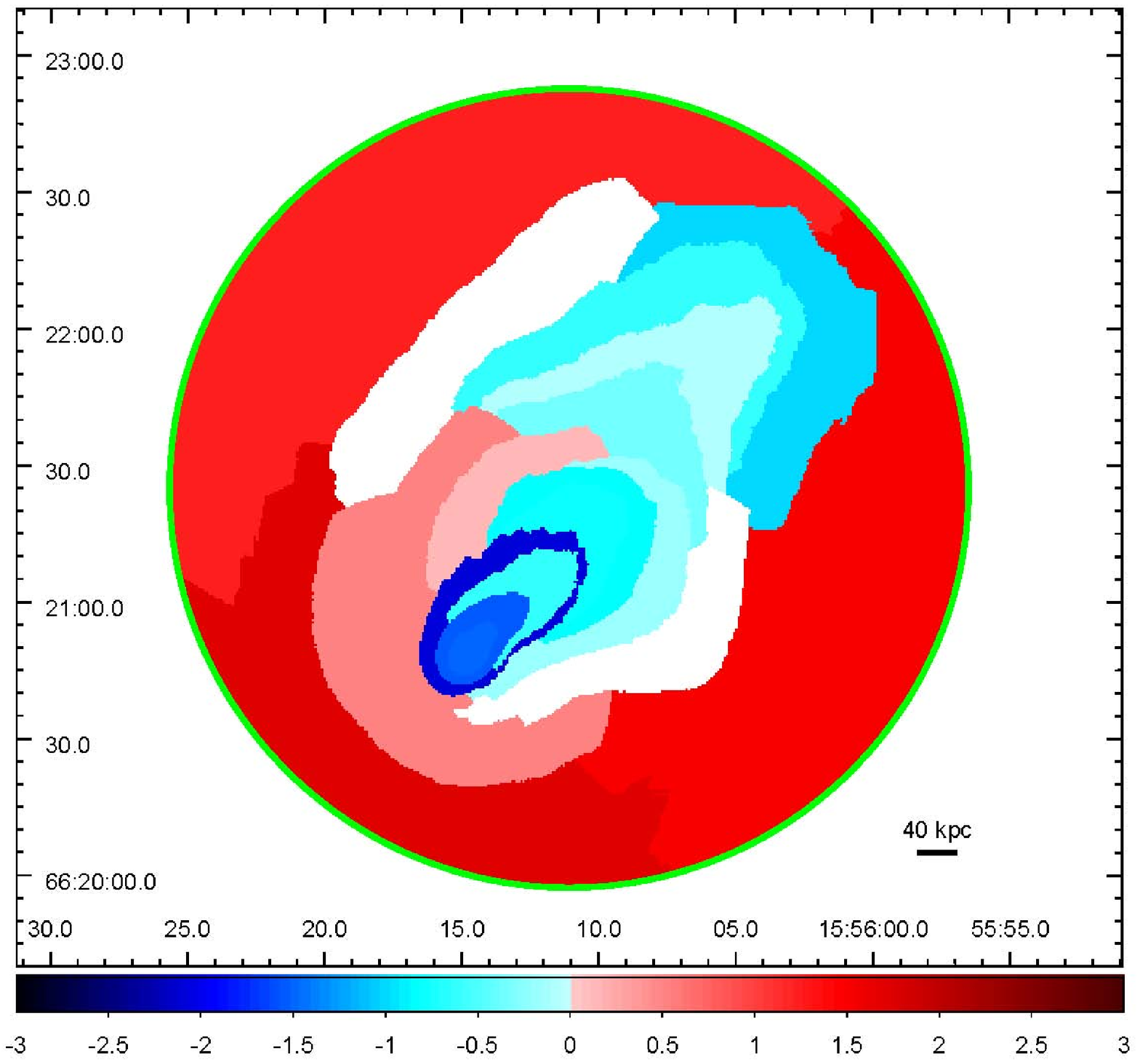}
\caption{Same as Figure \ref{a2142}, but for  Abell 2146. }
\label{a2146}
\vfill
\end{figure*}

We are able to obtain a reliable spectral fit in 17 out of the 19 regions originally selected in the X-ray image.
The best-fit $z_{\rm X}$ of the regions used in our analysis, along with their corresponding statistical and
systematic errors, are listed in Table \ref{table:a2146}.
The average redshift across the region is $\langle z_{\rm X} \rangle  = 0.2338$, which is consistent with
$z_{\rm Xtot}$ within 1$\sigma$.  We find $\chi^2 = 19.68$ for 16 dof, as listed in
Table \ref{chi2}, which is pretty much consistent with a constant redshift across the ICM. The histogram distribution of
$z_{\rm X}$, shown in Figure \ref{histo_A2146}, is consistent with
a Gaussian dominated by noise and shows no clear evidence of bulk motions.
The average bulk motion is formally $v_{\rm BM} = 480 \pm 490\ {\rm km\ s^{-1}}$.  The projected X-ray
redshift map and significance map are shown in Figure \ref{a2146}.
The maximum redshift difference is  $\Delta v_{\rm max} =8200 \pm 3600 \ {\rm km\ s^{-1}}$ and is obtained
between two distinct regions with homogeneous redshift: the ``bullet" and its trail, along the direction of
the merger, which are blueshifted with respect to the average redshift,
and all of the surrounding ICM emission, which is redshifted.  Despite the low S/N, the redshift map depicts a clear dynamical situation where the central ICM is dominated by a halo moving
toward the observer, while the larger scale emission has, on average, a larger redshift.
This provides support for the presence of bulk motions that are still dominated by the infall velocity
of the merging halos.

\subsection{Abell 1689}

Abell 1689 is a typical relaxed cluster at $z\sim 0.183$ \citep{hiroi2013}.  We reduced and analyzed
four ACIS-I observations for a total of 151.3\ ${\rm ks}$ (see Table \ref{tabsample2}).
The total number of net counts in the 0.5--10 keV
band within a circle of 1 arcmin is about $1.86 \times 10^5$ in the ACIS-I merged image.
The best-fit redshift obtained fitting the total X-ray emission is $z_{\rm Xtot} = 0.1814^{+0.0028}_{-0.0008}$.
To evaluate the systematic uncertainty on $z_{\rm X}$, we consider
ranges from 5 to 15 keV and 0.3 to 0.9 $Z/Z_\odot$ to explore the temperature and metal abundance
throughout the ICM, respectively.

\begin{figure*}
\centering
\includegraphics[width=7.8cm]{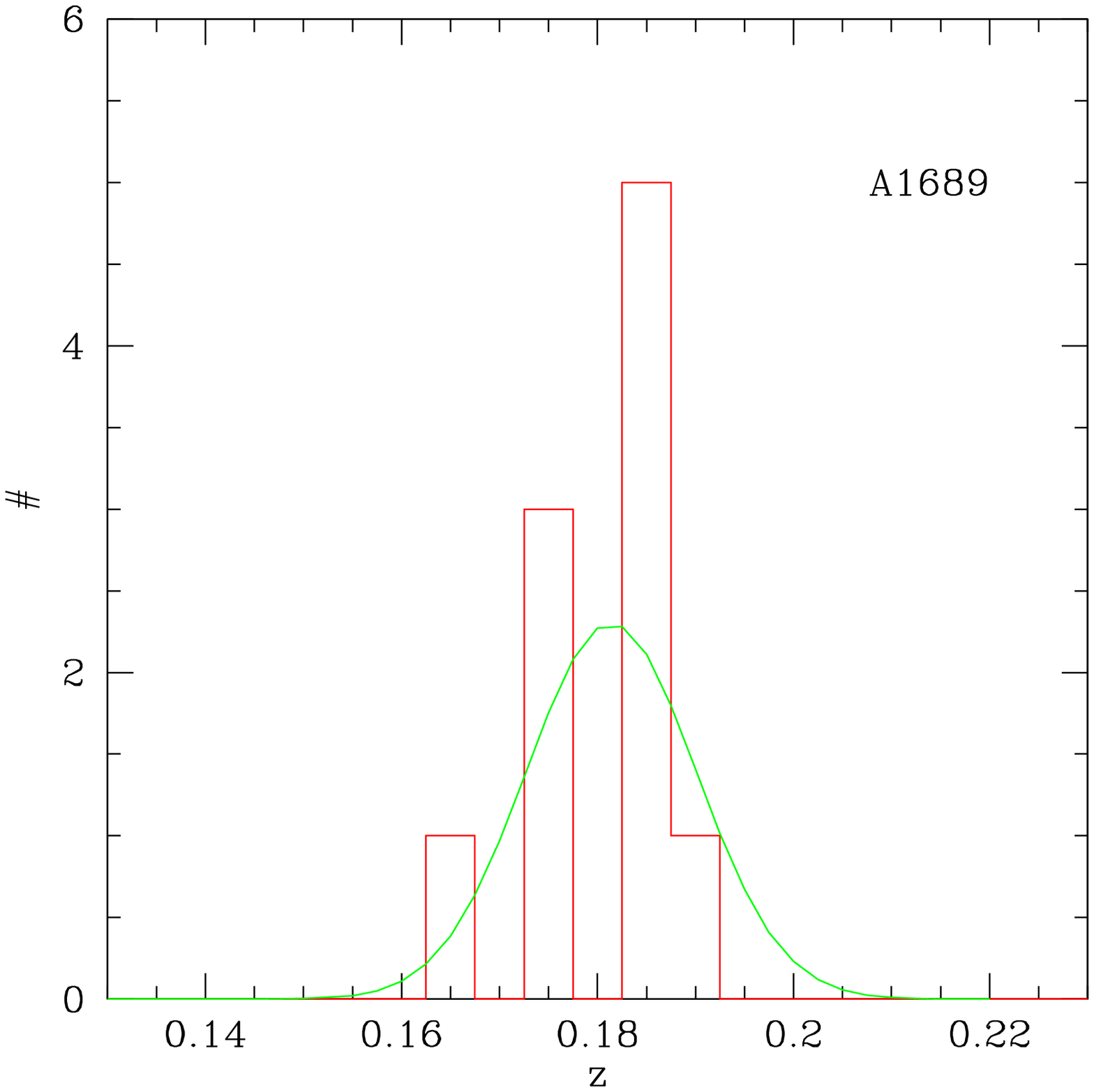}
\caption{Histogram distribution of the best-fit $z_{\rm X}$ for the 10 regions of A1689 with reliable spectral fit.
Lines are as in Figure \ref{histo_A2142}. }
\label{histo_A1689}
\vfill
\end{figure*}

We are able to obtain a reliable spectral fit in all 10 of the regions originally selected in the X-ray image.
The best-fit $z_{\rm X}$ of the regions used in our analysis with the corresponding statistical and
systematic errors are listed in Table \ref{table:a1689}.  The average redshift across the region is
$\langle z_{\rm X} \rangle  = 0.1814$.  The histogram distribution of
$z_{\rm X}$, shown in Figure \ref{histo_A1689}, is consistent with a Gaussian distribution whose $\sigma$ is
the average statistical error on $z_{\rm X}$.  We find $\chi^2 = 9.31$ for 9 dof, as
listed in Table \ref{chi2}.  Therefore, we do not find any hint of bulk motion, as expected, and find a hard
upper limit of $v_{\rm BM} < 1600\ {\rm km\ s^{-1}}$.

\subsection{Abell 1835}

Abell 1835 is a relaxed cluster at $z\sim 0.234$ \citep{girardi2014}.  We reduced and analyzed
three ACIS-I observations for a total of $193.7\ {\rm ks}$ (see Table \ref{tabsample2}).
The total number of net counts in the 0.5--10 keV
band within a circle of 1 arcmin is about $2.33 \times 10^5$  in the ACIS-I merged data.
The best-fit redshift obtained while fitting the total X-ray emission is $z_{\rm Xtot} = 0.2478^{+0.0021}_{-0.0012}$, which is
significantly larger than the optical value.
To evaluate the systematic uncertainty on $z_{\rm X}$ we consider
ranges from 4 to 14 keV and 0.2 to 0.8 $Z/Z_\odot$ to explore the temperature and metal abundance
throughout the ICM, respectively.

\begin{figure*}
\centering
\includegraphics[width=7.8cm]{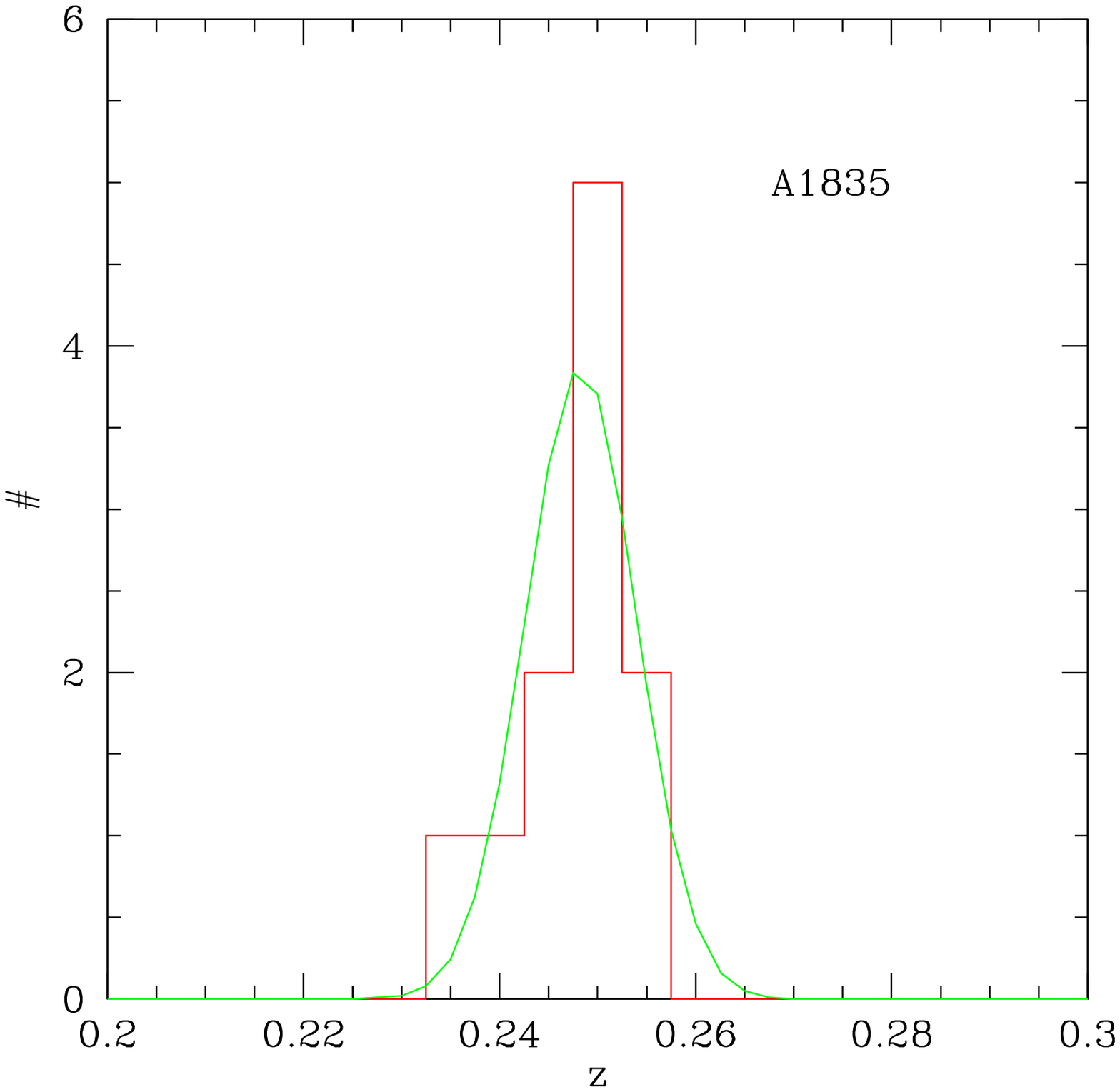}
\caption{Histogram distribution of the best-fit $z_{\rm X}$ for the 11 regions of A1835 with reliable spectral fit.
Lines are as in Figure \ref{histo_A2142}. }
\label{histo_A1835}
\vfill
\end{figure*}

We are able to obtain a reliable spectral fit in all 11 of the regions originally selected in the X-ray image.
The best-fit $z_{\rm X}$ of the regions used in our analysis with the corresponding statistical and
systematic errors are listed in Table \ref{table:a1835}.
The average redshift across the region is $\langle z_{\rm X} \rangle  = 0.2483$. The histogram distribution of
$z_{\rm X}$, shown in Figure \ref{histo_A1835}, is consistent with a Gaussian distribution whose $\sigma$ is
the average statistical error on $z_{\rm X}$.  We find $\chi^2 = 11.5$ for 10 dof, as listed in
Table \ref{chi2}.  Therefore, we do not find any hint of bulk motion, as expected, and find a hard
upper limit of $v_{\rm BM} < 1350\ {\rm km\ s^{-1}}$.

\section{Statistical properties of the cluster sample}

\label{sec5}

\begin{figure*}
\centering
\epsscale{1}
\includegraphics[width=12cm]{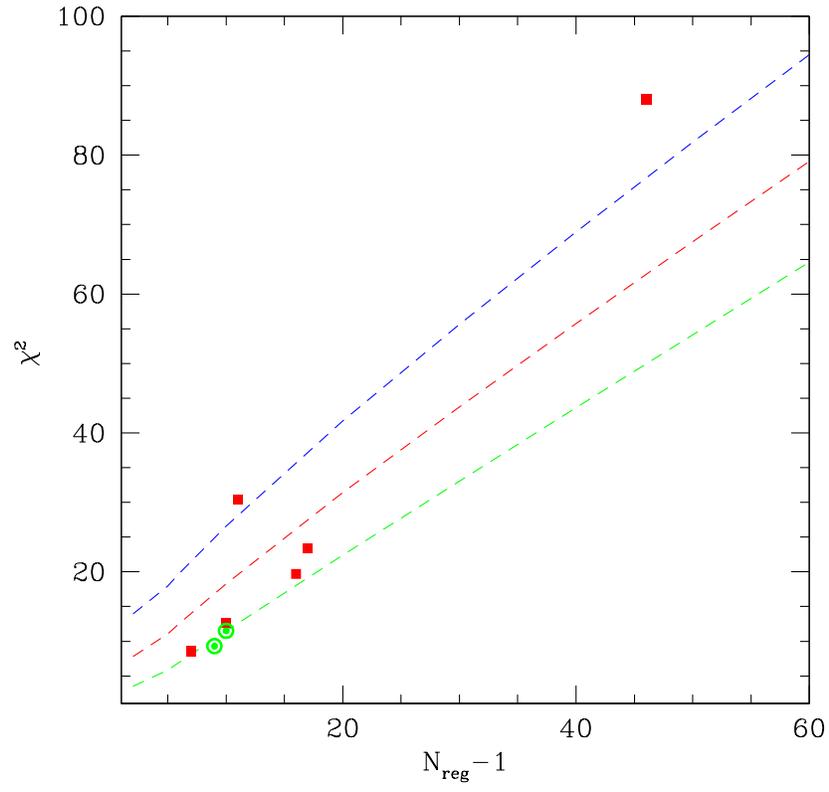}
\caption{Red squares show the $\chi^2$ values  versus the degrees of freedom
(equal to $N_{\rm reg}-1$) for merger clusters,
while circled green points shows the two relaxed clusters.  Dashed lines show the $\chi^2$ values
as a function of the degrees of freedom corresponding to the
1, 2, and 3$\sigma$ confidence levels, from the lowest to the highest. }
\label{chisquare}
\end{figure*}

In Figure \ref{chisquare}, we summarize our results, showing the $\chi^2$ values versus the dof for all of the clusters analyzed in this work and listed in Table \ref{chi2}. Comparison
with the lines corresponding to the 1, 2, and 3$\sigma$ confidence levels shows that, in general,
merging clusters always have a larger probability of showing bulk motions
in their ICM, but only in two cases do we have a statistical significance larger than 3$\sigma$.
In two cases, instead, the probability is less than 2$\sigma$. Finally, in the last two cases,
the global $\chi^2$ is not significantly different from that of the two relaxed clusters, which show no signs
of bulk motions.  Globally, the merging cluster sample always shows a larger probability of hosting
bulk motions, but it is hard to reach definitive conclusions in single cases.

However, we note that the analysis based simply on the $\chi^2$ including all of the ICM regions
tends to dilute the signal, since the majority of the ICM regions are expected to show no significant
bulk motions. In fact, the bulk motion signal is created by only a few, large masses of ICM
displaced at large velocities during the merger, and, in addition, can be partially erased by projection effects.
On the bright side, our global evaluation
of bulk motion is free from the ``look-elsewhere effect" \citep{2008Lyons}.
We conclude that, under our most conservative assumptions, only a fraction of the major mergers shows
unambiguous signatures of bulk motions in the {\sl Chandra} CCD data.
Specifically, we measure  $v_{\rm BM} = 1400 \pm 300 $ and $4600 \pm 1100\ {\rm km\ s^{-1}}$ in A2142 and
A115, respectively.  In A2034 and A520 we find $v_{\rm BM} = 1700\pm 840$ and $1800\pm 900\ {\rm km\ s^{-1}}$,
respectively, although this measurement is not significant (less then 2$\sigma$).
In the cases of 1RXS\ J0603.3+4214 and A2146, we find hard (corresponding to 3$\sigma$)
upper limit of 2100 and $\sim$1500\ ${\rm km\ s^{-1}}$, respectively.
These limits are similar to those found for the two relaxed clusters A1689 and A1835,
where we find $v_{\rm BM} < 1600 $ and $v_{\rm BM} < 1350\ {\rm km\ s^{-1}}$, respectively.

Local velocity differences between ICM regions may achieve much larger values with
corresponding larger errors. In the eighth column of Table \ref{chi2}, we list the values of $\Delta v_{\rm max}$ found by identifying
the ICM regions (or groups of regions) with the largest difference in $z_{\rm X}$.  Clearly,
this does not constitute a detection of bulk motion, however, it does provide a rough estimate of the
maximum velocity difference we can have across the ICM in our data.

Visual inspection of the redshift and significance maps also allows us to provide a rough
classification in terms of pre-merger or ongoing/post-merger.
The first case is when two comparable-mass halos are about to begin a head-on merger with large speed
along the line of sight, so that in the redshift map the two halos appear to have clearly distinct
average redshifts (see, for example, the case of 1RXS\ J0603.3+4214).
The ongoing or post-merger case is when shocks in the ICM are clearly present and
the first pericentric passage is recent (i.e., it happened less than one dynamical time ago, see A2034 and
A115).  Given the uncertainties on the bulk motion
velocity, we are not able to explore the regime where the residual kinetic energy of the merger event has
already decreased to a more diffuse and chaotic turbulent velocity field of the order of few $\times
100\ {\rm km\ s^{-1}}$.  Our qualitative classification is also provided in the last columns of Table \ref{chi2}.

\begin{deluxetable*}{ccccccccc}
\centering
\tablewidth{\textwidth}
\tablecaption{The results of the $\chi^{2}$ test on the distribution of redshift of the clusters, and corresponding
constraints on $v_{\rm BM}$ and $\Delta v_{\rm max}$ in ${\rm km\ s^{-1}}$.  }

\tablehead{
\colhead{Cluster Name}  & \colhead{$z_{\rm Xtot}$}  & \colhead{$\langle z_{\rm X}\rangle$} & \colhead{Degrees of Freedom}    &
\colhead{$\chi^{2}$} & \colhead{Prob of BM} &  \colhead{$v_{\rm BM}$} & \colhead{$\Delta v_{\rm max}$} &
\colhead{Classification} \\
	 & 	 & 	 & \colhead{($N_{\rm reg} - 1$)}    &   &  &  \colhead{${\rm km\ s^{-1}}$} & \colhead{${\rm km\ s^{-1}}$} &
}
\startdata
A2142 & $ 0.0852_{-0.0009}^{+0.0012}$ &  $0.0855$  & 46 & 88.02 & 99.98\%  & $ 1400\pm 300$ & $7400 \pm 2200$   &
post-merger\\
A2034 & $0.1124^{+0.0039}_{-0.0044}$ & 0.1122 & 10 & 12.70 & 76\%   & $1700\pm 840$ & $6200 \pm 3300$  &  pre/ongoing
merger \\
A115  & $0.1892^{+0.0030}_{-0.0040}$ & 0.1979 & 11 & 30.4 & 99.9\%   & $4600\pm 1100$ & $8600\pm 3000$    &
pre/ongoing merger \\
A520 & $0.2082^{+0.0046}_{-0.0049}$ &  0.2083 & 17 & 23.4 & 86\%    & $1800\pm 900$ & $5900\pm 4000$   & post-merger
\\
1RXS\ J0603.3+4214 & $0.2316^{+0.0029}_{-0.0033}$ & 0.2303 & 7 & 8.6 & 72\%    & $<2100$  &   $7200 \pm 2800$   &
pre-merger\\
A2146 & $0.2310^{+0.0022}_{-0.0010}$ &  0.2338 & 16 & 19.68  & 77\%    & $480\pm 490$ & $ 8200 \pm 3600$   &
pre/ongoing merger\\
A1689 & $0.1814^{+0.0028}_{-0.0008}$& 0.1814 & 9 & 9.3 & 59\%   &  $ < 1600$   &  $4600\pm 3000$    & relaxed \\
A1835 & $0.2478^{+0.0021}_{-0.0012}$ & 0.2483 & 10 & 11.5 & 38\%   & $ < 1350$ &  $4200 \pm 2600$   &  relaxed
\enddata
\tablenote{This information is combined with visual
inspection of the redshift maps to obtain
the qualitative classification listed in the last column.}
\label{chi2}
\end{deluxetable*}

\section{Discussion and future prospects}
\label{sec6}

In this work, we use the same conservative approach that we adopted in \citet{aliu2015}, where we
explored the presence of bulk motions in the ICM of 1E0657-56.  In particular, when estimating the
statistical significance of bulk motions, we used five conservative assumptions.

\begin{itemize}
\item[1] We considered only the Fe line emission in the hard band. The use of the soft band would have
significantly reduced the statistical error on $z_{\rm X}$, particulary for the cold regions.
However, since, in principle, gas with different temperatures can have different positions along the
line of sight and
different velocities, this may introduce a source of scatter in $z_{\rm X}$ which is hard to control.
\item[2] We computed the total error on $z_{\rm X}$ in each region, summing in quadrature the statistical
error and the 90\% percentile of the $z_{\rm X}$ distribution obtained by imposing a varying thermal
structure in the ICM.  This systematic component is often smaller than, or at best comparable to, the statistical
component.  The systematic component is likely to be slightly overestimated,
as we can verify by comparing the $\sigma_{\rm rms}$ of the $z_{\rm X}$ distribution with the average total error
$\langle \sigma_{\rm tot} \rangle$.
\item[3] We discarded all of the regions where the best-fit redshift is not identified clearly, rejecting all of the
spectra where the $C_{\rm stat}$ versus redshift plot shows a secondary minimum at a distance
$\Delta C_{\rm stat} < 6.6$ from the absolute minimum.
\item[4] Although the signature of bulk motion is presumably confined to a few, largely discrepant regions,
we evaluate the statistical significance with a simple $\chi^2$ test on all of the regions.  This of course would
dilute the effect, but provides a robust evaluation of the actual presence of bulk motions completely
free from the ``look-elsewhere effect" \citep{2008Lyons}.
\item[5] We use information from the spatial distribution of $z_{\rm X}$ only to provide a qualitative
description of the possible dynamical status of the cluster, and not to reinforce the statistical significance
of the $z_{\rm X}$ fluctuations.  In practice, we do not redefine regions to merge them together according to
their $z_{\rm X}$ value, a procedure that would reduce the statistical uncertainty on the final $z_{\rm X}$ maps.
\end{itemize}

Given our extremely conservative approach, we only hit the tip of the iceberg for signatures of
bulk motions in {\sl Chandra} CCD data. At the same time, our results show both the difficulty in
measuring bulk motions and also the widespread presence of bulk motions in massive, head-on mergers.

Based on these tantalizing but promising results, in the near future we plan to relax some of these
constraints.  In particular, we will include soft-band spectral information and quantitatively investigate the spatial
correlation of the $z_{\rm X}$ fluctuations, possibly merging projected ICM regions that have similar
redshifts on the basis of the spectral analysis and surface brightness features.  In addition,
we can also combine X-ray spectrally resolved analyses with dynamical studies based on the
optical redshift distribution of the member galaxies.  We are currently applying the caustic method
\citep{1999Diaferio,2005Diaferio,2013Serra,2015Yu} to a few nearby, massive clusters in the
Abell cluster catalog \citep{1989Abell}
with the aim of carefully comparing  the optical dynamical structure to the X-ray morphology and
spatially resolved spectral analysis (Yu et al. 2016, in preparation).

Clearly, any improvement that may be achieved with current facilities is limited by the
modest spectral resolution.  In particular, the inevitable superposition of ICM massive regions
with different redshifts
along the line of sight heavily blurs the redshift map.  Sometimes the presence of more than
one minima in the CCD spectra may be interpreted, in principle, as the superposition of two ICM regions
moving at different redshift.  However, it is impossible to reach this kind of conclusion using the
current coarse spectral resolution.  In the near future, the X-ray bolometer on board Astro-H
\citep{2014Takahashi}
will provide ICM spectra with very high resolution, so that the velocity
regime probed in the ICM will reach few $\times 100\ {\rm km\ s^{-1}}$ as opposed to the current limit of
$\sim$2000\ ${\rm km\ s^{-1}}$.  However, the lack of adequate spatial resolution will still make the {\sl Chandra} data
a crucial complementary source of information for exploring the ICM dynamics in the coming years.

\section{Conclusions}
\label{sec7}

We search for bulk motions in the ICM in a sample of six clusters at $0.1<z<0.3$
classified as massive mergers on the basis or their X-ray morphology, diffuse radio emission, or
dynamical optical substructure, plus two relaxed clusters. By performing spatially resolved spectral
analyses on {\sl Chandra} CCD data, we obtain the distribution of the redshift $z_{\rm X}$ across their ICM
after a careful and extremely conservative evaluation of the total
error on each $z_{\rm X}$ value.  We accurately investigate the distribution of $z_{\rm X}$, the
$z_{\rm X}$ map, and finally obtain a robust estimate of bulk motions
in the ICM. Considering the global distribution of $z_{\rm X}$ across the ICM,
our analysis shows that for two of the six merging clusters, there is significant evidence
of bulk motions at more than 3$\sigma$ at the level of $v_{\rm BM} = 1400 \pm 300 $ and
$4600 \pm 1100\ {\rm km\ s^{-1}}$ in A2142 and A115, respectively.
We also obtain measurements at lower significance (less than 2$\sigma$) of
$v_{\rm BM} = 1700\pm 840$ and $1800\pm 900\ {\rm km\ s^{-1}}$ in A2034 and A520, respectively.
Finally, for two clusters (1RXS\ J0603.3+4214 and A2146), the global analysis of the $z_{\rm X}$ distribution
is not significantly different from that of two relaxed clusters (A1689 and A1835),
where we do not find any signs of bulk motion, as expected.

Eventually, we consider specific regions by combining the $z_{\rm X}$ distribution with the redshift maps,
and we are able to identify regions that show relevant bulk motion velocities in all of the merger clusters.
We identify local velocity differences between ICM regions ranging from
5900 to 8600\ ${\rm km\ s^{-1}}$,
with typical uncertainties of 30\%--50\%. Whenever possible, we use the spatial distribution of $z_{\rm X}$
to infer a qualitative description of the dynamical status of the clusters,
tentatively classifying them as pre-merger, ongoing, or post-merger.  Our results are summarized
in Table \ref{chi2}.

We conclude that {\sl Chandra} data can be successfully used to detect ICM bulk motions in massive
merging clusters at $0.1<z<0.3$.  Unfortunately, the CCD coarse spectral resolution makes it impossible
to provide a meaningful reconstruction of the velocity field of the ICM in single clusters.  However,
as soon as high-resolution, spatially resolved X-ray spectroscopy is available, the ICM dynamics
will become a key diagnostic for understanding the dynamics of whole clusters, the physics of the
ICM itself, for achieving accurate measurement of the total mass, and, ultimately, for performing accurate
cosmological tests based on cluster abundances.  We are
currently planning a refined strategy to carefully relax some of our assumptions, and may
reach a lower velocity regime with current CCD data.  The hard limit imposed by the spectral
resolution can be partially overcome by near-future X-ray bolometers.  However, due to the low
angular resolution of X-ray bolometers, {\sl Chandra} data will be a key complementary probe
for ICM dynamics in the coming years.

\acknowledgments
This work was supported by the Chinese Ministry of Science and Technology
National Basic Science Program (Project 973) under grants No.2012CB821804 and 2014CB845806,
the Strategic Priority Research Program ``The Emergence of Cosmological Structure" of the Chinese
Academy of Sciences (No. XDB09000000), the National Natural Science Foundation of China under
grants Nos. 11373014, 11073005, and 11403002, and
the Fundamental Research Funds for the Central Universities and Scientific Research Foundation
of Beijing Normal University.
P.T. is supported by the Recruitment Program of High-end Foreign
Experts and he gratefully acknowledges the hospitality of Beijing Normal University.
A.L. is supported by NSF grant AST-1413056.

\newpage

\begin{figure*}
\centering
\includegraphics[width=0.4\textwidth]{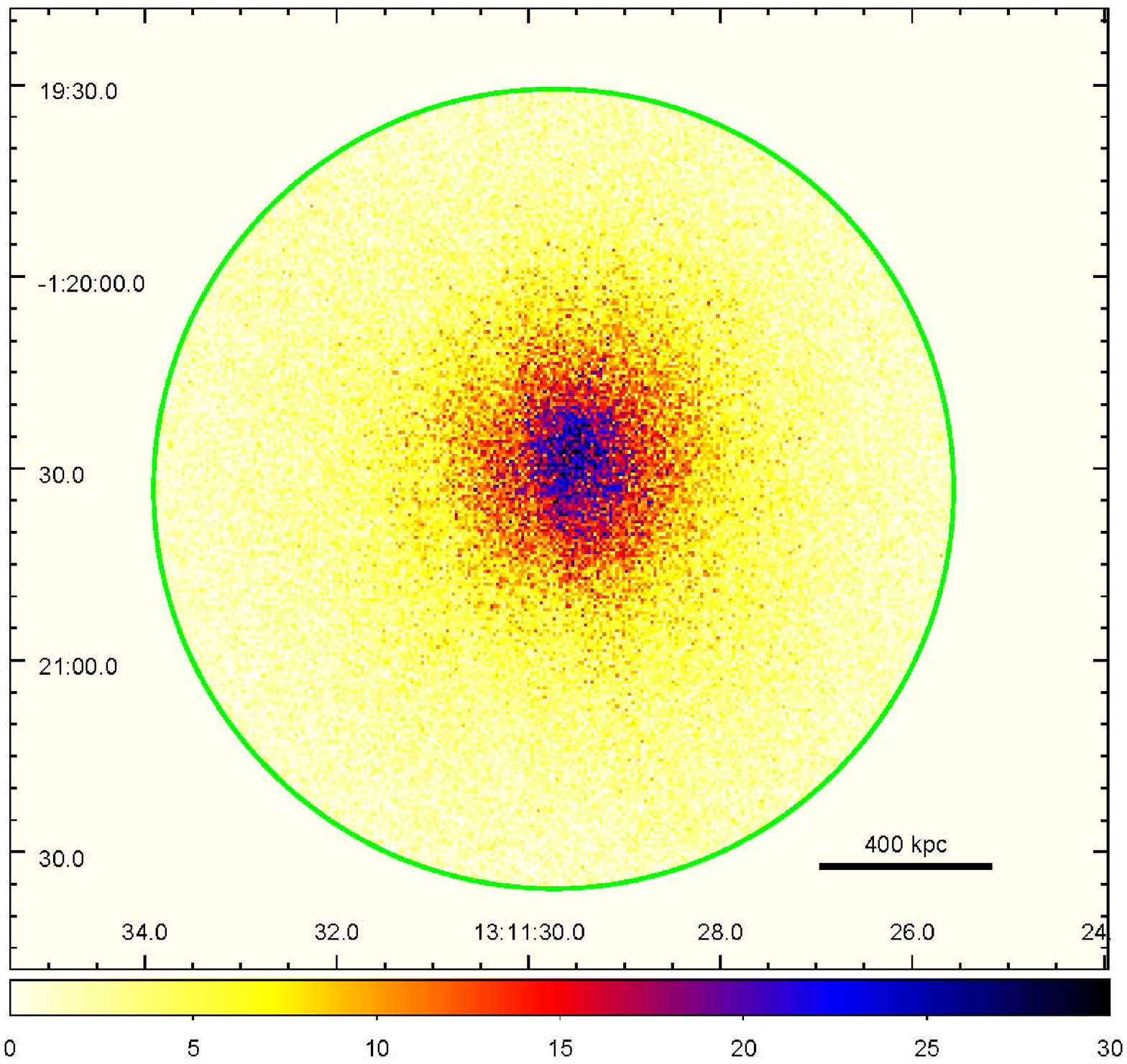}
\includegraphics[width=0.4\textwidth]{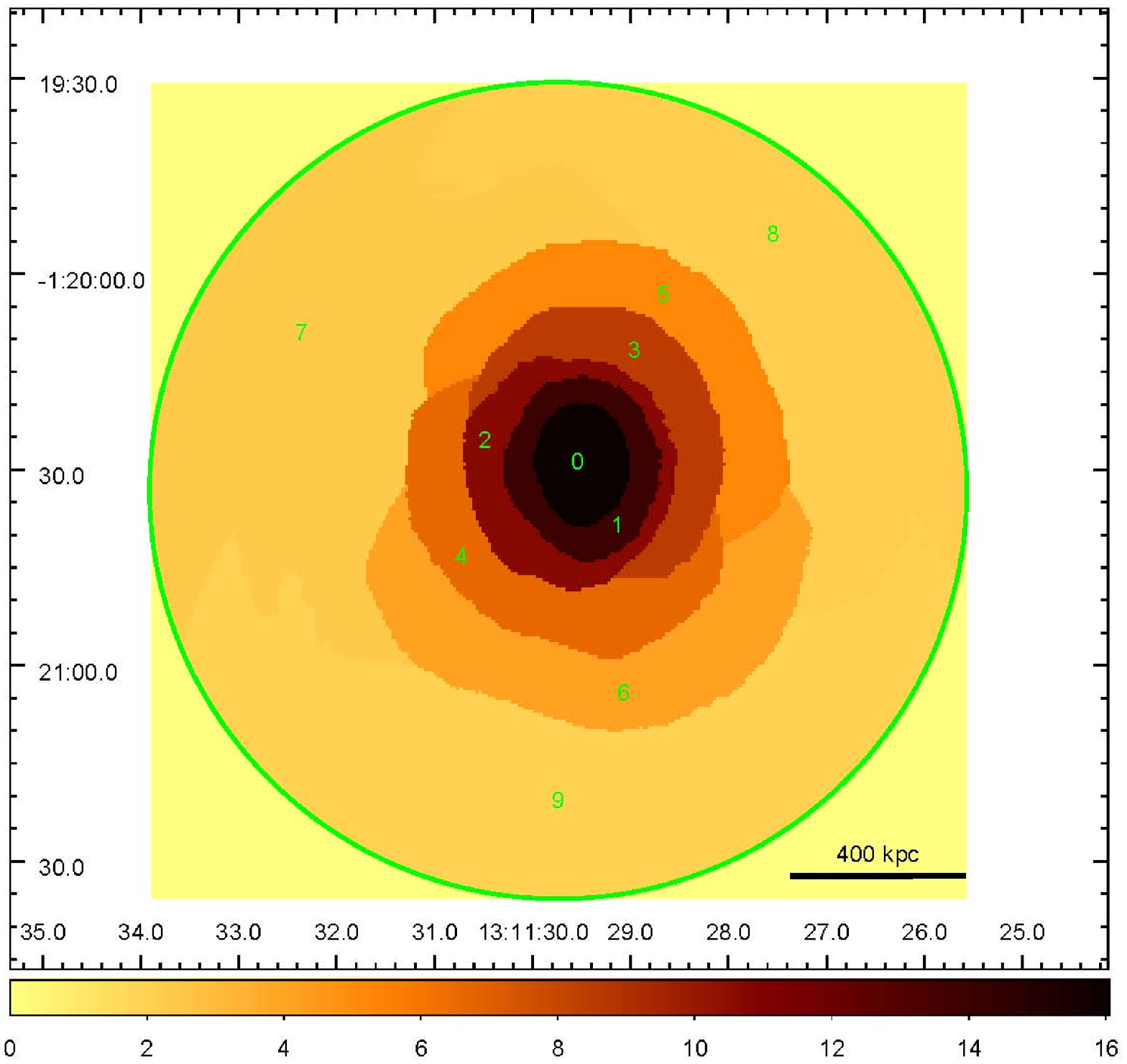}
\includegraphics[width=0.4\textwidth]{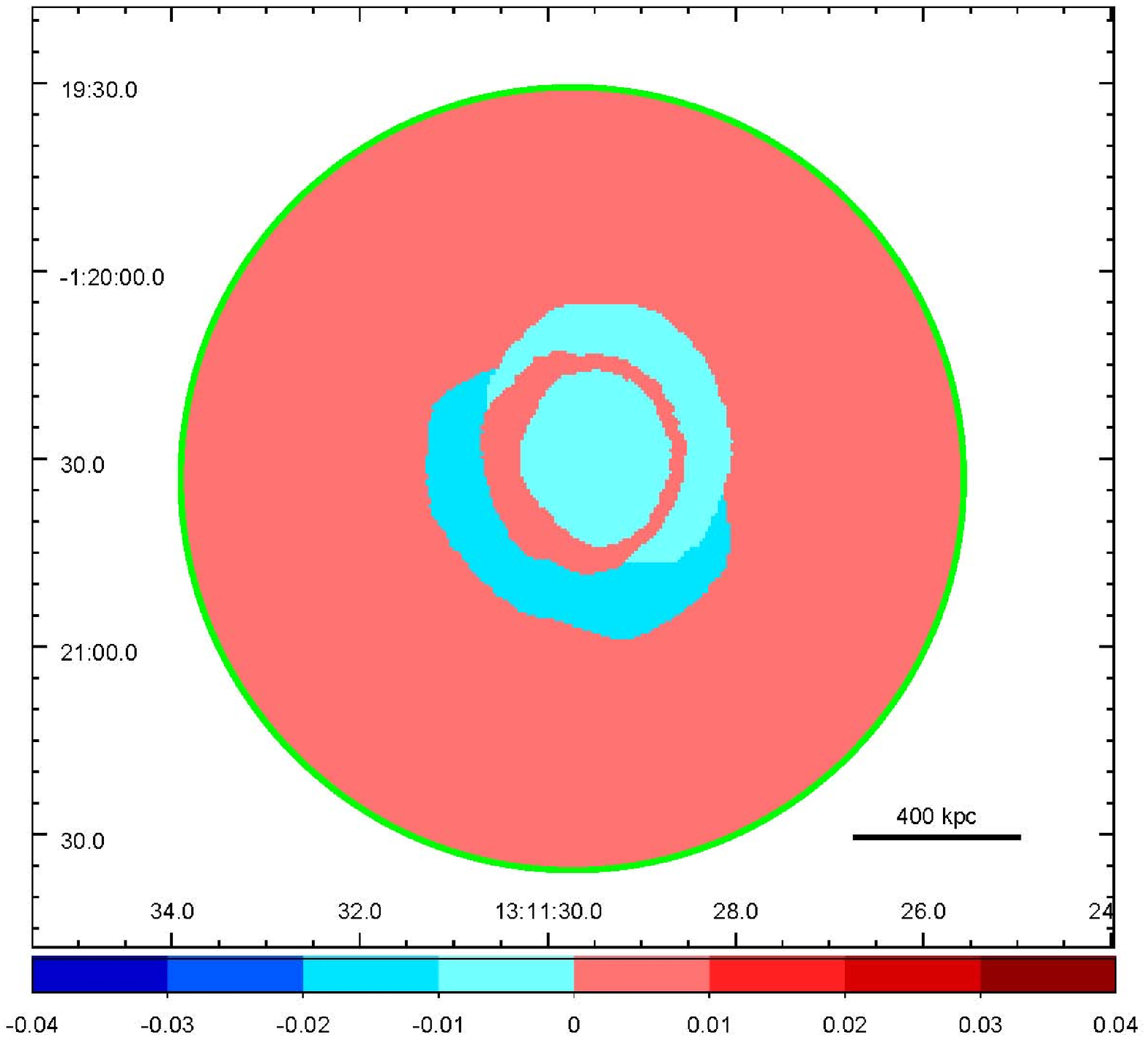}
\includegraphics[width=0.4\textwidth]{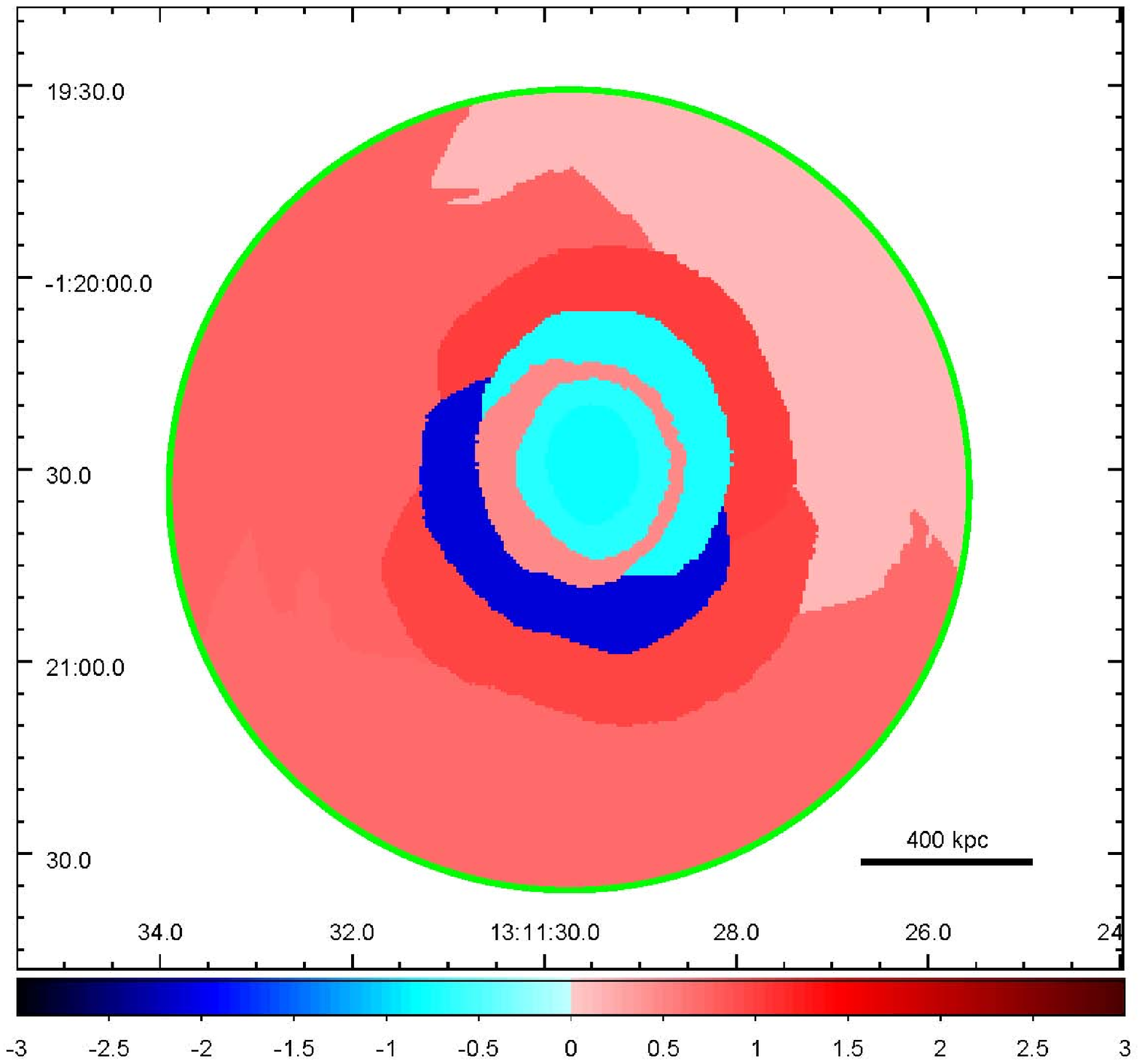}
\caption{Same as Figure \ref{a2142}, but for Abell 1689. }
\label{a1689}
\vfill
\end{figure*}

\begin{figure*}
\centering
\includegraphics[width=0.4\textwidth]{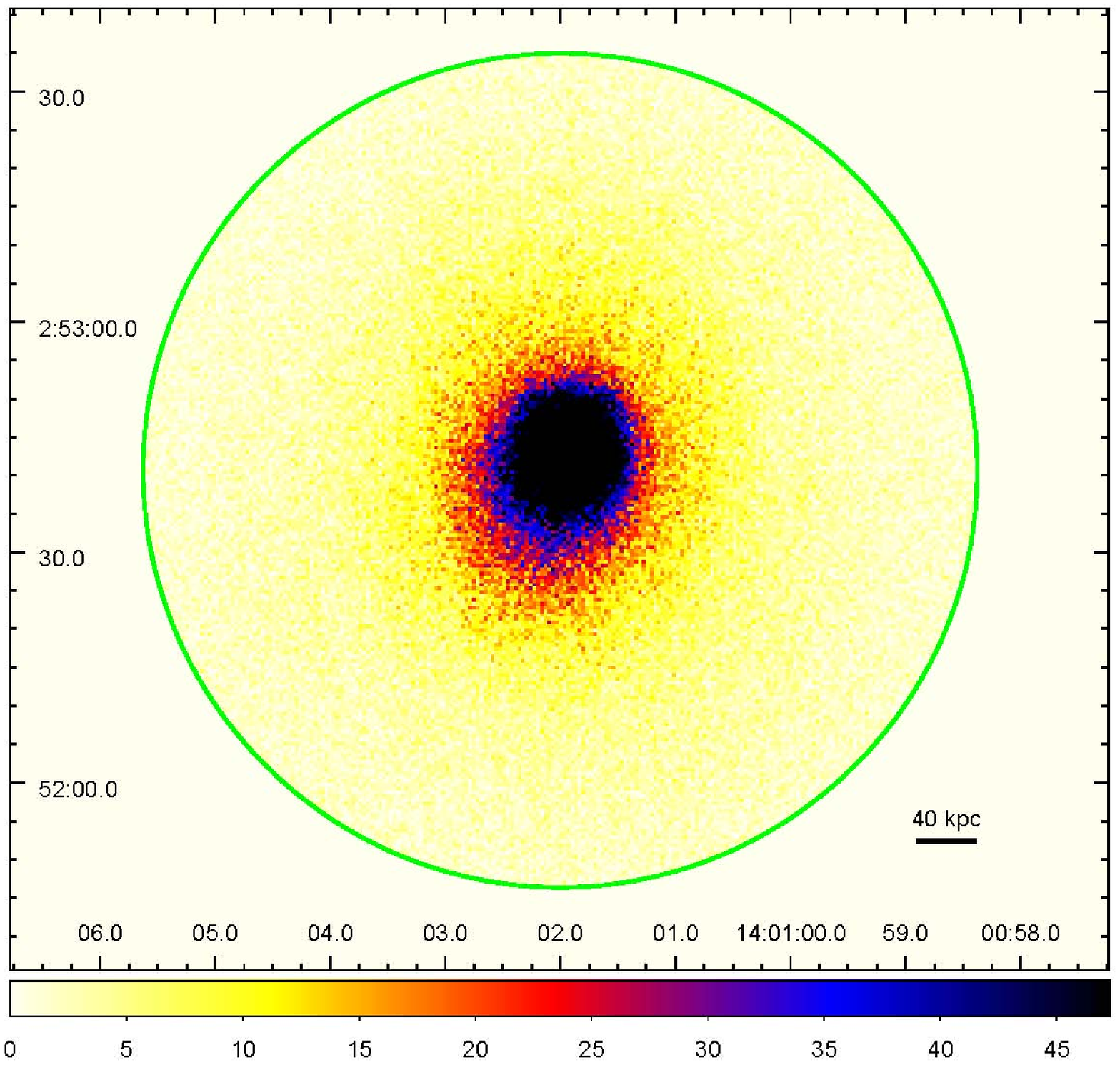}
\includegraphics[width=0.4\textwidth]{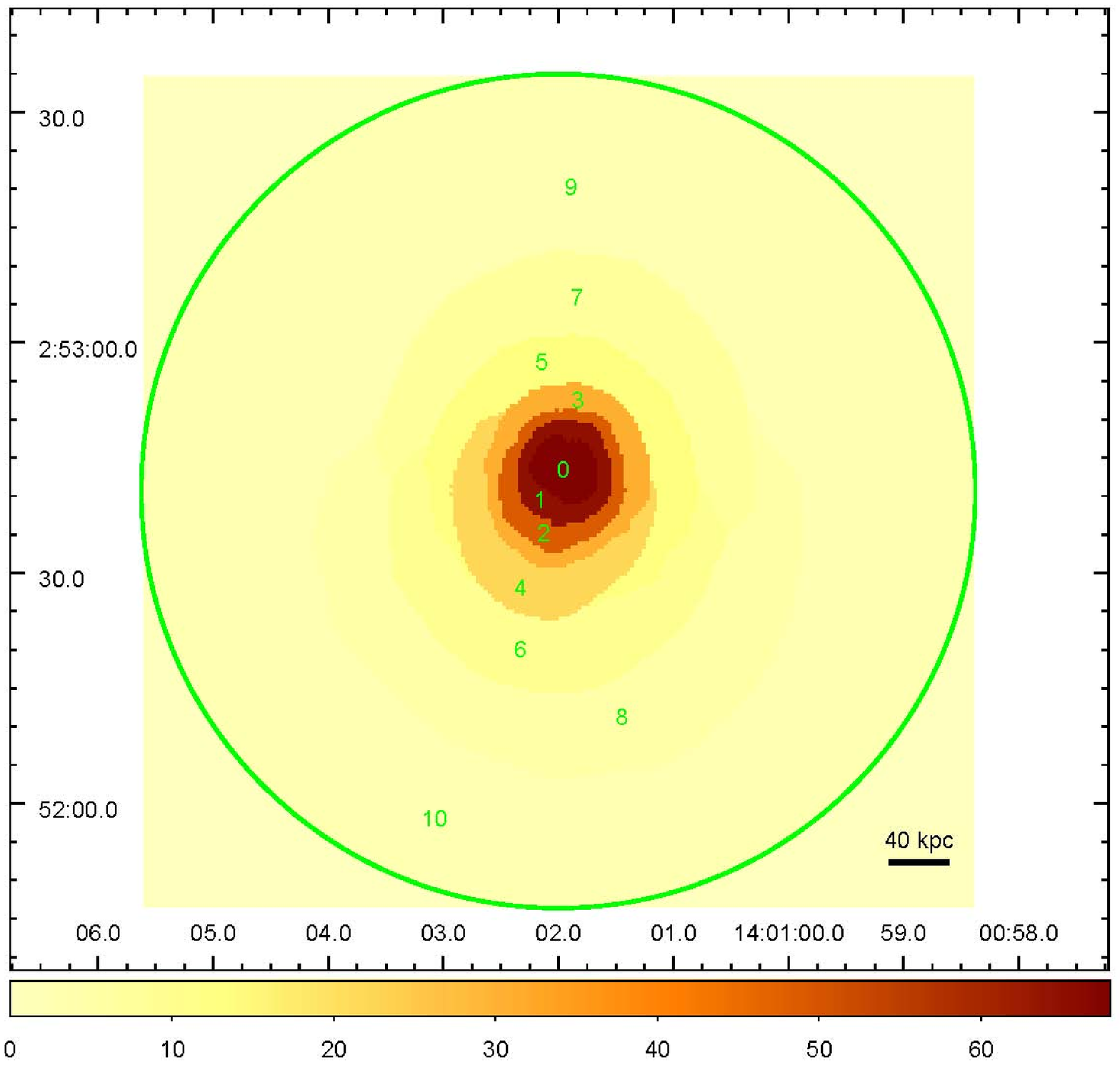}
\includegraphics[width=0.4\textwidth]{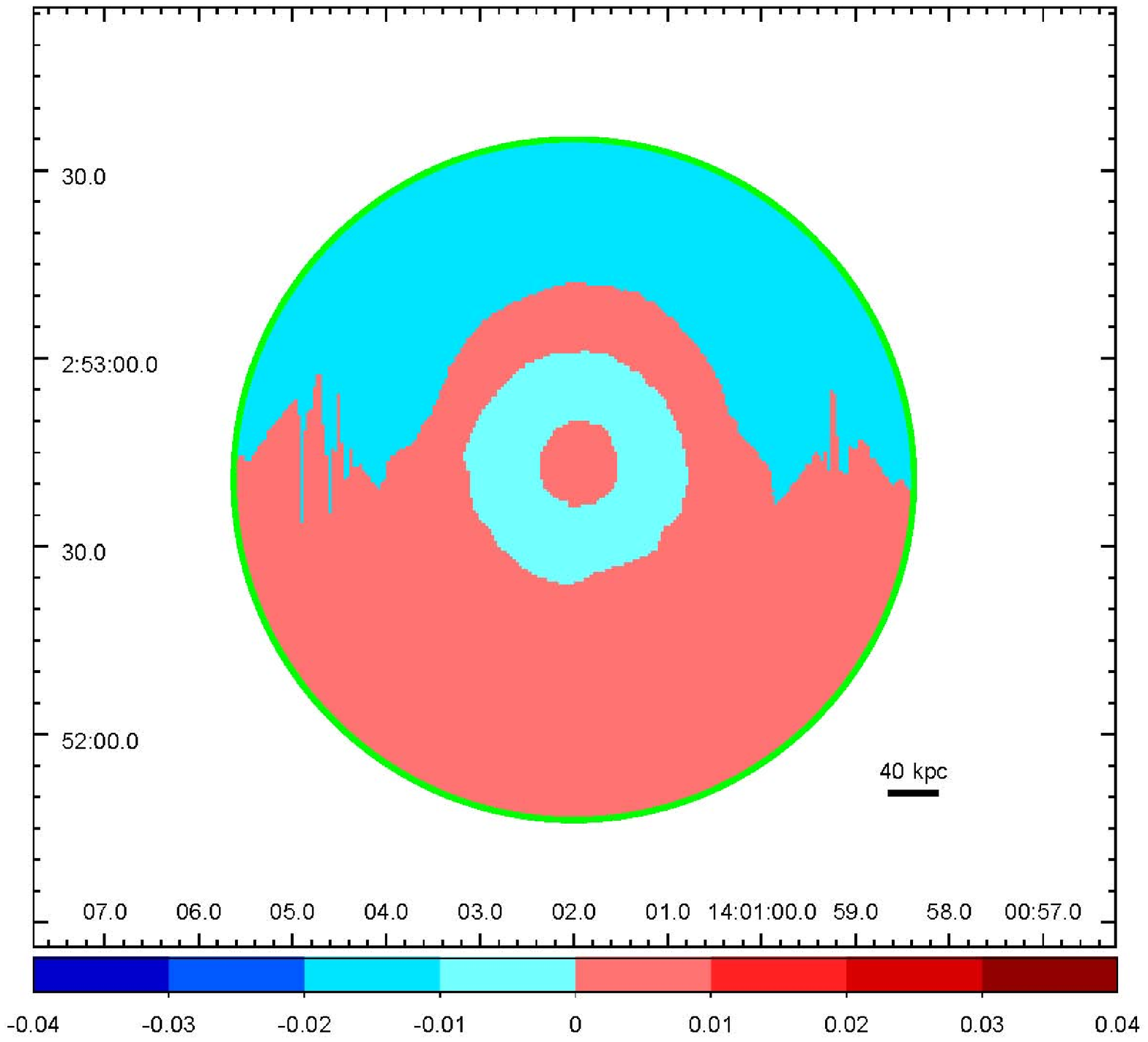}
\includegraphics[width=0.4\textwidth]{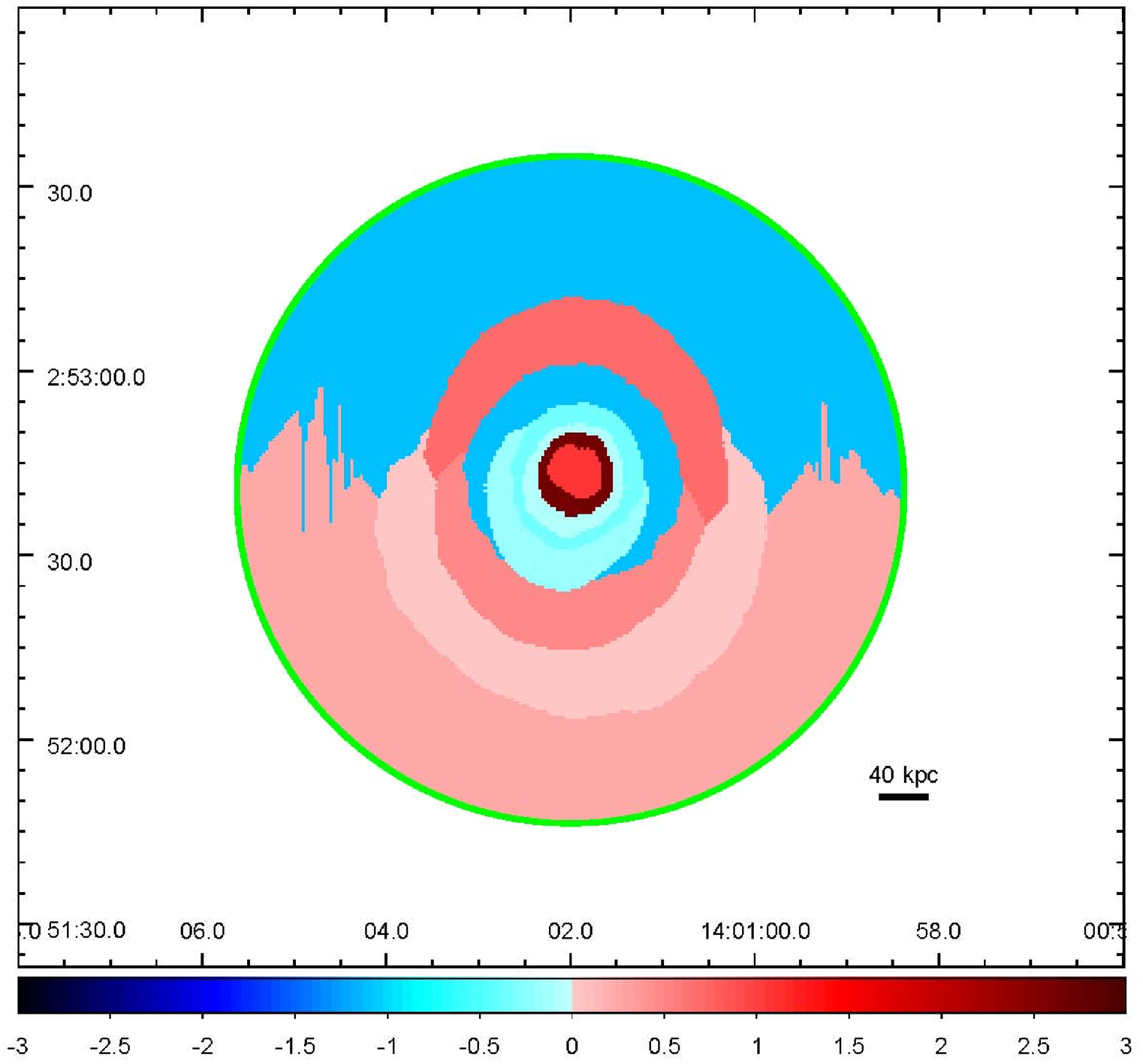}
\caption{Same as Figure \ref{a2142}, but for  Abell 1835. }
\label{a1835}
\vfill
\end{figure*}

\clearpage

\bibliography{ref}

\appendix
\label{appendix}
\section{Spectral analysis results of the clusters}

For each cluster analyzed in this work we report the best fit redshift $z_{\rm X}$ along with the statistical,
systematic and total error in all the selected regions with reliable spectral fit.

\begin{deluxetable*}{cccccccc}
\centering
\tablewidth{\textwidth}

\tablecaption{Spectral analysis results of Abell 2142.
 }
\tablehead{
\colhead{Region}           & \colhead{$z$}      &
\colhead{$\sigma_{\rm stat_{b}}$} & \colhead{$\sigma_{\rm stat_{u}}$} &
\colhead{$\sigma_{\rm syst_{b}}$} & \colhead{$\sigma_{\rm syst_{u}}$} &
\colhead{$\sigma_{\rm tot_{b}}$}  & \colhead{$\sigma_{\rm tot_{u}}$}}
\startdata
0	 &  0.0839 & 	-0.0030 & 0.0033 & 	-0.0009 & 	0.0002 &	-0.0031 & 0.0034 \\
1	 &  0.0847 & 	-0.0021 & 0.0026 & 	-0.0007 & 	0.0013 &	-0.0022 & 0.0029 \\
2	 &  0.0674 & 	-0.0048 & 0.0050 & 	-0.0014 & 	0.0005 &	-0.0051 & 0.0050 \\
3	 &  0.0881 & 	-0.0013 & 0.0029 & 	-0.0001 & 	0.0018 &	-0.0014 & 0.0034 \\
4	 &  0.0762 & 	-0.0064 & 0.0061 & 	-0.0040 & 	0.0001 &	-0.0076 & 0.0061 \\
5	 &  0.0863 & 	-0.0065 & 0.0058 & 	-0.0054 & 	0.0026 &	-0.0084 & 0.0063 \\
6	 &  0.0917 & 	-0.0022 & 0.0024 & 	-0.0001 & 	0.0013 &	-0.0022 & 0.0027 \\
7	 &  0.0866 & 	-0.0031 & 0.0032 & 	-0.0001 & 	0.0015 &	-0.0031 & 0.0035 \\
8	 &  0.0612 & 	-0.0064 & 0.0075 & 	-0.0001 & 	0.0455 &	-0.0064 & 0.0461 \\
10 & 	0.0808 & 	-0.0039 & 0.0039 & 	-0.0001 & 	0.0040 &	-0.0039 & 0.0055 \\
11 & 	0.0843 & 	-0.0030 & 0.0032 & 	-0.0001 & 	0.0020 &	-0.0030 & 0.0038 \\
12 & 	0.0885 & 	-0.0042 & 0.0098 & 	-0.0001 & 	0.0010 &	-0.0042 & 0.0099 \\
13 & 	0.0946 & 	-0.0034 & 0.0036 & 	-0.0001 & 	0.0014 &	-0.0034 & 0.0039 \\
14 & 	0.0787 & 	-0.0043 & 0.0036 & 	-0.0007 & 	0.0013 &	-0.0044 & 0.0039 \\
15 & 	0.0874 & 	-0.0043 & 0.0036 & 	-0.0001 & 	0.0016 &	-0.0043 & 0.0040 \\
16 & 	0.0860 & 	-0.0036 & 0.0038 & 	-0.0001 & 	0.0030 &	-0.0036 & 0.0049 \\
17 & 	0.0888 & 	-0.0029 & 0.0031 & 	-0.0020 & 	0.0001 &	-0.0035 & 0.0031 \\
18 & 	0.0908 & 	-0.0032 & 0.0024 & 	-0.0001 & 	0.0013 &	-0.0032 & 0.0027 \\
20 & 	0.0844 & 	-0.0029 & 0.0028 & 	-0.0004 & 	0.0015 &	-0.0030 & 0.0031 \\
21 & 	0.0885 & 	-0.0061 & 0.0083 & 	-0.0015 & 	0.0025 &	-0.0063 & 0.0087 \\
22 & 	0.0852 & 	-0.0033 & 0.0033 &  -0.0012 & 	0.0020 &	-0.0036 & 0.0039 \\
23 & 	0.0841 & 	-0.0037 & 0.0038 & 	-0.0001 & 	0.0020 &	-0.0037 & 0.0043 \\
25 & 	0.0848 & 	-0.0028 & 0.0027 & 	-0.0001 & 	0.0022 &	-0.0028 & 0.0034 \\
26 & 	0.0811 & 	-0.0027 & 0.0029 & 	-0.0010 & 	0.0010 &	-0.0028 & 0.0031 \\
27 & 	0.0882 & 	-0.0040 & 0.0038 & 	-0.0010 & 	0.0001 &	-0.0041 & 0.0038 \\
28 & 	0.0831 & 	-0.0029 & 0.0024 & 	-0.0010 & 	0.0001 &	-0.0031 & 0.0024 \\
29 & 	0.0777 & 	-0.0044 & 0.0045 & 	-0.0017 & 	0.0043 &	-0.0047 & 0.0062 \\
31 & 	0.0914 & 	-0.0036 & 0.0040 & 	-0.0001 & 	0.0025 &	-0.0036 & 0.0047 \\
32 & 	0.0869 & 	-0.0023 & 0.0022 & 	-0.0010 & 	0.0010 &	-0.0025 & 0.0024 \\
33 & 	0.0878 & 	-0.0071 & 0.0100 &  -0.0028 & 	0.0132 &	-0.0076 & 0.0166 \\
34 & 	0.0847 & 	-0.0027 & 0.0027 & 	-0.0007 & 	0.0007 &	-0.0028 & 0.0028 \\
35 & 	0.0966 & 	-0.0083 & 0.0058 & 	-0.0087 & 	0.0043 &	-0.0120 & 0.0072 \\
36 & 	0.0846 & 	-0.0027 & 0.0026 & 	-0.0006 & 	0.0024 &	-0.0028 & 0.0035 \\
37 & 	0.0821 & 	-0.0053 & 0.0050 & 	-0.0010 & 	0.0060 &	-0.0054 & 0.0078 \\
38 & 	0.0808 & 	-0.0045 & 0.0046 & 	-0.0013 & 	0.0032 &	-0.0047 & 0.0056 \\
39 & 	0.0853 & 	-0.0034 & 0.0035 & 	-0.0001 & 	0.0016 &	-0.0034 & 0.0038 \\
40 & 	0.0832 & 	-0.0033 & 0.0035 & 	-0.0007 & 	0.0023 &	-0.0034 & 0.0042 \\
41 & 	0.0826 & 	-0.0055 & 0.0041 & 	-0.0001 & 	0.0034 &	-0.0055 & 0.0053 \\
42 & 	0.0876 & 	-0.0024 & 0.0027 & 	-0.0001 & 	0.0030 &	-0.0024 & 0.0040 \\
43 & 	0.0820 & 	-0.0049 & 0.0053 & 	-0.0001 & 	0.0050 &	-0.0049 & 0.0073 \\
44 & 	0.0848 & 	-0.0027 & 0.0023 & 	-0.0008 & 	0.0012 &	-0.0028 & 0.0026 \\
45 & 	0.0884 & 	-0.0054 & 0.0051 & 	-0.0024 & 	0.0036 &	-0.0059 & 0.0062 \\
47 & 	0.0823 & 	-0.0037 & 0.0036 & 	-0.0001 & 	0.0010 &	-0.0037 & 0.0037 \\
48 & 	0.0885 & 	-0.0029 & 0.0031 & 	-0.0001 & 	0.0010 &	-0.0029 & 0.0033 \\
49 & 	0.0866 & 	-0.0028 & 0.0030 & 	-0.0001 & 	0.0014 &	-0.0028 & 0.0033 \\
50 & 	0.1124 & 	-0.0052 & 0.0063 & 	-0.0025 & 	0.0045 &	-0.0057 & 0.0077 \\
51 & 	0.0881 & 	-0.0049 & 0.0044 & 	-0.0001 & 	0.0052 &	-0.0049 & 0.0068

\label{table:a2142}
\tablenote{Column 1: region number;
Column 2: best-fit redshift $z_{\rm X}$  obtained fitting the 2.0--10 keV energy range;
Columns 3 \& 4: lower and upper $1\sigma$ error bars from fit statistics;
Columns 5  \& 6:  lower and upper $1\sigma$ error bars from systematics associated to
the ICM temperature structure;
Columns 7\& 8: total lower and upper $1\sigma$ error bars computed as
$\sigma_{\rm tot} = \sqrt{\sigma_{\rm stat}^{2}+\sigma_{\rm syst}^{2}}$. }
\enddata
\end{deluxetable*}

\begin{deluxetable*}{cccccccc}
\centering
\tablewidth{\textwidth}

\tablecaption{
Spectral analysis results of Abell 2034.   }
\tablehead{
\colhead{Region}           & \colhead{$z$}      &
\colhead{$\sigma_{\rm stat_{b}}$} & \colhead{$\sigma_{\rm stat_{u}}$} &
\colhead{$\sigma_{\rm syst_{b}}$} & \colhead{$\sigma_{\rm syst_{u}}$} &
\colhead{$\sigma_{\rm tot_{b}}$}  & \colhead{$\sigma_{\rm tot_{u}}$}}
\startdata
0	 & 0.1056	 & -0.0057 & 0.0056	 & -0.0001 & 0.0033 & -0.0057 & 	0.0065 \\
1	 & 0.1085	 & -0.006	 & 0.0068	 & -0.0016 & 0.0014 & -0.0062 & 	0.0069 \\
2	 & 0.1004	 & -0.0062 & 0.0088	 & -0.0001 & 0.0054 & -0.0062 & 	0.0103 \\
3	 & 0.1089	 & -0.0074 & 0.008	 & -0.0017 & 0.0053 & -0.0075 & 	0.0095 \\
4	 & 0.1318	 & -0.0093 & 0.0076	 & -0.0035 & 0.0001 & -0.0099 & 	0.0076 \\
6	 & 0.1116	 & -0.0038 & 0.0055	 & -0.0001 & 0.0018 & -0.0038 & 	0.0057 \\
9	 & 0.1301	 & -0.0067 & 0.0058	 & -0.0049 & 0.0001 & -0.0083 & 	0.0058 \\
10 & 0.1124	 & -0.0055 & 0.0059	 & -0.0018 & 0.0001 & -0.0057 & 	0.0059 \\
11 & 0.1099	 & -0.0077 & 0.0061	 & -0.0051 & 0.0001 & -0.0092 & 	0.0061 \\
12 & 0.1142	 & -0.0075 & 0.0102	 & -0.0035 & 0.0018 & -0.0082 & 	0.0103 \\
13 & 0.1011	 & -0.0068 & 0.0095	 & -0.0001 & 0.0040 & -0.0068 & 	0.0103
\label{table:a2034}
\tablenote{The
columns are the same as in Table \ref{table:a2142}. }
\enddata
\end{deluxetable*}

\begin{deluxetable*}{cccccccc}
\centering
\tablewidth{\textwidth}

\tablecaption{
Spectral analysis results of Abell 115.  }
\tablehead{
\colhead{Region}           & \colhead{$z$}      &
\colhead{$\sigma_{\rm stat_{b}}$} & \colhead{$\sigma_{\rm stat_{u}}$} &
\colhead{$\sigma_{\rm syst_{b}}$} & \colhead{$\sigma_{\rm syst_{u}}$} &
\colhead{$\sigma_{\rm tot_{b}}$}  & \colhead{$\sigma_{\rm tot_{u}}$}}
\startdata
0 & 0.1980 & -0.0019 & 0.0036 & -0.0000 & 0.0025 & -0.0019 & 0.0044 \\
1 & 0.1998 & -0.0027 & 0.0028 & -0.0000 & 0.0002 & -0.0027 & 0.0028  \\
2 & 0.2026 & -0.0036 & 0.0039 & -0.0000 & 0.0037 & -0.0036 & 0.0053 \\
4 & 0.1988 & -0.0051 & 0.0042 & -0.0000 & 0.0061 & -0.0051 & 0.0074 \\
6 & 0.1922 & -0.0067 & 0.0083 & -0.0032 & 0.0018 & -0.0074 & 0.0085 \\
8 & 0.1897 & -0.0124 & 0.0123 & -0.0037 & 0.0053 & -0.0129 & 0.0134 \\
10 & 0.2090 & -0.0082 & 0.0086 & -0.0040 & 0.0040 & -0.0091 & 0.0095 \\
11 & 0.2367 & -0.0060 & 0.0048 & -0.0087 & 0.0000 & -0.0106 & 0.0048 \\
12 & 0.1679 & -0.0094 & 0.0113 & -0.0049 & 0.0020 & -0.0106 & 0.0115  \\
14 & 0.1591 & -0.0064 & 0.0094 & -0.0000 & 0.0210 & -0.0064 & 0.0230  \\
15 & 0.2263 & -0.0177 & 0.0107 & -0.0000 & 0.0000 & -0.0177 & 0.0107 \\
17 & 0.1943 & -0.0069 & 0.0074 & -0.0038 & 0.0107 & -0.0079 & 0.0130

\label{table:a115}
\tablenote{ The
columns are the same as in Table \ref{table:a2142}.}
\enddata
\end{deluxetable*}

\begin{deluxetable*}{cccccccc}
\centering
\tablewidth{\textwidth}
\tablecaption{
Spectral analysis results of Abell 520.  }
\tablehead{
\colhead{Region}           & \colhead{$z$}      &
\colhead{$\sigma_{\rm stat_{b}}$} & \colhead{$\sigma_{\rm stat_{u}}$} &
\colhead{$\sigma_{\rm syst_{b}}$} & \colhead{$\sigma_{\rm syst_{u}}$} &
\colhead{$\sigma_{\rm tot_{b}}$}  & \colhead{$\sigma_{\rm tot_{u}}$}}
\startdata
0 	& 0.2102	 & -0.0037	 & 0.0065	 & -0.0001	 & 0.0021	 & -0.0037 & 0.0068 \\
1 	& 0.2069	 & -0.0047	 & 0.0043	 & -0.0001	 & 0.0027	 & -0.0047 & 0.0050 \\
2 	& 0.1933	 & -0.0085	 & 0.0096	 & -0.0001	 & 0.0074	 & -0.0085 & 0.0121 \\
3 	& 0.1962	 & -0.0075	 & 0.0076	 & -0.0001	 & 0.0024	 & -0.0075 & 0.0079 \\
5 	& 0.2276	 & -0.0155	 & 0.0147	 & -0.0025	 & 0.0049	 & -0.0157 & 0.0154 \\
6 	& 0.2207	 & -0.0146	 & 0.0079	 & -0.0122	 & 0.0001	 & -0.0190 & 0.0079 \\
8 	& 0.1937	 & -0.0128	 & 0.0133	 & -0.0018	 & 0.0056	 & -0.0129 & 0.0144 \\
9 	& 0.2048	 & -0.0087	 & 0.0075	 & -0.0006	 & 0.0027	 & -0.0087 & 0.0079 \\
10	& 0.2271	 & -0.0164	 & 0.0138	 & -0.0039	 & 0.0001	 & -0.0168 & 0.0138 \\
11	& 0.2152	 & -0.0106	 & 0.0096	 & -0.0012	 & 0.0058	 & -0.0106 & 0.0112 \\
12	& 0.2138	 & -0.0046	 & 0.0053	 & -0.0001	 & 0.0044	 & -0.0046 & 0.0068 \\
13	& 0.2282	 & -0.0149	 & 0.0126	 & -0.0175	 & 0.0006	 & -0.0229 & 0.0126 \\
14	& 0.2217	 & -0.0153	 & 0.0166	 & -0.0066	 & 0.0001	 & -0.0166 & 0.0166 \\
15	& 0.1921	 & -0.0088	 & 0.0085	 & -0.0083	 & 0.0001	 & -0.0120 & 0.0085 \\
16	& 0.195	   & -0.0124	 & 0.0131	 & -0.0081	 & 0.0001	 & -0.0148 & 0.0131 \\
17	& 0.1837	 & -0.0085	 & 0.0086	 & -0.0001	 & 0.0038	 & -0.0085 & 0.0094 \\
18	& 0.208	   & -0.0102	 & 0.0068	 & -0.0057	 & 0.0005	 & -0.0116 & 0.0068 \\
19	& 0.2104	 & -0.0099	 & 0.0093	 & -0.0001	 & 0.0114	 & -0.0099 & 0.0147
\label{table:a520}
\tablenote{The
columns are the same as in Table \ref{table:a2142}. }
\enddata
\end{deluxetable*}

\begin{deluxetable*}{cccccccc}
\centering
\tablewidth{\textwidth}

\tablecaption{
Spectral analysis results of 1RXS\ J0603.3+4214.   }
\tablehead{
\colhead{Region}           & \colhead{$z$}      &
\colhead{$\sigma_{\rm stat_{b}}$} & \colhead{$\sigma_{\rm stat_{u}}$} &
\colhead{$\sigma_{\rm syst_{b}}$} & \colhead{$\sigma_{\rm syst_{u}}$} &
\colhead{$\sigma_{\rm tot_{b}}$}  & \colhead{$\sigma_{\rm tot_{u}}$}}
\startdata
0 &	0.2317 & 	-0.0062	& 0.0065	&  -0.0030	&  0.0004	 &  -0.0068  &  0.0065  \\
1 &	0.2206 & 	-0.0069	& 0.0072	&  -0.0021	&  0.0065	 &  -0.0072  &  0.0097  \\
2 & 0.2347 &    -0.0075	& 0.0119	&  -0.0008	&  0.0071	 &  -0.0075	 &  0.0138  \\
4 &	0.2493 & 	-0.0067	& 0.0091	&  -0.0069	&  0.0001	 &  -0.0096  &  0.0091  \\
5 &	0.2198 & 	-0.0070	& 0.0063	&  -0.0045	&  0.0001	 &  -0.0083  &  0.0063  \\
6 &	0.2254 & 	-0.0108	& 0.0078	&  -0.0055	&  0.0001	 &  -0.0121  &  0.0078  \\
7 &	0.2294 & 	-0.0094	& 0.0066	&  -0.0044	&  0.0021	 &  -0.0103  &  0.0069  \\
8 &	0.2316 & 	-0.0091	& 0.0075	&  -0.0001	&  0.0042	 &  -0.0091  &  0.0085
\label{table:rxsj}
\tablenote{The
columns are the same as in Table \ref{table:a2142}.}
\enddata
\end{deluxetable*}

\begin{deluxetable*}{cccccccc}
\centering
\tablewidth{\textwidth}

\tablecaption{
Spectral analysis results of Abell 2146.   }
\tablehead{
\colhead{Region}           & \colhead{$z$}      &
\colhead{$\sigma_{\rm stat_{b}}$} & \colhead{$\sigma_{\rm stat_{u}}$} &
\colhead{$\sigma_{\rm syst_{b}}$} & \colhead{$\sigma_{\rm syst_{u}}$} &
\colhead{$\sigma_{\rm tot_{b}}$}  & \colhead{$\sigma_{\rm tot_{u}}$}}
\startdata
0 & 0.2292 & -0.0045 & 0.00312787  & -0.0000 & 0.0000 & -0.0045 & 0.0031 \\
1 & 0.2282 & -0.0035 & 0.00311993  & -0.0017 & 0.0018 & -0.0039 & 0.0036 \\
2 & 0.2277 & -0.0038 & 0.00571345  & -0.0000 & 0.0083 & -0.0038 & 0.0100 \\
3 & 0.2282 & -0.0051 & 0.0026654   & -0.0027 & 0.0000 & -0.0057 & 0.0027 \\
4 & 0.2294 & -0.0037 & 0.00436681  & -0.0000 & 0.0036 & -0.0037 & 0.0057 \\
5 & 0.2276 & -0.0065 & 0.00615287  & -0.0001 & 0.0044 & -0.0065 & 0.0076 \\
6 & 0.2325 & -0.0074 & 0.00676923  & -0.0055 & 0.0000 & -0.0092 & 0.0068\\
7 & 0.2354 & -0.0060 & 0.00619396  & -0.0114 & 0.0000 & -0.0129 & 0.0062\\
8 & 0.2473 & -0.0125 & 0.0069235   & -0.0223 & 0.0000 & -0.0255 & 0.0069\\
9 & 0.2436 & -0.0072 & 0.0066689   & -0.0036 & 0.0000 & -0.0080 & 0.0067\\
10 & 0.2277 & -0.0083 & 0.00881529 & -0.0000 & 0.0140 & -0.0083 & 0.0165\\
12 & 0.2372 & -0.0062 & 0.00682796 & -0.0010 & 0.0010 & -0.0063 & 0.0069\\
13 & 0.2326 & -0.0088 & 0.00909447 & -0.0025 & 0.0074 & -0.0091 & 0.0117\\
14 & 0.2256 & -0.0077 & 0.00950732 & -0.0025 & 0.0086 & -0.0081 & 0.0128\\
15 & 0.2495 & -0.0089 & 0.00691656 & -0.0015 & 0.0000 & -0.0090 & 0.0069\\
16 & 0.2226 & -0.0114 & 0.0110643  & -0.0060 & 0.0000 & -0.0129 & 0.0110\\
18 & 0.2505 & -0.0075 & 0.00798135 & -0.0080 & 0.0000 & -0.0110 & 0.0080

\label{table:a2146}
\tablenote{The
columns are the same as in Table \ref{table:a2142}.}
\enddata
\end{deluxetable*}

\begin{deluxetable*}{cccccccc}
\centering
\tablewidth{\textwidth}

\tablecaption{
Spectral analysis results of Abell 1689.  }
\tablehead{
\colhead{Region}           & \colhead{$z$}      &
\colhead{$\sigma_{\rm stat_{b}}$} & \colhead{$\sigma_{\rm stat_{u}}$} &
\colhead{$\sigma_{\rm syst_{b}}$} & \colhead{$\sigma_{\rm syst_{u}}$} &
\colhead{$\sigma_{\rm tot_{b}}$}  & \colhead{$\sigma_{\rm tot_{u}}$}}
\startdata
0	 & 0.1767	 & -0.0061	 & 0.0048	 & -0.0001	 & 0.0037	 & -0.0061 & 0.0060 \\
1	 & 0.1774	 & -0.0047	 & 0.0061	 & -0.0014	 & 0.0001	 & -0.0049 & 0.0061 \\
2	 & 0.1864	 & -0.0103	 & 0.0109	 & -0.0001	 & 0.0066	 & -0.0103 & 0.0127 \\
3	 & 0.1774	 & -0.0065	 & 0.0049	 & -0.0019	 & 0.002	 & -0.0067 & 0.0052 \\
4	 & 0.1675	 & -0.0074	 & 0.0066	 & -0.0019	 & 0.0001	 & -0.0076 & 0.0066 \\
5	 & 0.1859	 & -0.0045	 & 0.0038	 & -0.0001	 & 0.0016	 & -0.0045 & 0.0041 \\
6	 & 0.1895	 & -0.0084	 & 0.0087	 & -0.0001	 & 0.0385	 & -0.0084 & 0.0394 \\
7	 & 0.1859	 & -0.006	   & 0.0078	 & -0.0001	 & 0.0041	 & -0.0060 & 0.0088 \\
8	 & 0.1826	 & -0.0087	 & 0.0073	 & -0.0001	 & 0.005	 & -0.0087 & 0.0088 \\
9	 & 0.1847	 & -0.0047	 & 0.0071	 & -0.0001	 & 0.0031	 & -0.0047 & 0.0077
\label{table:a1689}
\tablenote{The
columns are the same as in Table \ref{table:a2142}. }
\enddata
\end{deluxetable*}

\begin{deluxetable*}{cccccccc}
\centering
\tablewidth{\textwidth}

\tablecaption{
Spectral analysis results of Abell 1835.  }
\tablehead{
\colhead{Region}           & \colhead{$z$}      &
\colhead{$\sigma_{\rm stat_{b}}$} & \colhead{$\sigma_{\rm stat_{u}}$} &
\colhead{$\sigma_{\rm syst_{b}}$} & \colhead{$\sigma_{\rm syst_{u}}$} &
\colhead{$\sigma_{\rm tot_{b}}$}  & \colhead{$\sigma_{\rm tot_{u}}$}}
\startdata
0	   & 0.2524	 & -0.0031	 & 0.0041	 & -0.0024	 & 0.0001	 & -0.0039 & 0.0041 \\
1	   & 0.2547	 & -0.0024	 & 0.003	 & -0.0001	 & 0.0025	 & -0.0024 & 0.0039 \\
2	   & 0.2478	 & -0.003	   & 0.0052	 & -0.0001	 & 0.0021	 & -0.0030 & 0.0056 \\
3	   & 0.2474	 & -0.003	   & 0.0027	 & -0.002	   & 0.0003	 & -0.0036 & 0.0027 \\
4	   & 0.2472	 & -0.0048	 & 0.006	 & -0.002	   & 0.0018	 & -0.0052 & 0.0062 \\
5	   & 0.2408	 & -0.0028	 & 0.0045	 & -0.0001	 & 0.0052	 & -0.0028 & 0.0068 \\
6	   & 0.2538	 & -0.01	   & 0.0065	 & -0.0057	 & 0.0016	 & -0.0115 & 0.0066 \\
7	   & 0.2519	 & -0.0046	 & 0.0064	 & -0.0028	 & 0.0001	 & -0.0053 & 0.0064 \\
8	   & 0.2486	 & -0.006	   & 0.0052	 & -0.0001	 & 0.0046	 & -0.0060 & 0.0069 \\
9	   & 0.2372	 & -0.0065	 & 0.0073	 & -0.0002	 & 0.007	 & -0.0065 & 0.0101 \\
10	 & 0.2496	 & -0.0068	 & 0.0078	 & -0.0019	 & 0.0023	 & -0.0070 & 0.0081
\label{table:a1835}
\tablenote{The
columns are the same as in Table \ref{table:a2142}. }
\enddata
\end{deluxetable*}

\end{document}